\documentclass[11pt]{article}
\usepackage{jheppub}

\usepackage[usenames,dvipsnames,table]{xcolor}
\usepackage{graphicx,amsmath,amssymb,multirow,array,bm,mathrsfs}
\usepackage{epsf,amsfonts}
\usepackage[numbers,sort&compress]{natbib}

\newcommand{\prn}[1]{\left ( #1 \right )}
\newcommand{\brk}[1]{\left [ #1 \right ]}
\newcommand{\bigbr}[1]{\Bigl\{ #1 \Bigr\} }
\newcommand{\half}{\frac{1}{2}}
\newcommand{\quarter}{\frac{1}{4}}
\newcommand{\rad}{r}
\newcommand{\form}[1]{\bm{#1}}

\newcommand{\hodge}{{}^\star}

\newcommand{\hodgeCFT}{ {\hodge}^{\text{\tiny{CFT}}}}

\newcommand{\fL}{\form{L}}
\newcommand{\Sp}{\Sigma}

\newcommand{\TMax}{\mathrm{T}_{_M}}

\newcommand{\JH}{\mathrm{J}_{_H}}
\newcommand{\fJH}{\form{\mathrm{J}}_{_H}}

\newcommand{\SpH}{\mathrm{\Sp}_{_H}}
\newcommand{\fSpH}{\form{\mathrm{\Sp}}_{_H}}

\newcommand{\THall}{\mathrm{T}_{_H}}
\newcommand{\fTHall}{\form{\mathrm{T}}_{_H}}

\newcommand{\fu}{\form{u}}

\newcommand{\fomega}{\form{\omega}}

\newcommand{\fA}{\form{A}}
\newcommand{\fF}{\form{F}}
\newcommand{\fB}{\form{B}}

\newcommand{\fGamma}{\form{\Gamma}}
\newcommand{\fR}{\form{R}}

\newcommand{\fP}{{\form{\mathcal{P}}}}
\newcommand{\ICS}{{\form{I}}_{CS}}

\newcommand{\PhiT}{\Phi_{_T}}
\newcommand{\rH}{r_{_H}}
\newcommand{\tV}{\text{(\tiny{V})}}

\newcommand{\bea}{\begin{eqnarray}}
\newcommand{\eea}{\end{eqnarray}}
\newcommand{\be}{\begin{equation}}
\newcommand{\ee}{\end{equation}}
\newcommand{\tr}{\mbox{tr}}
\newcommand{\R}[1]{{}_{_R}\form{#1}}
\newcommand{\F}[1]{{}_{_F}\form{#1}}
\newcommand{\B}[1]{{}_{_{\mathcal{B}}}\form{#1}}
\newcommand{\fchi}{\form{\chi}}
\newcommand{\oR}[1]{\overline{{}_{_R}\form{#1}}}

\def\Einstein{R_{ab}-\frac{1}{2} G_{ab}R}

\title{Holographic Thermal Helicity}

\author[a,b]{Tatsuo Azeyanagi,}
\author[b,c]{R. Loganayagam,}
\author[b]{Gim Seng Ng}
\author[b,d]{and Maria J. Rodriguez}

\affiliation[a]{
D\'{e}partement de Physique, Ecole Normale Sup\'{e}rieure, CNRS, 24 rue Lhomond,
75005 Paris, France.}
\affiliation[b]{Center for the Fundamental Laws of Nature, Harvard University, Cambridge, MA 02138, USA.}
\affiliation[c]{Junior Fellow, Harvard Society of Fellows, Harvard University, Cambridge, MA 02138,USA.}
\affiliation[d]{Institut de Physique Th\'eorique, CEA Saclay, CNRS URA 2306 , F-91191 Gif-sur-Yvette, France.}

\emailAdd{tatsuo.azeyanagi@phys.ens.fr}
\emailAdd{nayagam@gmail.com}
\emailAdd{nggimseng@post.harvard.edu}
\emailAdd{mjrodri@physics.harvard.edu}

\abstract{
We study the thermal helicity, defined in
\href{http://arxiv.org/abs/1211.3850}{\tt arXiv:1211.3850 [hep-th]}, 
of a conformal field theory with anomalies in the context of AdS$_{2n+1}$/CFT$_{2n}$.
To do so, we consider large charged rotating AdS black holes in the Einstein-Maxwell-Chern-Simons theory with a negative cosmological constant
using fluid/gravity expansion.
We compute the anomaly-induced current and stress tensor of the dual CFT in leading order of the fluid/gravity derivative expansion
and show their agreement with the field theoretical replacement rule for the thermal helicity.
Such replacement rule is reflected in the bulk by new replacement rules obeyed by the Hall currents around the black hole.}

\begin{document}

\maketitle

\section{Introduction}
Recently, a thermal observable called the thermal helicity  was defined  for relativistic quantum field theories
in even-dimensional spacetime\cite{Loganayagam:2012zg} and a variety of evidence now suggests that this quantity
is closely related to anomalies. In a QFT$_{d}$ (with $d=2n$), thermal helicity is the thermal expectation value of products
of all $(n-1)$ mutually commuting angular momentum operators $\hat{\mathfrak{L}}_{ij}$
and the spatial momentum operator $ \mathcal{\hat{P}}_{i}$ in the remaining direction. More explicitly, 
let $\hat{\mathfrak{L}}_{2k-1,2k}$ denote the angular momentum operators in the $(x^{2k-1},x^{2k})$-plane 
for each $k=1,\ldots,(n-1)$ in $\mathbb{R}^{2n-1,1}$ where the QFT$_{2n}$ lives. 
Then one can define the thermal helicity as:
\be
\left\langle
\prn{\prod_{k=1}^{n-1} \hat{\mathfrak{L}}_{2k-1,2k} }
 \hat{\mathcal{P}}_{2n-1}
\right\rangle%
=
\langle \hat{\mathfrak{L}}_{12} \hat{\mathfrak{L}}_{34}\ldots \hat{\mathfrak{L}}_{2n-3,2n-2} \hat{\mathcal{P}}_{2n-1} \rangle\,.
\ee
where $\langle \ldots \rangle $ is the expectation value in the thermal ground state
with temperature $T$ and chemical potential $\mu$. 

Thermal helicity per unit volume is conveniently written in terms of another function of $T$ and $\mu$,
denoted by $\mathfrak{F}_{Anom}[T,\mu]$ :
\begin{equation}\label{eq:heliF}
\begin{split}
\frac{1}{\text{Vol}_{2n-1}} \langle \hat{\mathfrak{L}}_{12} \hat{\mathfrak{L}}_{34}\ldots \hat{\mathfrak{L}}_{2n-3,2n-2} \hat{\mathcal{P}}_{2n-1} \rangle
\equiv -(n-1)! (2T)^{n-1} \mathfrak{F}_{anom}[T,\mu]\, , 
\end{split}
\end{equation} where $\text{Vol}_{2n-1}$ is the spatial volume.
Furthermore, Ref.~\cite{Loganayagam:2012zg}  conjectured a deep relation between the thermal helicity per unit volume
and the anomaly polynomial $\fP_{QFT}[\fF,\fR]$ which can be written in terms of a remarkable replacement rule  
\be\label{eq:RepRuleThermalHelicityPontryagin}
\mathfrak{F}_{Anom}[T,\mu]=\fP_{QFT}\left[
\fF\rightarrow \mu,\quad
p_1(\fR)\rightarrow -T^2,\quad
p_{k> 1}(\fR)\rightarrow 0
\right]\, , 
\ee where $p_k(\fR)$ is the $k$-th Pontryagin class (see Ref.~\cite{Loganayagam:2012zg} for definitions and conventions).
The equation above has to be interpreted thus as :  to obtain $\mathfrak{F}_{Anom}[T,\mu]$, we take the anomaly polynomial 
written in terms of the gauge field strength 2-form $\fF$ and various Pontryagin classes of Riemann curvature 2-form
$\fR$, replace $\fF$ with the corresponding chemical potential $\mu$ and the first Pontryagin class with $-T^2$ and set the  higher Pontryagin classes to zero. Thus, the 
replacement rule completely fixes the thermal helicity per unit volume of a QFT to be a homogeneous polynomial
in $T$ and $\mu$ with the polynomial coefficients directly related to the anomaly coefficients of the theory. 
For our purposes, it is useful to rewrite the replacement rule in terms of traces of Riemann products:
\be\label{eq:RepRuleThermalHelicity}
\mathfrak{F}_{Anom}[T,\mu]=\fP_{QFT}\left[
\fF\rightarrow \mu,\quad
\tr\ \fR^{2k}\rightarrow 2(2\pi T)^{2k}
\right]\, . 
\ee

This conjecture suggests that thermal helicity is a protected quantity when we turn on finite temperature and chemical potential
in a relativistic field theory. This surprising result is not obvious at all even in thermal perturbation theory
and probably hints at deeper aspects of thermal field theory. For example, for $d=2$ CFTs, the replacement rule is a 
corollary of the Cardy formula and its significance lies in the fact that it is a weaker corollary of the Cardy formula
which nevertheless generalizes to the larger class of even dimensional QFTs. Furthermore, this conjectured rule is closely related to
and is motivated by advances in the role of anomalies in finite temperature and chemical potential 
\cite{Bhattacharyya:2007vs,Erdmenger:2008rm,Banerjee:2008th,Torabian:2009qk,Son:2009tf,Kharzeev:2009p,
Lublinsky:2009wr,Neiman:2010zi,Bhattacharya:2011tra,Kharzeev:2011ds,Loganayagam:2011mu,Neiman:2011mj,Dubovsky:2011sk,
Kimura:2011ef,Lin:2011aa,Manes:2012hf,Banerjee:2012iz,Jensen:2012jy,Jain:2012rh,Valle:2012em,Jensen:2012kj,
Golkar:2012kb,Bhattacharyya:2012xi,Loganayagam:2012zg,Valle:2013aia,Bhattacharyya:2013ida,Megias:2013uua,Jensen:2013kka}.
A formal argument made by the authors of \cite{Jensen:2012kj} at the level of equilibrium partition function in
$d=2$ and $d=4$ spacetime dimensions can be directly adopted into an argument for the thermal helicity conjecture. 
Recently, this formal argument has been extended to higher dimensions\cite{Jensen:Nov2013}.

Until now, various attempts at explicit diagrammatic calculations in interacting field theories to study this result have faced 
difficulties because of the technical complexities of thermal perturbation theory\cite{Golkar:2012kb,Hou:2012xg}.
As shown by \cite{Landsteiner:2011cp,Loganayagam:2012pz,Loganayagam:2012zg}, even in free theories, 
the replacement rule in Eq.~\eqref{eq:RepRuleThermalHelicity} holds true because of somewhat mysterious cancellations
between various Fermi-Dirac and Bose-Einstein integrals.\footnote{The subtlety of these cancellations is demonstrated 
by free field theories with chiral gravitini where such cancellations
no longer happen leading to a violation of the replacement rule\cite{Loganayagam:2012zg}.}Given these difficulties,
it is natural to turn to holography which provides an explicit testing ground for this conjecture. In particular, 
we seek general insights as to how one should understand the origins of the replacement rule and holography is a 
promising avenue to look for such insights.  

The main aim of this paper is then to understand how a rule of the form given in Eq.~\eqref{eq:RepRuleThermalHelicity} 
arises in a holographic setting. As emphasized in \cite{Loganayagam:2012zg}, often the most convenient way to compute
thermal helicity is to place the theory under question  on a sphere, turn on chemical potentials for angular 
momenta on the sphere and then study its free energy in the limit where the rotational chemical potentials are small
and the radius of the sphere is large. The AdS dual of this procedure is to study large charged rotating AdS black holes
and their thermodynamics. To do this, we will begin by adopting the familiar tools of fluid/gravity correspondence
for the problem at hand (for some of the related works which study fluid/gravity correspondence in the context
of anomalies, see \cite{Bhattacharyya:2007vs,Erdmenger:2008rm,Banerjee:2008th,Torabian:2009qk,Kharzeev:2011ds,
Landsteiner:2011iq,Chapman:2012my,Megias:2013uua}). 

\subsection*{Basic setup}
We will begin by describing the basic setup on the AdS side. In holography, conformal field theories
with anomalies in $d=2n$ spacetime dimensions are dual to gravitational theories with Chern-Simons terms
in $d+1=2n+1$ spacetime dimensions (with the bulk Chern-Simons coefficients being determined by the anomaly coefficients of the CFT as we will describe below).  Thus we study the simplest class of gravitational
theories with  Chern-Simons terms : the Einstein-Maxwell-Chern-Simons theory with a negative cosmological constant.
In every odd spacetime dimension, we allow the most general gauge/gravitational/mixed
Chern-Simons term that can be formed out of a single Maxwell field $\fA$ and the Christoffel connection $\fGamma$.
Realistic holographic examples with Chern-Simons terms often also contain other higher derivative 
corrections and we believe many of our arguments are sufficiently general to apply to that general case. 
Thus, our motivation in considering  Einstein-Maxwell-Chern-Simons theory
is mainly to provide the simplest (if somewhat contrived) context in which our arguments can be 
clearly articulated. 

We thus focus on the black hole solutions of the Einstein-Maxwell-Chern-Simons system. The Lagrangian is of the form 
\begin{equation}\label{eq:action}
\begin{split}
\fL &=d^{d+1} x \,\sqrt{-G} \brk{ \frac{1}{16\pi G_N} \prn{R-2\Lambda} -\frac{1}{4g^2_{_{EM}}}  F_{ab} F^{ab} }+  \ICS\, ,\\
\end{split}
\end{equation}
where  the Chern-Simons part of the Lagrangian is denoted as $\ICS$. The fields are subject to the equations of motion following from (\ref{eq:action})
\begin{equation}\label{eq:EinsteinMaxwellEq}
\begin{split}
R_{ab}-\frac{1}{2}\prn{R-2\Lambda} G_{ab} &=8\pi G_{_N} \brk{ (\TMax)_{ab} +  (\THall)_{ab} }\ ,\\
\nabla^b F_{ab} &= g^2_{_{EM}} (\JH)_a\,,
\end{split}
\end{equation}
where $G_{ab}$ is an asymptotically AdS$_{d+1}$ metric with $d=2n$, $F_{ab}$ is the Maxwell field strength, $G_{_N}$  
and $g_{_{EM}}$ are the Newton and Maxwell couplings respectively. The cosmological constant is taken to be negative and is given by $\Lambda\equiv  -d(d-1)/2$ where the AdS radius is set to one. For future use, recall that the normalized $U(1)$ gauge coupling $\kappa_q$ is related to magnetic permeability
 $g^2_{_{EM}}$ by $16\pi G_{_N}/g^2_{_{EM}} =  \kappa_q\ (d-1)/(d-2)\,$.\footnote{It is sometimes convenient to normalize the gauge field such that  $\kappa_q\ (d-1)/(d-2) =1 $ which is equivalent to setting $g^2_{_{EM}} = 16 \pi G_{_N}$.} 
We now turn to the Einstein and Maxwell sources. The Maxwell energy-momentum tensor $ (\TMax)^{ab}$ becomes
\begin{equation}
\begin{split}
 (\TMax)^{ab}
&\equiv
 \frac{\kappa_q}{16\pi G_N}\ \frac{d-1}{d-2}\  \brk{ F^{ac}F^b{}_c-\quarter G^{ab} F_{cd} F^{cd} }\, ,\\
\end{split}
\end{equation}
whereas $(\THall)_{ab}$ and $(\JH)_a$ are the energy-momentum tensor and the Maxwell charge current
obtained by varying the Chern-Simons part of the action. We will call these currents as Hall currents since
in the case of Hall insulators, the only terms in the generating function is just the Chern-Simons term.
Thus, we can think of the system under study as made of three components : metric (with a negative
cosmological constant), a Maxwell field and a Hall insulator coupled to them via the currents $(\THall)_{ab}$ and $(\JH)_a$.

The bulk Hall currents on the AdS$_{d+1}$ side encode the anomaly coefficients of the CFT$_d$ 
and play a central role in the rest of our paper. Therefore, we will begin by describing the
basic structure of these currents that we will need for the rest of this introduction. 
Let $\ICS[\fA,\fF,\fGamma,\fR]$ be the Chern-Simons $(d+1)$-form appearing in the bulk 
Lagrangian density (\ref{eq:action}). The bulk Hall currents  are more conveniently written in terms of
the formal $(d+2)$ form $\fP_{CFT}[\fF,\fR]= d\ICS$. More concretely, we can define 
the spin Hall current $(\SpH)^{cb}{}_a$ and the charge Hall current $(\JH)^c$ corresponding to $\ICS$ via\footnote{
We refer the reader to Ref.~\cite{Jensen:2013kka} for our conventions for spin currents, differential forms etc.
Note that our spin current $(\SpH)^{cb}{}_a$ is twice the tensor $(L_{_H})^{cb}{}_a$ ,viz., 
$(\SpH)^{cb}{}_a=2(L_{_H})^{cb}{}_a$. 
We can convert our convention to the one commonly used in the gravity literatures
by replacing $\hodge\form{U}$ with $(-1) \hodge\form{U}$ (for every bulk form $\form{U}$) and sending the bulk $\varepsilon_{abcd\ldots}$ to
$- \varepsilon_{abcd\ldots}$, while keeping $\hodgeCFT$ unchanged. See Ref.~\cite{Azeyanagi:2013} for more detail. 
}
\begin{equation}
\begin{split}\label{eq:defsHall}
(\hodge \fSpH)^b{}_a &\equiv (\SpH)^{cb}{}_a\  \hodge dx_c  \equiv -2 \frac{\partial \fP_{CFT}}{ \partial \fR^a{}_b}\,, \\
\hodge \fJH &\equiv (\JH)^c\  \hodge dx_c  \equiv -\frac{\partial \fP_{CFT}}{ \partial \fF}.  \\
\end{split}
\end{equation}
The energy-momentum tensor associated with the Hall currents can in turn be written in terms of the 
spin-Hall currents : $(\THall)^{ab}=  \nabla_c (\SpH)^{(ab)c}$. 

We now turn to describing the procedure that we adopt to solve the Einstein-Maxwell-Chern-Simons system. 
We begin with the CFT$_{2n}$ thermal state living in flat spacetime in the absence of any external 
electric or magnetic fields. This thermal state is dual to the Anti de Sitter Reissner-Nordstrom (AdS-RN)
solution of the above equations.\footnote{We note that the Hall currents evaluate to zero in the AdS-RN solution. 
Hence, the AdS-RN solution is an exact solution of the Einstein-Maxwell-Chern-Simons system.}  In ingoing Eddington-Finkelstein co-ordinates, we have
\begin{equation}
\begin{split}
ds^2 &= 2\ dt\ dr - r^2 f(r,m,q)\ dt^2\ + r^2\ d\vec{x}_{d-1}^2\ ,\qquad
\fA = - \Phi(r,q) \ dt\, , 
\end{split}
\end{equation}
where
\begin{equation}
\begin{split}
f &\equiv 1- \frac{m}{r^d} + \half \kappa_q \frac{q^2}{r^{2(d-1)}} \ ,\qquad
\Phi \equiv \frac{q}{r^{d-2}}\, . 
\end{split}
\end{equation}
In order to study the replacement rule, we need to construct an explicit  gravitational solution in the fluid/gravity 
derivative expansion starting from a boosted counterpart of the solution above and compute
directly the anomaly-induced currents and stress tensor. 

More concretely, to compute the helicity of this thermal state, we turn on a small rotation on
the thermal state letting $u^\mu$ correspond to a rotating fluid configuration.
This is dual to bulk solutions in the fluid/gravity expansion to the appropriate orders
(which will be made precise later on). The thermal helicity is contained in terms proportional 
to the following pseudo-vector appearing in the fluid/gravity expansion of metric and the gauge field\footnote{
Our convention for $\varepsilon$ will be given in Sec. \ref{sec:sol}.
}
\begin{equation}\label{eq:ThermalHelicity}
V^\mu \equiv \varepsilon^{\mu\nu\lambda_1 \sigma_1 \lambda_2 \sigma_2 \ldots \lambda_{n-1} \sigma_{n-1}}
u_\nu (\nabla_{\lambda_1} u_{\sigma_1}) (\nabla_{\lambda_2} u_{\sigma_2}) \ldots (\nabla_{\lambda_{n-1}} u_{\sigma_{n-1}})\, ,
\end{equation}
or in terms of forms $\hodgeCFT \form{V} = \fu \wedge (d\fu)^{n-1} \equiv \fu \wedge (2\fomega)^{n-1}$ where
$\fu\equiv u_{\mu}dx^{\mu}$ is the velocity one-form and 
$\fomega \equiv (1/2)d\fu = (1/2)\omega_{\mu\nu}dx^{\mu}\wedge dx^{\nu}$ is the vorticity 2-form with 
$\omega_{\mu\nu}\equiv P_\mu{}^\alpha P_\nu{}^\beta (\partial_\alpha u_\beta-\partial_\beta u_\alpha)/2$.
Here,  $P^\mu{}_\nu\equiv \delta^\mu_{\,\nu}+u^\mu u_\nu$ is the transverse projector which projects a vector to 
its component normal to $u^\mu$ and $\omega_{\mu\nu}$ is the vorticity of the rotating fluid configuration .
Thus, the relevant fluid/gravity solution takes the form
 \begin{eqnarray}
ds^2 &=& -2 u_\mu dx^\mu\ \otimes_{sym} dr - r^2 f(r,m,q)\ u_\mu u_\nu dx^\mu \otimes_{sym} dx^\nu\ 
+ r^2\ P_{\mu\nu} dx^\mu \otimes_{sym} dx^\nu +\ldots \nonumber \\
&& +
2 g_{_V}(r,m,q) u_\mu V_\nu\ dx^\mu\ \otimes_{sym} dx^\nu + \ldots \nonumber \, , \\
\fA &=&  \Phi(r,q)\ u_\mu\ dx^\mu + \ldots
\ldots+ a_{_V}(r,m,q)   V_\mu\ dx^\mu +\ldots\ \, , 
\end{eqnarray}
where we will find useful for later to define the function $\PhiT \equiv \half r^2 \frac{df}{dr} $.
The functions $g_{_V}(r,m,q)$ and $a_{_V}(r,m,q)$ are obtained by solving Einstein-Maxwell-Chern-Simons
equations and are proportional to the Hall currents. They describe how the Hall currents 
dress the original AdS Kerr-Newman solution. 

Once these solutions are constructed, we compute the  currents and stress tensor of the dual CFT.\footnote{ We note that computation of   the  currents and stress tensor are complicated by 
the higher derivative nature of Chern-Simons terms in the bulk Lagrangian. The procedure for
computing dual currents and stress tensor (say by employing counterterms) is not well-understood
for higher derivative actions and this problem is worsened by the fact that Chern-Simons
terms are not covariant. We will address this issue in a subsequent paper \cite{Azeyanagi:2013} in sufficient detail. For purposes
of this paper, we will content ourselves with directly using the results of that subsequent paper 
with some heuristic motivation to guide the reader.} In fluid/gravity expansion, the 
leading order effect of the cloud of Hall currents that envelops the blackhole  
is to shift the total energy, angular momentum and the electric charge of the rotating blackhole
under question. These leading order shifts are all that is needed to compute the shift in the free 
energy on the sphere and in turn, the thermal helicity in the zero rotation limit. 
In the rest of this introduction, we will describe how the replacement rule emerges out
of our computations followed by an outline of the paper.

\subsection*{Replacement rule in AdS}
We will now describe the elegant way in which holography realizes the replacement rule.
The CFT replacement rule in \eqref{eq:RepRuleThermalHelicity} is explained by the following
insight which follows from our computations  : \textit{the CFT replacement rule  gets 
reflected in the bulk by the replacement rules obeyed by the Hall currents around the blackhole}, i.e.,
we find that Hall currents evaluated on the blackhole background assume a very simple form which 
holds for arbitrary Chern-Simons terms. Once this fact is realized, a direct application of 
AdS/CFT dictionary then translates this result into the required CFT replacement rule.
We now describe a set of rules which summarize our computations.

The significance of these rules is the following : we have derived them within a particular holographic setup
in  the classical gravity limit. However, given the way these rules are
closely linked to the  CFT replacement rule (which is a non-perturbative statement), we expect 
our rules to be robust features in AdS more generally. 

The relevant part of the currents for formulating such rules are the components 
$(\JH)_\mu$ and $(\THall)_{r\mu}$ that describe the Hall flow of the radial momentum
and the Abelian charge along the boundary directions. We find that, in the fluid/gravity expansion,
the leading order contribution to these flows  are along the pseudovector $V_\mu$ introduced
in \eqref{eq:ThermalHelicity}, viz., $(\JH)_\mu =\JH^\tV(r,m,q) V_\mu+\ldots $ and 
$(\THall)_{r\mu}=\THall^\tV(r,m,q) V_\mu+\ldots$. Similarly, the component $(\SpH)_{(r\mu)}{}^r$
of the spin Hall current has an expansion $(\SpH)_{(r\mu)}{}^r=\SpH^\tV(r,m,q) V_\mu+\ldots $.
These components $\JH^\tV(r,m,q),\THall^\tV(r,m,q)$ and $\SpH^\tV(r,m,q)$ are apriori general functions
of their arguments which need to be computed case by case. The intent of the rules that we give 
below is to let us simply write down these functions by looking at the Chern-Simons Lagrangian 
under question.

To formulate these rules, we begin by focusing on the following two functions associated with AdS-RN blackbrane
solution :
\begin{equation}
\begin{split}
\PhiT &\equiv \half r^2 \frac{df}{dr} = \frac{1}{2r^{d-1}} \brk{ m d- \kappa_q(d-1)\frac{q^2}{r^{d-2}} }\quad \text{and}\\
\Phi &\equiv \frac{q}{r^{d-2}}.
\end{split}
\end{equation}
These functions satisfy $\PhiT(r=\rH) =2\pi T$ and $\Phi(r=\rH) =\mu $ where
$\rH$ is the radius of the outer horizon, $T$ is the Hawking temperature and
$\mu$ is the chemical potential of the black brane. In the next step, we define 
the functional $\mathbb{G}^\tV$ via the following rule 
\begin{equation}\label{eq:repwB}
\begin{split}
\mathbb{G}^\tV 
\equiv\fP_{CFT}\left[
\fF\rightarrow \Phi;\quad
\tr[ \fR^{2k}] \rightarrow 2\PhiT^{2k}\right].\
\end{split}
\end{equation}
Then, we have
\begin{equation}
\begin{split}
\JH^\tV & =  \frac{1}{r^{d-3}}\frac{d}{dr}
\frac{\partial \mathbb{G}^\tV }{\partial \Phi}\, ,  \\
 \THall^\tV & 
 =\frac{1}{r^{d-2}}\frac{d}{dr}\left[
r^{d-2}\SpH^\tV
\right]
  = -\frac{1}{2r^{d-1}}\frac{d}{dr}r^2\frac{d}{dr}\frac{\partial \mathbb{G}^\tV }{\partial \PhiT},
 \end{split} 
\end{equation}
where $\SpH^\tV$  is defined by $(\SpH)_{(r\mu)}{}^r \equiv  \SpH^\tV V_\mu$ and satisfies
 \be\label{eq:sigmaV}
 \SpH^\tV=-\frac{1}{2 r^{d-2}} \frac{d}{dr}\left[ r\frac{\partial  \mathbb{G}^{\tV}
}{\partial \Phi_T}
\right].
 \ee
These expressions then describe how the Hall currents are turned on by 
charged rotating blackhole background. To see how this gets reflected in the 
dual CFT, we first work out the change in the metric and gauge field (the functions
$g_{_V}$ and $a_{_V}$) due to these Hall currents and then use them to compute 
how they shift the energy/angular momentum/charge of the blackhole. We then use these
shifts to derive the thermal helicity that we are after. The bulk replacement rules
above then get translated in the CFT to the field theoretical replacement rule.

\subsection*{Outline}

In Sec.~\ref{sec:sol}, we briefly 
describe the equations of motion of the Einstein-Maxwell-Chern-Simons action with a negative cosmological constant.
Then, we obtain expressions for the anomaly-induced currents and stress tensors in terms of Hall sources derived 
from the Chern-Simons terms in the equations of motion. Next, in Sec.~\ref{sec:abelianCS} we present the 
computation of the source term in the Abelian Chern-Simons case and then proceed to Sec.~\ref{sec:GeneralCS}
to compute the general sources. As described in the introduction, the general structure of the sources
satisfies a new set of replacement rules deduced from the anomaly polynomial. Having obtained the sources
explicitly, in Sec.~\ref{sec:compTabJa} we finally complete the computation of the currents and stress tensor.
The results indeed verify the replacement rules in Eq.~(\ref{eq:RepRuleThermalHelicity}).
We then conclude the paper with a brief discussion on the implications of our results as well as future directions.

In the appendices, we provide details of computations deriving the results in the main text. In particular,
in Appendix \ref{sec:AdSKerrNewman2ndOrder}, we discuss the AdS$_{d+1}$-Kerr-Newman metric which we have
computed up to second order in fluid/gravity expansion. Then, in Appendix \ref{sec:productRiemanns}, we 
examine in detail various products (and traces of products) of Riemann curvature
two-forms evaluated on our solutions. These results are then used in Appendices \ref{sec:MaxwellSources}
and \ref{sec:EinsteinSources} to prove the general structure of the Maxwell and Einstein sources respectively.

\section{Solution of Einstein-Maxwell equations}
\label{sec:sol}

We write the dual gravitational solution to this rotating fluid configuration as
\begin{eqnarray}\label{eq:fluidGravityAnsatz}
ds^2 &=& -2 u_\mu dx^\mu\ \otimes_{sym} dr - r^2 f(r, m, q)\ u_\mu u_\nu dx^\mu \otimes_{sym} dx^\nu\ + r^2\ P_{\mu\nu} dx^\mu \otimes_{sym} dx^\nu +\ldots \nonumber \\
&&\qquad +
2 g_{_V}(r ,m , q) u_\mu V_\nu\ dx^\mu\ \otimes_{sym} dx^\nu + \ldots \, , \nonumber \\
\fA &=&  \Phi(r, q)\ u_\mu\ dx^\mu +
\ldots+ 
a_{_V}(r, m, q)   V_\mu\ dx^\mu +\ldots \, , 
\end{eqnarray}
where $\fF=d\fA$. We note that throughout this paper we shall employ the Latin letters $a,b,\ldots$ 
to denote indices running through all spacetime coordinates and the Greek letters $\mu,\nu,\ldots$ 
for all the boundary coordinates. In particular, we use the coordinates $x^a=\left\{r,x^\mu\right\}=\left\{r,t,\ldots \right\}$
with the convention for the bulk epsilon tensor being $\varepsilon_{rt\ldots }=-\sqrt{-G}$ in a $(d+1)$-dimensional bulk spacetime
with bulk metric $G_{ab}$. We also refer the readers to Appendix \ref{sec:Riemanns0thmetric} for the raising/lowering 
of the Greek indices. We note that $f,g_V, \Phi$ and $a_V$ depend on $r$ but not on the CFT$_{2n}$ coordinates $x^{\mu}$,  
while the rest of the variables only depend on $x^{\mu}$.
The pseudovector $V_\mu$ is of the form $\hodgeCFT \form{V} = \fu \wedge (d\fu)^{n-1}$ 
which appears in the anomaly induced transport at the $(n-1)$-th order in the derivative expansion.
The terms preceding the first $(\ldots)$ give boosted RN black brane metric about 
which we will perform the fluid/gravity expansion. We have dropped all the subsequent terms 
except the anomaly induced transport terms.

The Einstein-Maxwell contributions in the equations of motion is then evaluated to give the homogeneous parts
\begin{equation}
\begin{split}
&\frac{1}{8\pi G_{_N}}\bigbr{-\left(\Einstein\right)+\half d(d-1) G_{ab}+ 8\pi G_{_N}
\prn{\TMax}_{ab} }  dx^a \otimes_{sym} dx^b\\
&\quad = \ldots+ 
 \frac{1}{8\pi G_{_N} } \times \frac{1}{2r^{d-1}}\frac{d}{dr} \brk{ r^{d+1} \frac{d}{dr}\prn{\frac{g_{_V}}{r^2}}
- \kappa_q  (d-1)\ q a_{_V}}  \\
&\qquad \qquad\qquad \times 2\prn{dr+ r^2 f\ u_\mu dx^\mu} \otimes_{sym} V_\nu\ dx^\nu +\ldots \\
\end{split}
\end{equation}
and
\begin{equation}
\begin{split}
&-\frac{1}{g^2_{_{EM}}}(\nabla^b F_{ab}) dx^a \\
&\qquad = \ldots+
 \frac{1}{g^2_{_{EM}} r^{d-3}}\frac{d}{dr} \brk{ r^{d-1} f\ \frac{d a_{_V}}{dr}-(d-2) q \prn{\frac{g_{_V}}{r^2}} }
 V_\mu\ dx^\mu +\ldots\\
&\qquad = \ldots+ \frac{1}{8\pi G_{_N} } \times  \kappa_q\ \frac{d-1}{d-2}\times
\frac{1}{2r^{d-3}}\frac{d}{dr} \brk{ r^{d-1} f\ \frac{d a_{_V}}{dr}-(d-2) q \prn{\frac{g_{_V}}{r^2}} }
 V_\mu\ dx^\mu +\ldots \, ,\\
\end{split}
\end{equation}
where we have exhibited only the anomaly induced transport terms. For the metric and gauge field in \eqref{eq:fluidGravityAnsatz}
to be a solution, we have to choose the functions $\{g_{_V},a_{_V}\}$ such that the
terms above exactly cancel the contribution from the Hall currents arising from the Chern-Simons terms. We shall solve for $\{g_{_V},a_{_V}\}$ explicitly in the next step.

If we parametrize the contribution from the Hall currents as
\begin{equation}
\begin{split}
\prn{\THall}_{ab}   dx^a \otimes_{sym} dx^b &= 
\THall^\tV\  2\prn{dr+ r^2 f\ u_\mu dx^\mu} \otimes_{sym} V_\nu\ dx^\nu +\ldots \, , \\
\prn{\JH}_a   dx^a  &=
\JH^\tV V_\mu\ dx^\mu +\ldots \, , \\
\end{split}
\end{equation}
then we get the Einstein-Maxwell equations with Hall current sources as
\begin{equation}\label{eq:EinMax}
\begin{split}
-\half\frac{d}{dr} \brk{ r^{d+1} \frac{d}{dr}\prn{\frac{g_{_V}}{r^2}}
- \kappa_q\  (d-1)\ q\ a_{_V}} &= 8\pi G_{_N} r^{d-1} \THall^\tV\, ,  \\
-\frac{d}{dr} \brk{ r^{d-1} f\ \frac{d a_{_V}}{dr}-(d-2) q \prn{\frac{g_{_V}}{r^2}} }
 &= g^2_{_{EM}} r^{d-3} \JH^\tV\, . 
\end{split}
\end{equation}

To solve the above two equations, let us now define the Hall current induced total mass function $\mathcal{M}^\tV(r)$ and
the total charge function $\mathcal{Q}^\tV(r)$ (as is familiar
from say the discussions of Tolman-Oppenheimer-Volkoff equation) via
\begin{equation} \label{eq:QandM}
\begin{split}
\mathcal{Q}^\tV(r) &\equiv \int_{\rH}^r dr' \brk{ (r')^{d-3} \JH^\tV (r') } \, , \\
\mathcal{M}^\tV(r) &\equiv \int_{\rH}^r dr' \brk{
(r')^{d-1}\ f(r')\ \THall^\tV (r') - \mathcal{Q}^\tV(r')\ \ \frac{d\Phi(r')}{dr'}}\, , \\
\end{split}
\end{equation}
which measure the total  Hall charge and Hall mass per unit area until a radius $r$
in an ingoing Eddington-Finkelstein tube. We have defined our mass and charge functions such that they vanish at the horizon. More precisely $\mathcal{Q}^\tV$ has a simple zero at $r= \rH$
and $\mathcal{M}^\tV$ has a double zero at $r= \rH$. Note that the total mass includes the electrostatic contribution proportional to the 0th order electric field $-\frac{d\Phi}{dr}$ times the Hall charge current.

These mass and charge functions can be used to solve the Einstein-Maxwell equations
\eqref{eq:EinMax} . The first integral of the Maxwell equation is given by
\begin{equation}\label{eq:MaxOrder1}
\begin{split}
 r^{d-1} f\ \frac{d a_{_V}}{dr}
 -(d-2) q \prn{\frac{g_{_V}}{r^2}}
 + g^2_{_{EM}} \mathcal{Q}^\tV =0 \, . 
\end{split}
\end{equation}
Here we have fixed the constant of integration by demanding that $g_{_V}$
vanishes at the horizon $r= \rH$ : we can always do this by an appropriate choice
of velocity definition in dual hydrodynamics.

Next, we turn to the Einstein equations. Subtracting one of the identities in  Eq.~\eqref{eq:dfdrIdentity}, we find
\begin{equation}
\begin{split}
&\frac{d}{dr} \brk{ r^{d+1} \frac{d}{dr}\prn{\frac{g_{_V}}{r^2}} - \kappa_q\  (d-1)\ q\ a_{_V} }-
\frac{d}{dr} \brk{ r^{d+1} \frac{df}{dr}+ \kappa_q (d-1) q\Phi }\prn{\frac{g_{_V}}{r^2f}} \\
 &\quad
= \frac{1}{f}\bigbr{ \frac{d}{dr} \brk{ r^{d+1}f^2 \frac{d}{dr}\prn{\frac{g_{_V}}{r^2f}} }
 - \frac{\kappa_q\ (d-1) q}{r^{d-1}} \brk{r^{d-1}f\frac{d a_{_V}}{dr} - (d-2)q \prn{\frac{g_{_V}}{r^2}}  }  } \\
 &\quad = \frac{1}{f} \bigbr{ \frac{d}{dr} \brk{ r^{d+1}f^2 \frac{d}{dr}\prn{\frac{g_{_V}}{r^2f}} }
 -16\pi G_{_N} \mathcal{Q}^\tV\frac{d\Phi}{dr} } \, , \\
\end{split}
\end{equation}
where in the last line we have used \eqref{eq:MaxOrder1}. Thus, Einstein equation simplifies to
\begin{equation}
\begin{split}
\frac{d}{dr} \brk{ r^{d+1}f^2 \frac{d}{dr}\prn{\frac{g_{_V}}{r^2f}} }
 &= -16\pi G_{_N} \brk{ r^{d-1}f\ \THall^\tV - \mathcal{Q}^\tV\frac{d\Phi}{dr}  }
 = -16\pi G_{_N}\frac{d \mathcal{M}^\tV}{dr} \, . \\
\end{split}
\end{equation}
This gives the solution for the metric as
\begin{equation} \label{eq:gv}
\begin{split}
& 
g_{_V} 
= 16\pi G_{_N} r^2 f\int_r^{\infty}dr'  \frac{
\mathcal{M}^\tV(r') }{(r')^{d+1}f^2(r')}\, .
\\
\end{split}
\end{equation}

In order to solve the gauge field $a_{_V}$, we rewrite \eqref{eq:MaxOrder1} in the form
\begin{equation}
\begin{split}
0 &= \frac{d a_{_V}}{dr}
 + \prn{\frac{g_{_V}}{r^2f}}  \frac{d \Phi}{dr}
 + g^2_{_{EM}} \frac{\mathcal{Q}^\tV }{r^{d-1}f}\\
 &=\frac{d }{dr} \brk{a_{_V}+\Phi\frac{g_{_V}}{r^2f} }
 -\Phi  \frac{d}{dr} \prn{\frac{g_{_V}}{r^2f}}
 + g^2_{_{EM}} \frac{\mathcal{Q}^\tV }{r^{d-1}f}\\
  &=\frac{d }{dr} \brk{a_{_V}+\Phi\frac{g_{_V}}{r^2f} }
 +16\pi G_{_N}\ q\  \frac{\mathcal{M}^\tV }{r^{2d-1}f^2}
 + g^2_{_{EM}} \frac{\mathcal{Q}^\tV }{r^{d-1}f} \, , \\
\end{split}
\end{equation}
which gives
\begin{equation} \label{eq:av}
\begin{split}
a_{_V} &= g^2_{_{EM}}\int_r^{\infty}dr'  \frac{\mathcal{Q}^\tV(r') }{(r')^{d-1}f(r')}
- 16\pi G_{_N} \frac{q}{r^{d-2}}\int_r^{\infty}dr'  \frac{\mathcal{M}^\tV(r') }{(r')^{d+1}f^2(r')}
\\
&\qquad
+16\pi G_{_N} q\int_r^{\infty}dr'  \frac{\mathcal{M}^\tV(r') }{(r')^{2d-1}f^2(r')} \, .  \\
\end{split}
\end{equation}
Thus we conclude that our final solution is given by Eq. \eqref{eq:gv} and \eqref{eq:av} with Eqs. \eqref{eq:QandM}.

The asymptotic expansion of the metric and gauge field is hence given by\footnote{
Note that at this point we have assumed  
$ \mathcal{M}^\tV(r) $ and
$\mathcal{Q}^\tV(r)$ are ${\cal O}(r^0)$ when $r\rightarrow \infty$.
Physically, this assumption is equivalent to saying that $g_{_V}$ and $a_{_V}$ have fall-off near infinity just like
the appropriate multiple-moments of the fields, i.e. $g_{_V}$ and $a_{_V}$ both fall off as $\sim r^{2-d}$.
We will see in Eq.~(\ref{eq:MQCFTlogan}) that indeed both
$ \mathcal{M}^\tV(r) $ and $\mathcal{Q}^\tV(r)$ are of ${\cal O}(r^0)$ as $r\rightarrow \infty$
after we have obtained 
$ \JH^\tV (r)$ and $ \THall^\tV (r)$ explicitly. 
}
\begin{equation}
\begin{split}
g_{_V} &= \frac{16\pi G_{_N}}{r^{d-2}d} \mathcal{M}^\tV(r=\infty) + \ldots \, , \\
a_{_V}
  &= \frac{g^2_{_{EM}}}{r^{d-2}(d-2)} \mathcal{Q}^\tV(r=\infty) + \ldots \, . \\
\end{split}
\end{equation}

From this we obtain the anomaly-induced part of 
the renormalized Brown-York stress tensor and current:\footnote{ We assume $d>2$. The analysis for $d=2$ involves additional contributions to the charges from the Chern-Simons terms.
See, for e.g., Ref~\cite{Kraus:2006wn} for the analysis in $d=2$.
}
\begin{eqnarray}
(T_{\alpha\beta})_{anom}  
&=&
 -\lim_{r\rightarrow \infty} \frac{r^{d-2}}{8 \pi G_N} \left[
 K_{\alpha\beta}
-K   (G_{\alpha\beta}-n_{\alpha}n_{\beta})
+(d-1) (G_{\alpha\beta}-n_{\alpha}n_{\beta})
\right]_{anom}
\nonumber\\
&=&
 -\prn{ V_\alpha u_\beta + V_\beta u_\alpha }
\lim_{r\to\infty}  \frac{r^{d+1}}{16\pi G_{_N}}\frac{d}{dr}\prn{\frac{g_{_V}}{r^2}} \nonumber \\
&=& 
 \prn{ V_\alpha u_\beta + V_\beta u_\alpha }\mathcal{M}^\tV(r=\infty) \nonumber \\
&=& 
 \prn{ V_\alpha u_\beta + V_\beta u_\alpha } \int_{\rH}^{\infty} dr'\brk{
 (r')^{d-1}f(r') \THall^\tV (r')-\frac{d\Phi}{dr} \int_{\rH}^{r'} dr'' (r'')^{d-3}\ \JH^\tV(r'')    }\, , \nonumber \\
(J_\alpha)_{anom}  &=& - \lim_{r\rightarrow \infty} \frac{r^{d-1}}{g^2_{EM}} g_{\mu\alpha}\left( F^{r\mu} \right)_{anom}
\nonumber\\
&=& 
- V_\alpha \lim_{r\to\infty} \frac{(d-1)}{(d-2)} \frac{\kappa_q}{16\pi G_{_N}}
r^{d-1} \frac{da_{_V}}{dr} 
\nonumber \\ 
&=&V_\alpha \mathcal{Q}^\tV(r=\infty) 
\nonumber \\
&=& 
V_\alpha  \int_{\rH}^{\infty} dr' (r')^{d-3}\ \JH^\tV(r') \, , 
\label{eq:TabAndJa}
\end{eqnarray}
where we have assumed that the integrals under question converge to a finite value.
The extrinsic curvature $K_{ab}$ is defined in the standard way
as $K_{ab}\equiv (1/2){\cal L}_{n} G_{ab}$ where $n^a\partial_a$ is the 
normal vector of surfaces of constant-$r$.
In Ref.~\cite{Azeyanagi:2013}, we show that the Chern-Simons contributions to the charges vanish (for AdS$_{2n+1}$ with $n\ge2$) which is consistent with the above assumption that the Chern-Simons contribution to the stress tensor is zero.

The intuitive meaning behind these expressions is clear : the anomaly-induced currents of the boundary
are obtained by integrating from horizon to asymptotic infinity,
the contribution to energy-momentum and charge due to the bulk Hall currents.
The energy-momentum integral has a `gravitational potential energy' part proportional to $ \THall^\tV$
and the `electric potential energy'  part proportional to $\JH^\tV$.

At this point, the problem of computing $(T_{\alpha\beta})_{anom} $ and $(J_\alpha)_{anom}$ reduces to that of computing the sources $\THall^\tV$ and $\JH^\tV$, which we will carry out in the next sections.

\section{Abelian Chern-Simons term}
\label{sec:abelianCS}
In this section, we explain the evaluation of the Hall contribution for the Abelian Chern-Simons term.
This was first done in \cite{Kharzeev:2011ds} and here we repeat their derivation in our notation
for the convenience of the reader.  Evaluation of the Hall contribution for more general
Chern-Simons terms is illustrated explicitly in the case of AdS$_7$
while the general results are summarized in Sec.~\ref{sec:GeneralCS}.
The detail of the computation is given in Appendices \ref{sec:MaxwellSources} and \ref{sec:EinsteinSources}.

For the Abelian Chern-Simons term on AdS$_{d+1}$ with
$d=2n$ and its CFT$_{2n}$ dual, we take the anomaly polynomial as
$\fP_{_{CFT}}= c_{_A} \fF^{n+1}$
where $c_A$ is a constant.  We then have
the charge and spin Hall current as well as the Hall energy-momentum tensor as
\begin{equation}
\hodge \fJH = -(n+1) c_{_A} \fF^n 
\quad \text{and} \quad  (\hodge \fSpH)^a{}_b = 0, \qquad   (\hodge\fTHall)^{ab} = 0
\,.
\end{equation}
Using $\fA =  \Phi\ \fu + \fA_\infty+ \ldots$, we obtain the corresponding $U(1)$ field strength as
\begin{equation}
\begin{split}
\fF &=  \frac{d\Phi}{dr}dr \wedge \fu +\fF_\infty +\Phi\ d\fu +\ldots\, ,   \\
\end{split}
\end{equation}
where $\fF_{\infty} \equiv d \fA_\infty= \fB$ is the magnetic field.
Then, the charge Hall current is evaluated as
\begin{equation}\label{eq:JHAbelianTemp}
\begin{split}
\hodge \fJH &= n (n+1) c_{_A} \frac{d\Phi}{dr}\  \fu \wedge (\fF_\infty +\Phi d\fu)^{n-1}\wedge dr +\ldots\\
&= n (n+1) c_{_A} \frac{d\Phi}{dr}\  \fu \wedge (\fB +2\fomega\Phi )^{n-1}\wedge dr +\ldots \\
&=  \hodgeCFT \prn{ \frac{d}{dr}\frac{\partial \form{\mathbb{G}}}{\partial \Phi} } \wedge dr+\ldots\, ,  \\
\end{split}
\end{equation}
where we have defined the combination of pseudovectors $\mathbb{G}$ as
\begin{equation}
\begin{split}
 \hodgeCFT \form{\mathbb{G}} &\equiv  \hodgeCFT\prn{ \mathbb{G}_\mu  dx^\mu }\\
 &= \frac{\fu}{(2\fomega)^2}
\wedge \brk{ c_{_A} (\fB +2\fomega\Phi )^{n+1} - c_{_A} \fB^{n+1}- (n+1) c_{_A} \fB^n(2\fomega\Phi)}\\
&= \sum_{k=1}^{n} \binom{n+1}{k+1} c_{_A} \Phi^{k+1}\ \fu \wedge \fB^{n-k}\wedge (2\fomega)^{k-1}\, . \\
\end{split}
\end{equation}
In the presence of external magnetic fields\footnote{
We consider non-zero external magnetic fields only in 
this section (i.e. in the case of Abelian Chern-Simons terms). 
For the rest of the paper, we shall switch off the magnetic fields.
}
 $\fB\equiv d\fA_{\infty}$, additional pseudo-vectors of the 
form 
\begin{eqnarray}\label{eq:generalpseudoV}
\hodgeCFT \form{V} 
\equiv \fu \wedge (\fB)^{n-k}\wedge (d\fu)^{k-1},
\end{eqnarray}
appear for each $k$ satisfying $1\le k\le n$. 
Let us now define $\mathbb{G}\equiv\mathbb{G}_\mu dx^\mu$
and the corresponding coefficients of each pseudovector $V_\mu$ in $\mathbb{G}$  by $\mathbb{G}^\tV$ :
\begin{equation}
\begin{split}
\label{eq:calGvector}
\mathbb{G}_\mu  dx^\mu
 &= \sum_{\{V\}}  \mathbb{G}^{\tV} V_\mu dx^\mu.
\end{split}
\end{equation}
Then, we have
\begin{equation}
\begin{split}
\label{eq:GVabelian}
\mathbb{G}^{\scriptscriptstyle{B^{n-k}(2\omega)^{k-1}} }
 &= \binom{n+1}{k+1} c_{_A} \Phi^{k+1}\, .
\end{split}
\end{equation}
Note that the  second or higher order terms in the fluid/gravity
expansion of $\fF$ do not contribute at order $\fomega^{k-1}$ for any $k$.
This is a direct consequence of the fact that wedge products of two or more
zeroth order $\fF$'s are zero.

In the next step, we use the following result in \eqref{eq:JHAbelianTemp} : For a $p$-form $\form{N}$ with legs only along the boundary directions 
and completely transverse to the velocity vector $u_{\mu}$, the following relation
for the Hodge duals of the CFT and the bulk holds:
\begin{equation}
\begin{split}
 \prn{ \hodgeCFT \form{N}} \wedge dr
&= \frac{(-1)^d}{r^{d-1-2p}}\ \hodge  \form{N}\, .  \\
\end{split}
\end{equation}
Here we assumed one is working in the Eddington-Finkelstein coordinates.
This gives another expression for the charge Hall current which
is useful when we compare with the result of the bulk AdS calculation:
\begin{equation}
\begin{split}
\hodge \fJH &= \hodge \prn{ \frac{1}{r^{d-3}}\frac{d}{dr}\frac{\partial \form{\mathbb{G}}}{\partial \Phi} }+\ldots \, ,\\
\end{split}
\end{equation}
or
\begin{equation}
\begin{split}
(\JH)_a dx^a & = \frac{1}{r^{d-3}}\frac{d}{dr}\brk{\frac{\partial \mathbb{G}_\mu }{\partial \Phi} } dx^\mu +\ldots\,  . \\
\end{split}
\end{equation}
Picking out the coefficient of one of the pseudo-vectors $V_\mu$, we have the expression for the
Maxwell source (for convenience, we also denote the expression for the Einstein source, though
it is trivially zero in the case of Abelian Chern-Simons terms)
\begin{equation}
\begin{split}
\JH^\tV & =  \frac{1}{r^{d-3}}\frac{d}{dr}\brk{
\frac{\partial \mathbb{G}^\tV }{\partial \Phi} }\, , \\
 \THall^\tV & = 0\, .
 \end{split}
\end{equation}
In particular, we have
\begin{equation}\label{eq:AbelianSources}
\begin{split}
\JH^{\scriptscriptstyle{B^{n-k}(2\omega)^{k-1}} } & = \frac{1}{r^{d-3}}\frac{d}{dr}\brk{ (k+1)\binom{n+1}{k+1} c_{_A} \Phi^k } \, , \\
 \THall^{\scriptscriptstyle{B^{n-k}(2\omega)^{k-1}} } & = 0\, .
 \end{split}
\end{equation}

\section{General Chern-Simons terms: Maxwell and Einstein sources}
\label{sec:GeneralCS}

For the sake of clarity, in this section we will only present a summary of our results of the Hall contribution for more general Chern-Simons terms in AdS$_{2n+1}$ for $1 \le n\le4$. 
A more detailed computation for all the other dimensions can be found in Appendices \ref{sec:MaxwellSources} and \ref{sec:EinsteinSources}
. To exemplify the computations, we will however first present here the details of the AdS$_7$ ($n=3$) case. Using these findings, finally, we make a proposal for the general structure of the sources and replacement rule.

\subsection{Illustrative case: Chern-Simons terms in AdS$_{7}$}

For a theory with anomaly polynomial
\begin{equation}
\begin{split}
\fP_{_{CFT_6}}= c_{_A} \fF^4 + c_{_M} \fF^2 \wedge \text{tr}\ \fR^2
+ c_{_g} \text{tr}\ \fR^2 \wedge \text{tr}\ \fR^2 + \tilde{c}_{_g} \text{tr}\ \fR^4 \, ,
 \end{split}
\end{equation}
via Eq.~(\ref{eq:defsHall}) we find the corresponding charge Hall current
$(\JH)^a$ and the Hall energy-momentum tensor $ (\THall)^{ab}$,
\begin{equation}
\begin{split}
& (\JH)^a  = - \frac{1}{8} \varepsilon^{ap_1p_2p_3p_4p_5p_6} 
\brk{4c_{_A} F_{p_1p_2} F_{p_3p_4} F_{p_5p_6} + 2 c_{_M}  
R^b{}_{cp_1p_2} R^c{}_{bp_3p_4} F_{p_5p_6} },
  \\
&(\THall)^{ab} \  =-  \quarter c_{_M}  \nabla_c\brk{\varepsilon^{ap_1p_2p_3p_4p_5p_6}R^{bc}{}_{p_1p_2}F_{p_3p_4}
F_{p_5p_6}
 +\varepsilon^{bp_1p_2p_3p_4p_5p_6}R^{ac}{}_{p_1p_2}F_{p_3p_4}F_{p_5p_6}}\\
 &\quad -  \half c_{_g}  \nabla_c\brk{\varepsilon^{ap_1p_2p_3p_4p_5p_6}R^{bc}{}_{p_1p_2}R^d{}_{fp_3p_4}R^f{}_{dp_5p_6}
 +\varepsilon^{bp_1p_2p_3p_4p_5p_6}R^{ac}{}_{p_1p_2}R^d{}_{fp_3p_4}R^f{}_{dp_5p_6}}\\
 &\quad -  \half \tilde{c}_{_g}   \nabla_c\brk{\varepsilon^{ap_1p_2p_3p_4p_5p_6}R^b{}_{dp_1p_2}
 R^d{}_{fp_3p_4}R^{fc}{}_{p_5p_6}
 +\varepsilon^{bp_1p_2p_3p_4p_5p_6}R^a{}_{dp_1p_2}R^d{}_{fp_3p_4}R^{fc}{}_{p_5p_6}}.\\
 \end{split}
 \end{equation}
 Evaluating these sources (see Sec.~\ref{sec:ExplicitAdS7MaxwellText} and Appendix~\ref{sec:appendixHallAdS7Einstein}
 for the computations), we obtain
 \begin{equation}\label{eq:Source6d} 
\begin{split}
\JH^\tV & =  \frac{1}{r^3}\frac{d}{dr}
\frac{\partial \mathbb{G}^\tV }{\partial \Phi},  \\
 \THall^\tV &  = -\frac{1}{2r^5}\frac{d}{dr}r^2\frac{d}{dr}\frac{\partial \mathbb{G}^\tV }{\partial \PhiT},
 \end{split}
\end{equation}
where  $\mathbb{G}^\tV$ is the the coefficient of the pseudovector $V_\mu$ in
a pseudovector combination $\mathbb{G}_\mu$ with
\begin{equation}
\begin{split}
 \hodgeCFT \form{\mathbb{G}} &\equiv  \hodgeCFT\prn{ \mathbb{G}_\mu  dx^\mu }\\
 &= \frac{\fu}{(2\fomega)^2}
\wedge \Bigl[ c_{_A} (
2\fomega\Phi )^4+ 2 c_{_M} (
2\fomega\Phi )^2 (2\fomega\PhiT )^2 
+
\prn{2^2c_{_g}+2\tilde{c}_{_g}}(2\fomega\PhiT )^4
\ \Bigr]\\
&
=\brk{ c_{_A} \Phi^4 + 2 c_{_M} \Phi^2 \PhiT^2
+ \prn{2^2c_{_g}+2\tilde{c}_{_g}} \PhiT^4  }\ \fu \wedge (2\fomega)^2 \,. \\
\end{split}
\end{equation}
Thus, we have
\begin{equation}\label{eq:GV6d}
\begin{split}
\mathbb{G}^{(2\omega)^2}
 &=  c_{_A} \Phi^4 + 2 c_{_M} \Phi^2 \PhiT^2
+ \prn{2^2c_{_g}+2\tilde{c}_{_g}} \PhiT^4 \, . \\
\end{split}
\end{equation}
The Hall sources are obtained by substituting \eqref{eq:GV6d} into \eqref{eq:Source6d}.

\subsubsection{Explicit computation for the Maxwell source in AdS$_7$}
\label{sec:ExplicitAdS7MaxwellText}
In the case of AdS$_7$, only one type of the mixed term is allowed in the anomaly polynomial:
\be
 \fP_{CFT_6}= c_{_M} \form{F}^2 \wedge \tr[\form{R}^2]\, ,
\ee
from which we get the Maxwell source of the form
\be
{}^\star\form{J}_H= -2 c_{_M}  \form{F} \wedge \tr[\form{R}^2]\, .
\ee
For now let us consider the 0th and 1st order terms
in $\fF$ and $\fR$ only and forget the 2nd and higher order terms.
First, we shall explain some notations (which are explained in details in Appendix~\ref{sec:productRiemanns}). We denote the curvature 2-form  $\fR^a{}_b$ at $m$-th order
of the derivative expansion as $(\R{m})^{a}{}_b$.
Similarly, let us denote the $m$-th order term of  the $U(1)$ field strength $\form{F}$ 
by $(\F{m})$. Then the leading contribution to the Maxwell source is given by
\bea\label{eq:leadingAdS7Maxwell}
{}^\star\form{J}_H&=&
-  4c_{_M}
(\F{1})\wedge \tr[(\R{0}\R{1})]-  2c_{_M}
(\F{0})\wedge\tr[(\R{1}\R{1})]\nonumber\\
&=&\frac{d}{dr}\left(
 4c_{_M} \Phi  \Phi^2_T
\right)(\fu\wedge(2\fomega)^2 \wedge dr), \nonumber\\
J_H^{(2\omega)^2}&=&
  \frac{1}{r^3} \frac{d}{dr}\frac{\partial (2c_{_M}\Phi^2 \Phi^2_T)}{\partial \Phi},
\eea
where ${}^\star \form{J}_H$ is indeed of order $\fomega^2$.
At lower orders, we encounter 
\begin{eqnarray}
(\F{0})\wedge \tr[(\R{0}\R{0})], \quad (\F{0})\wedge \tr[(\R{0}\R{1})], \quad (\F{1})\wedge \tr[(\R{0}\R{0})], 
\end{eqnarray} 
but these are all zero as a result of $(\R{0}\R{0})=0$ (see Eq.~(\ref{eq:R0R0})) 
and the fact that both $(\F{0})$ and $\tr[(\R{0}\R{1})]$ are proportional
to $dr\wedge \fu$ (see Eq.~(\ref{eq:F0F1F2}) and Eq.~(\ref{eq:TracesofProductsRiemanns})).
Therefore the contribution in Eq.~(\ref{eq:leadingAdS7Maxwell}) is indeed the leading one.

Now we take into account the 2nd and higher order terms in $\fF$ and $\fR$. Then, up to $\fomega^2$ order, 
we encounter the following terms:
\begin{eqnarray}
(\F{0})\wedge \tr[(\R{0}\R{2})], \quad (\F{2})\wedge \tr[(\R{0}\R{0})]\, ,  
\end{eqnarray}
which also vanish, since $(\R{0}\R{0})=0$ and $\tr[(\R{0}\R{2})]=0$ (see Eq.~(\ref{eq:Tr02})). 

As an application of this result, let us compute the thermal helicity of the (2,0) theory using its anomaly 
polynomial.\footnote{Anomaly polynomial of (2,0) theory is given in say
Eq.~(2.1) and Eq.~(2.2) of \cite{Alday:2009qq} or
Eq.~(2.9) and Eq.~(2.10) of \cite{Bah:2012dg}. }
The thermal helicity of (2,0) theory with a gauge group G is then given by 
\begin{equation}
\mathfrak{F}_{Anom}=-\frac{2\pi r_{_G}}{48}\left[p_2(\mu_{SO(5)})+\frac{1}{4}(T^2+p_1(\mu_{SO(5)}))^2\right]-\frac{2\pi d_{_G} h_{_G}}{24}p_2(\mu_{SO(5)}) \, ,
\end{equation}
where $r_{_G},d_{_G}$ and $h_{_G}$ are the rank, dimension and Coxeter number of the gauge group $G$ respectively and 
\begin{eqnarray}
&& p_1(\mu_{SO(5)})\equiv -\frac{1}{(2\pi)^2} \left(\frac{1}{2} \text{Tr} \,\mu_{SO(5)}^2 \right)\,,\\
&& p_2(\mu_{SO(5)})\equiv -\frac{1}{(2\pi)^2} \left(\frac{1}{4} \text{Tr} \,\mu_{SO(5)}^4-\frac{1}{8} (\text{Tr} \,\mu_{SO(5)}^2)^2 \right) \, .
\end{eqnarray}
It would be interesting for future work to study the thermal partition function of (2,0) 
theory and match it against this prediction from holography.

\subsection{Summary of results: general structure of sources and replacement rule}

The computations for the sources  in various dimensions are performed in two different ways. On the one hand, 
we employed a straightforward {\it brute force} computation (till AdS$_9$ using the solutions up to second order given in
Appendix \ref{sec:AdSKerrNewman2ndOrder}) with \emph{Mathematica} of the 
Riemann tensor and field strength contributions to get the sources.\footnote{
Note that a priori the third order metric can contribute to the
 sources in AdS$_9$. However, as we argue in  Appendix C and Appendix
D, the second order metric suffices even in AdS$_9$.}
On the other hand, we also performed a more elegant though lengthy
computation order-by-order in the derivative expansion.
The details of the latter calculations can be found in Appendices \ref{sec:productRiemanns}, 
\ref{sec:MaxwellSources} and \ref{sec:EinsteinSources}.  These independent calculations
perfectly matched and served as crosschecks of our results that are contained
in Table \ref{table:results} with Maxwell and Einstein sources defined by
\begin{equation}\label{eq:SourceGen}
\begin{split}
\JH^\tV & =  \frac{1}{r^{d-3}}\frac{d}{dr}
\frac{\partial \mathbb{G}^\tV }{\partial \Phi}\, ,   \\
 \THall^\tV & 
=\frac{1}{r^{d-2}}\frac{d}{dr}\left[
r^{d-2}\SpH^\tV
\right]
  = -\frac{1}{2r^{d-1}}\frac{d}{dr}r^2\frac{d}{dr}\frac{\partial \mathbb{G}^\tV }{\partial \PhiT},
 \end{split} 
\end{equation}
 where $\SpH^\tV$ is given by equation (\ref{eq:sigmaV}).
 
\begin{table}[ht]
\centering 
\begin{tabular}{| c | c | c | } 
\hline\hline 
\#$n$ &  ${(V)} $& $\mathbb{G}^{\tV} $  \\ [0.5ex]
\hline
1 & $(2\omega)^0$ &   $ 2 c_{_g}  \PhiT^2$ \\
 \hline
 2 & $(2\omega)^1$ &   $c_{_A} \Phi^3 +2 c_{_M} \Phi\PhiT^2$ \\  \hline
3 & $(2\omega)^2$ &   $c_{_A} \Phi^4 + 2 c_{_M} \Phi^2 \PhiT^2
+ \prn{2^2c_{_g}+2\tilde{c}_{_g}} \PhiT^4$ \\
\hline
4 & $(2\omega)^3$ &   $c_{_A} \Phi^5 + 2 c_{_M} \Phi^3 \PhiT^2
+ \prn{2^2\tilde{c}_{_M}+2\tilde{\tilde{c}}_{_M}} \Phi \,\PhiT^4 $ \\
 \hline
\hline 
\end{tabular}
\caption{$\mathbb{G}^\tV$ polynomials for the Maxwell and Einstein sources (\ref{eq:SourceGen}) in AdS$_{2n+1}$ and $1 \le n\le 4$.}
\label{table:results}
\end{table}

Our results are consistent with the replacement rule for $\mathbb{G}^\tV$ as a polynomial in $\{\Phi,\PhiT \}$ in Eq.~(\ref{eq:repwB}).
We work out in the next section the implications of such a conjecture.

\section{General computations of \texorpdfstring{$(T_{\alpha\beta})_{anom} $}{anomalous stress tensor} 
and \texorpdfstring{$(J_{\alpha})_{anom}$}{charge current} }
\label{sec:compTabJa}
We now turn to solving the Einstein-Maxwell equations for sources of the form in (\ref{eq:SourceGen}).  To do this, we compute the mass and the charge integrals via \eqref{eq:QandM}. 
To evaluate these integrals explicitly, it is useful to define the following function which has a double zero at $r=\rH$ :
\begin{equation}
\begin{split}
\tilde{\mathbb{G}}^\tV &= \brk{\mathbb{G}^\tV }_{r=\rH}- \mathbb{G}^\tV
+ \brk{ \Phi -\Phi(\rH)}\  \brk{ \frac{\partial \mathbb{G}^\tV }{\partial \Phi} }_{r=\rH}
+ \brk{ \PhiT-\PhiT(\rH)}\  \brk{ \frac{\partial \mathbb{G}^\tV }{\partial \PhiT} }_{r=\rH}\, ,  
\end{split}
\end{equation}
so that we can write the sources  (\ref{eq:SourceGen}) as
\begin{equation}
\begin{split}
\JH^\tV =  -\frac{1}{r^{d-3}}\frac{d }{dr}\frac{\partial \tilde{\mathbb{G}}^\tV }{\partial \Phi} \, ,  \qquad
 \THall^\tV = \frac{1}{r^{d-1}}\frac{d}{dr}\brk{\frac{r^2}{2}\frac{d}{dr} \frac{\partial \tilde{\mathbb{G}}^\tV }{\partial \PhiT}  }\, . 
 \end{split}
\end{equation}
First of all, this gives the total charge function
\begin{equation}  \label{eq:charge_integrated}
\begin{split}
\mathcal{Q}^\tV &= - \frac{\partial \tilde{\mathbb{G}}^\tV }{\partial \Phi}\, . 
\end{split}
\end{equation}
In order to do the integral in the expression of  $\mathcal{M}^\tV$, we look at
\begin{equation}
\begin{split}
 r^{d-1} f\  \THall^\tV - \mathcal{Q}^\tV \frac{d\Phi}{dr}
 &= f\frac{d}{dr}\brk{\frac{r^2}{2}\frac{d}{dr} \frac{\partial \tilde{\mathbb{G}}^\tV }{\partial \PhiT}}
 + \frac{\partial \tilde{\mathbb{G}}^\tV }{\partial \Phi} \frac{d\Phi}{dr}\\
 &= \half\frac{d}{dr}\brk{r^2f^2\frac{d}{dr}\prn{\frac{1}{f}\frac{\partial \tilde{\mathbb{G}}^\tV }{\partial \PhiT} }\ }
 + \frac{\partial \tilde{\mathbb{G}}^\tV }{\partial \Phi} \frac{d\Phi}{dr}+ \frac{\partial \tilde{\mathbb{G}}^\tV }{\partial \PhiT}  \frac{d\PhiT}{dr}\\
 &=  \frac{d}{dr}\brk{\tilde{\mathbb{G}}^\tV
 +\half r^2f^2\frac{d}{dr}\prn{\frac{1}{f}\frac{\partial \tilde{\mathbb{G}}^\tV }{\partial \PhiT} }\ }\, ,\\
\end{split}
\end{equation}
so that we finally have the total mass function
\begin{equation} \label{eq:mass_integrated}
\begin{split}
\mathcal{M}^\tV &= \tilde{\mathbb{G}}^\tV
 +\half r^2f^2\frac{d}{dr}\prn{\frac{1}{f}\frac{\partial \tilde{\mathbb{G}}^\tV }{\partial \PhiT} }\, . 
\end{split}
\end{equation}

Thus, we finally obtain
\begin{equation}
\begin{split}
g_{_V} &= 16\pi G_{_N} r^2 f\int_r^{\infty}dr'  \frac{\mathcal{M}^\tV(r') }{(r')^{d+1}f^2(r')}\, , \\
a_{_V}
  &= g^2_{_{EM}}\int_r^{\infty}dr'  \frac{\mathcal{Q}^\tV(r') }{(r')^{d-1}f(r')}
- 16\pi G_{_N} \frac{q}{r^{d-2}}\int_r^{\infty}dr'  \frac{\mathcal{M}^\tV(r') }{(r')^{d+1}f^2(r')}\\
&\qquad + 16\pi G_{_N} q\int_r^{\infty}dr'  \frac{\mathcal{M}^\tV(r') }{(r')^{2d-1}f^2(r')}\, , \\
\end{split}
\end{equation}
with
\begin{equation}
\begin{split}
\mathcal{M}^\tV &= \tilde{\mathbb{G}}^\tV
 +\half r^2f^2\frac{d}{dr}\prn{\frac{1}{f}\frac{\partial \tilde{\mathbb{G}}^\tV }{\partial \PhiT} }\, , \\
 \mathcal{Q}^\tV &= - \frac{\partial \tilde{\mathbb{G}}^\tV }{\partial \Phi}\, . 
\end{split}
\end{equation}

By using these expressions, we calculate the total mass and charge correction due to the Hall sources.
Noticing that  
 as $r\to\infty$,  $\{\Phi,\PhiT\}\to 0$ also
\begin{equation}
\begin{split}
\prn{ \tilde{\mathbb{G}}^\tV }_{r=\infty} &=\prn{\mathbb{G}^\tV
 -\Phi\   \frac{\partial \mathbb{G}^\tV }{\partial \Phi}
-\PhiT\  \frac{\partial \mathbb{G}^\tV }{\partial \PhiT} }_{r=\rH}\, ,  \\
\prn{ \frac{\partial \tilde{\mathbb{G}}^\tV }{\partial \Phi} }_{r=\infty}
&=\prn{ \frac{\partial \mathbb{G}^\tV }{\partial \Phi} }_{r=\rH} \, , \\
\prn{ \frac{\partial \tilde{\mathbb{G}}^\tV }{\partial \PhiT} }_{r=\infty}
&= \prn{ \frac{\partial \mathbb{G}^\tV }{\partial \PhiT} }_{r=\rH} \, . \\
\end{split}
\end{equation}
 Then, the total mass and the charge correction due to the Hall sources is
\begin{equation}
\begin{split}
\mathcal{M}^\tV(r=\infty) &= \prn{\mathbb{G}^\tV
 -\Phi\   \frac{\partial \mathbb{G}^\tV }{\partial \Phi}
-\PhiT\  \frac{\partial \mathbb{G}^\tV }{\partial \PhiT} }_{r=\rH}\, ,  \\
\mathcal{Q}^\tV (r=\infty)&=- \prn{ \frac{\partial \mathbb{G}^\tV }{\partial \Phi} }_{r=\rH} \, . \\
\end{split}\label{eq:MQCFTlogan}
\end{equation}

From this,  we get the anomaly induced currents as
\begin{equation}
\begin{split}
(T_{\alpha\beta})_{anom} &=
\prn{\mathbb{G}_\alpha
 -\Phi\   \frac{\partial \mathbb{G}_\alpha }{\partial \Phi}
-\PhiT\  \frac{\partial \mathbb{G}_\alpha }{\partial \PhiT} }_{r=\rH} u_\beta \\
&\qquad +u_\alpha \prn{\mathbb{G}_\beta
 -\Phi\   \frac{\partial \mathbb{G}_\beta }{\partial \Phi}
-\PhiT\  \frac{\partial \mathbb{G}_\beta }{\partial \PhiT} }_{r=\rH}  \, , \\
(J_\alpha)_{anom}  &= - \prn{ \frac{\partial \mathbb{G}_\alpha }{\partial \Phi} }_{r=\rH} \, .\\
\end{split}\label{eq:TabCFTlogan}
\end{equation}

The physics behind Eq.~(\ref{eq:MQCFTlogan}) and Eq.~(\ref{eq:TabCFTlogan}) is very intuitive.
In terms of the energy $\mathcal{M}$, the standard Gibbs potential $G$ and the charge $\mathcal{Q}$ are given by thermodynamic relations
\begin{equation}
\mathcal{M}=G+\mu \,\mathcal{Q} + T \mathcal{S}=G-\mu \prn{\frac{\partial G}{\partial \mu}}_{T}-T \prn{\frac{\partial G}{\partial T}}_{\mu}\, ,
\quad \quad \mathcal{Q} = -\prn{\frac{\partial G}{\partial \mu}}_{T} \, ,
\end{equation}
which is just Eq.~(\ref{eq:MQCFTlogan})  after substituting $\Phi(r_H)=\mu$ and $\Phi_T(r_H)=2\pi T$.
Thus we conclude that the vector $\mathbb{G}_\alpha (r=r_{_H})$ is the anomaly-induced free-energy current. We refer the reader to Refs.~\cite{Loganayagam:2011mu,Loganayagam:2012pz,Banerjee:2012cr} for a more detailed
discussion of how the anomalous parts of charge/entropy/energy could be derived from the Gibbs current, which
gives exactly Eq.~(\ref{eq:TabCFTlogan}).

As described in \cite{Loganayagam:2012zg}, the anomaly induced free-energy can be exponentiated to compute the 
anomaly-induced partition function from which the correlator describing thermal helicity follows simply.
As advertised, it is straightforward to check that this procedure yields the correct thermal helicity as
conjectured in \cite{Loganayagam:2012zg}.

\section{Discussions and conclusions}
We will begin with a brief summary of our results. In this paper, we have studied in detail how
holography gives rise to the replacement rule for thermal helicity that was proposed in 
\cite{Loganayagam:2012zg}. Combined with the tests in free field theory that were performed
in \cite{Loganayagam:2012pz,Loganayagam:2012zg}, these results suggest 
that replacement rule holds in a wide class of field theories and complement the 
recent formal proofs of the replacement rule in \cite{Jensen:2012kj,Jensen:Nov2013}.
Let us remind the reader of some of the features of the replacement rule which are surprising from
a field theory viewpoint but find natural explanation in a holographic context. 

The first surprising feature  is that the gravitational anomalies contribute to thermal helicity at all in the first place. In
fact, if one merely constrains the equilibrium partition function of the field theory to have the 
correct anomalous transformation (or equivalently, imposes anomalous Ward identities on thermal correlators)
one would naively conclude that gravitational anomalies involve too
many derivatives to contribute to the thermal correlator that defines thermal helicity\cite{Bhattacharya:2011tra,Banerjee:2012iz,
Jensen:2012jy,Jain:2012rh,Valle:2012em,Banerjee:2012cr}. Hydrodynamic arguments involving studying 
the second law of thermodynamics in the presence of anomalies also lead to the 
same naive result\cite{Kharzeev:2009p,Loganayagam:2011mu}. Thus, replacement rule is a puzzling feature in thermal field theory
whereby anomalies impose an important constraint on observables and this constraint cannot be merely 
captured by anomalous Ward identities alone.\footnote{We remark that this issue shows up in a different
disguise in studies of thermal Hall effect, which complicates the derivation of thermal Hall effect
\cite{Stone:2012ud}.}

On the AdS side, this puzzle gets easily resolved. While it is true that gravitational Chern-Simons terms 
have too many derivatives of the metric, as can be gleaned from our computation, this does not prevent 
them from affecting the blackhole solution at a much lower order in the fluid/gravity expansion. 
This is a simple consequence of the fact that in AdS the extra derivatives can be soaked 
up by the radial derivatives thus letting the higher derivative terms 
to contribute with low number of boundary derivatives. This is an example of how a very puzzling
feature of the CFT finds an essentially trivial explanation on AdS side.

The second feature of the replacement rule is that the temperature appears in Eq.\eqref{eq:RepRuleThermalHelicity}
via the factors of $2\pi T$. These powers of $2\pi$ are another indication that any algebraic manipulation
of anomalous Ward identities cannot lead to the replacement rule. In holography, on the other hand, $2\pi T$
is just the surface gravity of the black brane and it is unsurprising that a gravity calculation involves
such factors.  

It is instructive to compare these features of the replacement rule to the Cardy formula
which also involves derivative-jumping by the conformal anomaly and additional factors of $2\pi$
neither of which can be established by examining Ward identities alone. In the case of Cardy formula,
it is well-known that any field-theory proof of it should necessarily invoke the modular property of
the underlying 2d CFT (or an equivalent thereof) to reproduce these features. On the other hand,
Cardy formula is of course routinely derived in the AdS$_3$ by a straightforward gravity computation.
Our calculations show that replacement rule in higher dimension shares this feature : while the formal
field-theory proofs of replacement rule involve subtle arguments regarding partition function
on cones\cite{Jensen:2012kj,Jensen:Nov2013}, on the gravity side, we have a straightforward gravity computation.

A third puzzling feature of the replacement rule is why only the first Pontryagin class $p_{_1}(\fR)$ contributes
to the thermal helicity. While we do not know of a simple field theoretic reason for this fact, we have 
a simple explanation on the gravity side - when Hall currents are computed in charged rotating blackhole
background, to leading order, terms in Hall currents involving the higher Pontryagin class vanish. We do
not yet know of a simple way to translate this observation to the CFT side. It would be interesting to 
see whether we can use such insights gleaned from holography to understand better how the replacement rule 
arises in thermal field theory.

We will now conclude by mentioning various future directions. Perhaps the simplest generalization of the 
computations in this paper is to turn on the boundary magnetic field. This was already done for 
pure Abelian Chern-Simons terms by the authors of \cite{Kharzeev:2011ds} as we reviewed in 
Sec.~\ref{sec:abelianCS}. We expect that most of our arguments for mixed Chern-Simons terms 
would continue to hold with some simple modifications. There are detailed predictions from field 
theory on how the magnetic field should appear in the anomaly induced transport
and it would be nice to check these predictions against holography.

A more involved exercise would be to study the response to turning on boundary Riemann curvature.
This necessarily involves proceeding to higher orders  in the fluid/gravity expansion and studying 
Hall currents and their backreaction in more detail. The field theory studies of these terms \cite{
Jensen:2013kka,Jensen:Nov2013} suggests that the new terms would be conveniently arranged
in terms of a spin chemical potential $(\mu_{_R})^\alpha{}_\beta$ defined
via $(\mu_{_R})^\alpha{}_\beta= T \nabla_\beta\prn{\frac{u^\alpha}{T}}$. It would be instructive 
to see how this happens in gravity.
In this paper, we have essentially used the standard holographic counter term prescription for 
the Einstein-Maxwell system to compute the currents and stress tensor of the dual CFT. We have done
it by assuming that the modifications due to Chern-Simons terms do not contribute to the computations
in this paper. This is a non-trivial assumption since the counter term procedure has never been
generalized to arbitrary higher dimensional Chern-Simons terms. AdS/CFT with Chern-Simons terms
have been studied extensively in the last decade mostly in the context of AdS$_3$/CFT$_2$\cite{Kraus:2006wn}.
In AdS$_3$/CFT$_2$ context, many proposals now exist for computing total charges in the presence of Chern-Simons
terms \cite{Deser:2003vh,Solodukhin:2005ah,Bouchareb:2007yx,Compere:2008cv,Skenderis:2009nt}. Most of them have not yet  been generalized, much less tested in higher dimensional holography even in the 
absence of Chern-Simons terms. 

For a manifestly covariant bulk Lagrangian, a more well-established and well-tested 
result in higher dimensions is the entropy proposal by Wald for higher derivative actions. In the last few years,
there have been various attempts to extend the famous analysis of Iyer-Wald to Chern-Simons terms in arbitrary 
dimensions\cite{Tachikawa:2006sz,Bonora:2011gz}. As far as we are aware of, however, there is no current proposal which gives manifestly
covariant charges/entropy for the dual anomalous CFT.\footnote{To be more precise, we remind the reader that
the usual currents and the energy-momentum tensor of a QFT with anomalies are not expected to be covariant.
However, this situation is easily resolved by adding certain state-independent non-gauge invariant terms (called Bardeen-Zumino terms). 
Such Bardeen-Zumino corrected currents and the energy-momentum tensor are indeed covariant and what we are seeking is 
a procedure to calculate such covariant currents of the CFT from the AdS side.} Hence, it would be nice 
to write down a covariant proposal for the charges of dual CFT from the gravitational solution.
This would also lead to a covariant proposal for entropy extending the constructions in
\cite{Tachikawa:2006sz,Bonora:2011gz} to arbitrary Chern-Simons terms.
This will be addressed in our subsequent work \cite{Azeyanagi:2013}.

Another generalization would be to study anomaly-induced transport slightly away from equilibrium,
where again recent field theory arguments \cite{Jensen:Nov2013} seem to suggest novel contributions to anomaly-induced transport.  More broadly, we hope that our computations here are a first step towards 
a deeper understanding of finite temperature holography in the presence of Chern-Simons terms.

\section*{Acknowledgements}
We would like to thank A.~Adams, J.~Bhattacharya, S.~Chapman, G.~Comp\`ere, K.~Jensen, S.~Minwalla, M.~Rangamani, S.~Wadia and A.~Yarom for valuable discussion. T.~A. was supported by 
the LabEx ENS-ICFP: ANR-10-LABX-0010/ANR-10-IDEX-0001-02 PSL*
and JSPS Postdoctoral Fellowship for
Research Abroad. R.~L. was supported by the Harvard Society of Fellows through a junior fellowship.
R.~L. would  like to thank various colleagues at the society for interesting discussions.
M.~J.~R. was supported by the European Commission - Marie Curie grant
PIOF-GA 2010-275082.
G.~N. was supported by DOE grant DE-FG02-91ER40654 and the Fundamental Laws Initiative at Harvard.

\appendix

\section{\texorpdfstring{AdS$_{d+1}$}{AdS}-RN black brane and \texorpdfstring{AdS$_{d+1}$}{AdS}-Kerr-Newman metric
}
\label{sec:AdSKerrNewman2ndOrder}
\subsection{\texorpdfstring{AdS$_{d+1}$}{AdS}-RN black brane solution}
AdS-RN black brane solution of the Einstein-Maxwell-Chern-Simons system takes the form
\begin{equation}
\begin{split}
ds^2 &= 2\ dt\ dr - r^2 f(r,m,q)\ dt^2\ + r^2\ d\vec{x}_{d-1}^2\ ,\qquad
\fA = - \Phi(r,q) \ dt\, , 
\end{split}
\end{equation}
where
\begin{equation}
\begin{split}\label{eq:f-onshell}
f(r, m , q) &\equiv 1- \frac{m}{r^d} + \half \kappa_q \frac{q^2}{r^{2(d-1)}} \ ,\qquad
\Phi(r, q) \equiv \frac{q}{r^{d-2}} \, . 
\end{split}
\end{equation}
The Einstein-Maxwell equations in
(\ref{eq:EinsteinMaxwellEq})
 then lead to the following relation between $f(r, m, q)$ and $\Phi(r, q)$ :
\begin{equation}\label{eq:dfdrIdentity}
\begin{split}
 \frac{d}{dr}\brk{r^{d+1}\frac{df}{dr} +\kappa_q (d-1) q\Phi } &= 0\, ,  \\
\frac{d}{dr} \brk{ r^{d-1} f\ \frac{d \Phi}{dr}+(d-2) q f } &=0\, . 
\end{split}
\end{equation}

For later convenience, we introduce the function $\PhiT(r, m, q)$ via
\begin{equation}
\PhiT(r, m , q) \equiv \half r^2 \frac{df}{dr} = \frac{1}{2r^{d-1}} \brk{ m d- \kappa_q(d-1)\frac{q^2}{r^{d-2}} }\, . 
\end{equation}
The function $\PhiT$  satisfies $\PhiT(r=\rH) =2\pi T$ where
$\rH$ is the radius of the outer horizon and $T$ is the Hawking temperature
of the black brane.

We will reparametrize the solution in terms of two other variables $\{\rH,Q\}$ where
$\rH$ is the radius of the outer horizon and $Q$ is a measure of the total charge of the AdS
RN solution:
\begin{equation}
\begin{split}
m &= \rH^d \brk{1+ \half \kappa_q Q^2 } \ ,\qquad
q =  \rH^{d-1}  Q\, . \\
\end{split}
\end{equation}
The thermodynamic charges and potentials of this solution are parametrized by 
($\mathcal{M}$: mass density, $\mathcal{S}$: entropy density, 
$\mathcal{Q}$: electric charge density, $\mu$: chemical potential)
\begin{eqnarray}
\mathcal{M} = \frac{(d-1)}{16\pi G_{_N}}\ m = \frac{(d-1)}{16\pi G_{_N}}\ \rH^d \brk{1+ \half \kappa_q Q^2 } \,,\qquad \mathcal{S} =  \frac{\rH^{d-1}}{4G_{_N}}\,,\\
\mathcal{Q} =  \frac{(d-1)\kappa_q}{16\pi G_{_N}}\  q =  \frac{(d-1)}{16\pi G_{_N}}\ \kappa_q \rH^{d-1}  Q \,,\qquad  \mu = \Phi(\rH) =  Q\ \rH \,,\\
T = \frac{\rH^2 f'(\rH) }{ 4\pi } = \frac{1}{2\pi}\PhiT(\rH) = \frac{\rH}{4\pi}\brk{d -\half (d-2) \kappa_q Q^2 } \,.
\end{eqnarray}

This AdS-RN black-brane solution can be boosted into a solution of the form
\begin{equation}
\begin{split}\label{eq:boostedRN}
ds^2&=-2 u_{\mu} \, dx^{\mu}(dr+ r \mathcal{W}_{\nu} dx^{\nu})
-r^2 f(r,m,q) \, u_{\mu} u_{\nu}  \, dx^{\mu} dx^{\nu}+r^2 P_{\mu\nu}\, dx^{\mu} dx^{\nu}\, , \\
\fA &= \Phi(r,q) \, u_{\mu} dx^{\mu} ,
\end{split}
\end{equation}
where $u^\mu$ is the constant velocity that defines the rest frame of the blackbrane  and
\begin{equation}\label{eq:PmnandWmu}
P_{\mu\nu} =g_{\mu\nu}+u_{\mu}u_{\nu}\ ,\quad
\mathcal{W}_{\mu}=(u^{\nu}\nabla_{\nu}) u_\mu-\frac{\nabla_{\nu}u^\nu}{d-1}u_\mu\ .
\end{equation}
Here $P_{\mu\nu} $ is the projection operator and
$\mathcal{W}_\mu$ is the hydrodynamic Weyl connection introduced in \cite{Loganayagam:2008is}. 

\subsection{\texorpdfstring{AdS$_{d+1}$}{AdS}-Kerr-Newman metric: second order }

We will now consider the Einstein-Maxwell system with Chern-Simons terms turned off.
The AdS$_{d+1}$ Kerr-Newman solution -- which is unknown -- of this system can be found in a derivative expansion
via the fluid/gravity correspondence by starting with the boosted AdS-RN black brane metric (\ref{eq:boostedRN}).
The metric and gauge field can be written in the form
\begin{equation}
\begin{split}
\label{eq:2ndOrderMetric}
ds^2 &=
-2 u_{\mu} dx^{\mu}\prn{ dr+ r\, \mathcal{W}_{\nu}\,  dx^{\nu}-\mathcal{S}_{\nu\lambda}u^\lambda dx^\nu } \\
& -r^2 f(r,m,q) \, u_{\mu} u_{\nu}  \, dx^{\mu} dx^{\nu}
+\brk{ r^2 P_{\mu\nu}-{\omega_{\mu}}^{\alpha}\omega_{\alpha\nu} }\, dx^{\mu} dx^{\nu}\\
&+g(r,m,q)\,\omega_{\alpha\beta}\omega^{\alpha\beta} \,u_{\mu}u_{\nu}\,dx^{\mu}dx^{\nu}
+h(r,m,q)\left[{\omega_{\mu}}^{\alpha}\,\omega_{\alpha\nu}+\frac{1}{d-1} \omega_{\alpha\beta}\omega^{\alpha\beta} P_{\mu\nu}\right]dx^{\mu}dx^{\nu}\,,\\
\fA &= \Phi(r,q) \brk{1- \frac{1}{2r^2} \omega_{\alpha\beta}\omega^{\alpha\beta} }\, u_{\mu} dx^{\mu} ,
\end{split}
\end{equation}
where $P_{\mu\nu},\mathcal{W}_{\mu},f(r, m, q)$ and $ \Phi(r, q)$ are given by
Eq.~(\ref{eq:PmnandWmu}) and Eq.~(\ref{eq:f-onshell}) while
\begin{equation}
g(r, m, q)= -\frac{m}{2 r^d} + \frac{\kappa_q}{2}\frac{q^2}{r^{2(d-1)}} \left[1-\frac{1}{(d-1)(d-2)}\right]\, , 
\end{equation}
and
\begin{equation}
h(r, m , q)=- \frac{d}{d-2}\ \kappa_q\ \frac{ r^2 q^2}{r_{_H}^{2d}}
\int_{r/r_{_H}}^{\infty}\frac{\zeta^d-1}{\zeta^{2d+1} f(\zeta\ r_{_H},m,q)}d\zeta .
\end{equation}

The Weyl covariantized Schouten tensor $\mathcal{S}_{\mu\nu} $ is a tensor defined  in terms of 
the Ricci tensor $R^{bdy}_{\mu\nu}$ and Ricci scalar $R^{bdy}$ of the boundary metric $g_{\mu\nu}$ and the hydrodynamic  Weyl connection $\mathcal{W}_\mu$. It is given by
\begin{equation}\label{eqn:Sdef}
\begin{split}
\mathcal{S}_{\mu\nu} &\equiv
\frac{1}{d-2}\prn{ R^{bdy}_{\mu\nu}-\frac{R^{bdy} }{2(d-1)} g_{\mu\nu} }
+\nabla_\mu \mathcal{W}_\nu + \mathcal{W}_\mu \mathcal{W}_\nu - \frac{1}{2} \mathcal{W}^2 g_{\mu\nu}\\
&\qquad+\frac{1}{d-2} \prn{\nabla_\mu \mathcal{W}_\nu-\nabla_\nu \mathcal{W}_\mu }\, . 
\end{split}
\end{equation}
We have ignored the non-stationary contributions to the fluid gravity metric which are zero for AdS-Kerr-Newman
solution. We have checked using Mathematica that the above metric and gauge field solve the Einstein
as well as the Maxwell equations up to second order and $d+1=11$ provided the $\{m,q,u^\mu\}$ solve the
ideal fluid equations. For the AdS-Kerr-Newman case, we want to further turn off non-stationary contributions
both of which can be implemented by demanding
\begin{eqnarray}
  \nabla_\mu m + d\ \mathcal{W}_\mu m &=0 \,,\qquad
  \nabla_\mu q + (d-1)\ \mathcal{W}_\mu q &= 0\, ,
\end{eqnarray}
  up to second order in a derivative expansion. This, in particular, implies
 \begin{equation}
 \begin{split}
  \nabla_\mu \nabla_\nu  m +  \prn{d\ \nabla_\mu \mathcal{W}_\nu - d^2 \ \mathcal{W}_\mu \mathcal{W}_\nu} m &=0\, ,  \\
  \nabla_\mu \nabla_\nu q +  \prn{(d-1)\ \nabla_\mu \mathcal{W}_\nu - (d-1)^2 \ \mathcal{W}_\mu \mathcal{W}_\nu} q &= 0\, ,\\
  \nabla_\mu \mathcal{W}_\nu-\nabla_\nu \mathcal{W}_\mu &= 0\, . 
 \end{split}
\end{equation}

We will present the fluid variables in a convenient coordinate system for comparison with the rest of the literature.
Consider the AdS$_{d+1}$-Kerr-Newman BHs : we begin by defining two integers
$n$ and $\sigma_{_{CFT}}$ via $d= 2 n+\sigma_{_{CFT}}$ with $\sigma_{_{CFT}}= d\ (\text{mod}\ 2)$ where
$n$ is the number of angular momenta in AdS$_{d+1}$.

We can then parametrize AdS$_{d+1}$ Kerr-Newman solution by a radial 
coordinate $r$, a time co-ordinate $t$ along with
$d-1= 2 n+\sigma_{_{CFT}}-1$ spheroidal coordinates on S$^{d-1}$. We choose
these spheroidal coordinates to be
\begin{itemize}
 \item $n+\sigma_{_{CFT}}$ number of direction cosines $\rad_k$,  obeying $\sum_{k=1}^{n+\sigma_{_{CFT}}}\rad_k^2 = R^2$
 and $\rad_k \geq 0$ where $R$ denotes the radius of the sphere  S$^{d-1}$ . We can thus take the first $n+\sigma_{_{CFT}}-1$
 direction cosines  as independent variables with
 $\rad_{n+\sigma_{_{CFT}} } = (R^2- \sum_{k=1}^{n+\sigma_{_{CFT}}-1} \rad_k^2)^{1/2}\ $.
 \item $n+\sigma_{_{CFT}}$ azimuthal angles $\varphi_i$ with $\varphi_{n+1}=0$ identically.
\end{itemize}
For the sake of simplicity, in what follows, we denote the sums
$\sum_{k=1}^{n+\sigma_{_{CFT}}}$ as just $\sum_k$. 
The angular velocities along the different $\varphi_k$'s
are denoted by $\Omega_k$ (with $\Omega_{n+1}=0$  identically).

It is then convenient to take the dual CFT velocity configuration and the boundary metric of
AdS Kerr-Newman solution as
\begin{equation}
\begin{split}
u^\mu \partial_\mu &\equiv \partial_{t}+ \sum_i \Omega_i \partial_{\varphi_i},  \\
g_{\mu\nu} dx^\mu dx^\nu &\equiv  \frac{1}{\Delta}\left[-dt^2 + \sum_i\left(d\rad_i^2 + \rad_i^2 d\varphi_i^2\right)\right]\, , \\
\Delta &\equiv 1 - \sum_k \Omega_k^2\rad_k^2\, . \\
\end{split}
\end{equation}
This corresponds to a purely rotating velocity configuration in a spacetime which is $\mathbb{R}\times S^{d-1}$ up to a
convenient conformal factor $\Delta$. With this choice of conformal factor, we have $\mathcal{W}_{\mu} = 0$ and
we can take the corresponding $m$ and $q$ to be constant independent of the boundary coordinates.

Using the above configuration, we can then evaluate various tensors appearing in the fluid/gravity metric
\begin{equation}
\begin{split}
\fu &\equiv u_\mu dx^\mu = - \frac{1}{\Delta} dt +\frac{1}{\Delta} \sum_i \Omega_i \rad_i^2 d\varphi_i\, , 
\\
dt+\fu &= \frac{1}{\Delta}\sum_i  \rad_i^2 \Omega_i  \prn{d\varphi_i-\Omega_i\ dt}\, ,  \\
d\varphi_k+\Omega_k \fu &= d\varphi_k-\Omega_k dt +\frac{\Omega_k}{\Delta}\sum_i  \rad_i^2 \Omega_i  \prn{d\varphi_i-\Omega_i\ dt} \, ,  \\
P_{\mu\nu} dx^\mu dx^\nu
&= -\frac{1}{\Delta}(dt+\fu)^2
+\frac{1}{\Delta} \sum_i\brk{ d\rad_i^2 + \rad_i^2 \prn{ d\varphi_i+\Omega_i \fu}^2 } \, , \\
\mathcal{W}_{\mu} &= 0.\\
\end{split}
\end{equation}

Further we define the following scalars and tensors which will be useful in what follows
{\small{
\begin{equation}
\begin{split}
\Upsilon_{\mu\nu} dx^\mu dx^\nu &\equiv \frac{1}{R^2\Delta}dt^2
-\frac{1}{\Delta}\sum_k \Omega_k^2 \prn{d\rad_k^2 + \rad_k^2 d\varphi_k^2}
- \prn{\frac{1}{\Delta}\sum_k \Omega_k^2\rad_k d\rad_k}^2 \, , \\
P_\mu{}^\alpha P_\nu{}^\beta \Upsilon_{\alpha\beta}dx^\mu dx^\nu &\equiv \frac{1}{R^2\Delta}(dt+\fu)^2
-\frac{1}{\Delta}\sum_k \Omega_k^2 \brk{d\rad_k^2 + \rad_k^2\ (d\varphi_k+\Omega_k \fu)^2}
- \prn{\frac{1}{\Delta}\sum_k \Omega_k^2\rad_k d\rad_k}^2\, , \\
\Delta_1 &\equiv \frac{1}{2\Delta}\brk{\frac{1}{R^2}- \sum_k \rad_k^2 \Omega_k^4 }\ ,\qquad
\Delta_2 \equiv \frac{1}{R^2}+ \sum_k \Omega_k^2\ , \\
\end{split}
\end{equation}
}}
which obey $ \Upsilon_{\mu\nu} u^\mu u^\nu  =  2 \Delta_1$
and  $ \Upsilon_{\mu\nu} P^{\mu\nu}  = 4 \Delta_1 - 2 \Delta_2$. In terms of these, we can write
\begin{equation}
\begin{split}
\mathcal{S}_{\mu\nu} &= \Delta_1\ g_{\mu\nu} + \Upsilon_{\mu\nu}\, ,  \\
\omega_\mu{}^\lambda \omega_{\lambda\nu} &= P_\mu{}^\alpha P_\nu{}^\beta \Upsilon_{\alpha\beta}\, ,  \\
\end{split}
\end{equation}
and 
\begin{eqnarray}
\omega_{\alpha\beta}\omega^{\alpha\beta} &= -P^{\alpha\beta}\Upsilon_{\alpha\beta}=
-4\Delta_1 +2 \Delta_2\,,\qquad
\end{eqnarray}
\begin{equation}
\begin{split}
&\brk{2 u_{\mu}\mathcal{S}_{\nu\lambda}u^\lambda  -\omega_{\mu}{}^{\lambda}\omega_{\lambda\nu} }
dx^\mu dx^\nu \\
&\qquad =  -\frac{1}{R^2\Delta}dt^2 +\sum_k \frac{\Omega_k^2}{\Delta}\prn{d\rad_k^2 + \rad_k^2 d\varphi_k^2}
+ \prn{\frac{1}{\Delta}\sum_k \Omega_k^2\rad_k d\rad_k}^2\, ,
\end{split}
\end{equation}
so that when one substitutes all these
into Eq.~(\ref{eq:2ndOrderMetric})
the AdS-Kerr-Newman metric becomes
that in Eddington-Finkelstein-like coordinates.


\section{Forms and its products in a derivative expansion}
\label{sec:productRiemanns}

In this Appendix, we summarize the expressions
and properties of the metric, the Christoffel connection 1-form and the curvature 2-form at different orders in the derivative expansion.\\
We also calculate various products of the curvature 2-form which we will use for the evaluation of the Einstein and Maxwell sources 
in the Appendices \ref{sec:MaxwellSources} and \ref{sec:EinsteinSources}.

At the 0th and 1st orders, we provide the explicit expression as a result of
tedious but straightforward calculation, while at the 2nd and higher orders, we just summarize
some properties of the curvature 2-form derived from the tensor and form structures as well as the
symmetries,  since they are sufficient for the purpose of this paper.

Throughout the text we will label the $m$-th order of the derivative expansion of a form $\mathcal{B}$ as  $(\B{m})^{a}{}_{b}$. Moreover, we denote the product of $k$ matrix-valued 2-forms, $(\B{m_1}),  (\B{m_2}), \cdots$ and  $(\B{m_k})$, as $(\B{m_1}\B{m_2}\ldots \B{m_{k}})$, so that the matrix-valued 2-forms inside the brackets are always multiplied through matrix multiplication. For example, $(\B{m_1}\B{m_2}\B{m_{3}})^a{}_b
=(\B{m_1})^a{}_{c_1}\wedge (\B{m_2})^{c_1}{}_{c_2}\wedge (\B{m_3})^{c_2}{}_{b}$
is a matrix-valued 6-form. We also abbreviate the $k$ product of $(\B{m})$ by $(\B{m}^k)$.

\subsection{Metric, connection 1-form and curvature 2-form}
\label{sec:Riemanns0thmetric}
We consider the metric in Eq.~(\ref{eq:fluidGravityAnsatz}) up to first order in the derivative expansion.
The metric $G_{ab}$ with $x^a=\{r,x^{\mu}\}$ and its inverse are given by
\bea
G_{rr}&=&0,\quad
G_{r\mu} = - u_\mu ,\quad
G_{\mu\nu}= -2 \Psi(r)u_{\mu} u_{\nu}
+r^2 P_{\mu\nu}, \nonumber\\
G^{rr}&=& 2\Psi(r)  ,\quad
G^{r\mu}= u^\mu,\quad
G^{\mu\nu}=r^{-2} P^{\mu\nu},
\eea where the $\Psi(r)\equiv r^2 f(r)/2$ and $f(r)$ as defined in Eq.(\ref{eq:fluidGravityAnsatz}). If we imposed the equations of motion, then the explicit expression of $r^2 f(r)/2$ takes the form in Eq.~(\ref{eq:f-onshell}).

Let us recall that in this paper we set the boundary metric to be flat, $g_{\mu\nu}=\eta_{\mu\nu}$ where $(\eta_{\mu\nu}) = {\rm diag}(-1,1,1,\cdots,1)$. Lowering and raising
of the Greek boundary indices indices ($\mu, \nu, \rho, \sigma, ...$) are done by $\eta_{\mu\nu}$
and its inverse $\eta^{\mu\nu}$ ($u^{\mu} = \eta^{\mu\nu}u_{\nu}$ and
$P^{\mu\nu} = \eta^{\mu\rho}\eta^{\nu\sigma}P_{\rho\sigma}$, for example).
Furthermore, we take the velocity vector $u^\mu$ (normalized as $u^2\equiv g^{\mu\nu} u_{\mu} u_{\nu}=-1$) corresponding to pure rotation, 
where  $u^\mu \omega_{\mu\nu}=0$ and $\partial_{(\mu}u_{\nu)}=0$. Here, $\omega_{\mu\nu}\equiv (\partial_{\mu} u_{\nu}-\partial_{\nu} u_{\mu})/2$ is the vorticity.
We also note that the projection operator $P_{\mu\nu} \equiv g_{\mu\nu}+u_{\mu}u_{\nu}$
satisfies $P^{\mu\nu} u_\nu=0$ and  $P_{\mu}^{\,\,\, \rho} P_{\rho\nu} =P_{\mu\nu}$.

By using the above metric, the components of the Christoffel connection are calculated as
\bea
\Gamma^a_{\ rr}&=&0,\quad
 \Gamma^r_{\ r\nu}= u_\nu \Psi'
,\quad
\Gamma^r_{\ \rho\nu}
=
 2\Psi\left[
 \Psi' u_{\rho}u_{\nu}
-r P_{\rho\nu}
\right]
,\quad
\Gamma^\mu_{\ r \nu}
= r^{-2}
\left[ r P^\mu{}_\nu
+\omega^\mu{}_{\nu}
\right], \nonumber\\
\Gamma^\mu_{\ \rho\nu}
&=&
u^{\mu}
\left[
  u_\rho u_\nu \Psi'
-r P_{\rho \nu}
\right]
-2 r^{-2}   (2\Psi-r^2)
u_{(\nu}\omega_{\rho)}{}^{\mu}\, ,
 \eea
or, in terms of the differential form, the connection 1-form defined
as $\fGamma^a{}_b\equiv \Gamma^a{}_{bc} dx^c$
(the latin indices label all the spacetime directions in the bulk, including $r$-direction) is
\bea
\fGamma^r{}_r&=&  \Psi' \fu ,\quad 
\fGamma^r{}_\rho= u_\rho \Psi' dr + 2\Psi\left( \Psi'u_\rho \fu - r P_{\rho\nu} dx^\nu\right), \quad
\fGamma^\mu{}_r=r^{-2}\left(
r P^\mu{}_\nu+\omega^{\mu}{}_\nu
\right)dx^\nu, \nonumber\\
\fGamma^\mu{}_\rho&=&
r^{-2}\left(r P^\mu{}_\rho+\omega^\mu{}_\rho\right)dr
+\left[u^\mu u_\rho \Psi' + r^{-2}(2\Psi-r^2)\omega^\mu{}_\rho \right]\fu
\nonumber\\
& &
-r u^\mu P_{\rho\nu}dx^\nu
+r^{-2}\left(2 \Psi-r^2\right) u_\rho \omega^\mu{}_\nu dx^\nu,
\eea where the velocity 1-form is $\fu\equiv u_\mu dx^\mu$. 

Employing the notation introduced already at the beginning of this Appendix \ref{sec:productRiemanns},
let us denote the $m$-th order term of  the $U(1)$ field strength $\form{F}$ by $(\F{m})$. 
In particular, the first three terms in the derivative expansion are
\be\label{eq:F0F1F2}
(\F{0})= \Phi' dr \wedge \fu ,\quad
(\F{1})= 2\Phi\fomega ,\quad
(\F{2})\sim  dr \wedge \fu \times \omega^{\alpha\beta}\omega_{\alpha\beta}\,.
\ee 
Note that $d\fu=2\fomega$ and $\fomega\equiv (1/2)\,\omega_{\mu\nu}dx^\mu \wedge dx^\nu$.
We remind the reader of Eq.~(\ref{eq:2ndOrderMetric}), from which $(\F{m})$ for $m\le2$ above are computed.

Next, we evaluate the curvature 2-form defined from the Riemann tensor $R^a{}_{bcd}$ as $\fR^a{}_b\equiv (1/2) R^a{}_{bcd} dx^c \wedge dx^d$ at 0th and 1st orders:\\
At 0th order, we have
\bea
(\R{0})^r{}_r
&=&\Psi''dr\wedge \fu,  \nonumber\\
(\R{0})^{\mu}{}_{r}
&=& r^{-1} \Psi'  dx^\mu \wedge \fu,
 \nonumber\\
 (\R{0})^{r}{}_{\rho}
 &=&
 r \Psi' P_{\rho\nu}dx^\nu\wedge dr
-2\Psi \Psi'' u_\rho  \fu\wedge dr,
\\
(\R{0})^\mu{}_{\rho}
&=&
-r\left(\partial_r \left(r^{-1} \Psi'\right)\right) u^\mu u_\rho
\fu\wedge dr
+r^{-1} \Psi' u_\rho dx^\mu \wedge dr
\nonumber\\
& &
-r \Phi_T u^\mu
P_{\rho\nu} dx^\nu \wedge \fu
+2 r^{-1} \Psi dx^\mu \wedge\left[
\Psi' u_\rho \fu
-r P_{\rho\nu} dx^\nu
\right], \nonumber
\eea
and at 1st order,
\bea
(\R{1})^r{}_{r} &=&(2 \Phi_T \fomega), \nonumber\\
(\R{1})^\mu{}_{r} &=&
-r^{-2}   \fu
\wedge (\Phi_T \omega^\mu{}_{\nu} dx^\nu), \\
(\R{1})^r{}_{\rho} &=&
  dr\wedge(\Phi_T \omega_{\rho\nu} dx^\nu)
+2 \Psi \left[
u_\rho (2\Phi_T\fomega)
+  \fu\wedge (\Phi_T\omega_{\rho\nu} dx^\nu)
\right], \nonumber\\
(\R{1})^\mu{}_{\rho} &=&
-2 r^{-2} \Phi_T \omega^\mu{}_{\rho}\fu \wedge dr
+r^{-2}  u_\rho dr\wedge  (\Phi_T\omega^\mu{}_{\nu} dx^\nu)
+ u^\mu\left[
  u_\rho (2\Phi_T \fomega)
+ \fu \wedge (\Phi_T\omega_{\rho\nu} dx^\nu )\right].
 \nonumber
\eea
Here we defined
$\Phi_T
\equiv  (1/2)  r^2 \frac{df}{dr}=r^2 \partial_r(r^{-2} \Psi)=\Psi'-2 \Psi r^{-1} $.

At 2nd order the curvature 2-form has non-trivial contributions
coming purely from the 0th order metric (and its derivatives).
To distinguish from the whole 2nd order curvature 2-form $(\R{2})$, we denote these by $(\R{2}')$ and they are given by
\bea\label{eq:R2prime}
(\R{2}')^r{}_{r} &=&0, \nonumber\\
(\R{2}')^\mu{}_{r} &=&
 -
r^{-4} \left(\omega^\mu{}_\nu \omega^\nu{}_\sigma dx^\sigma\right)\wedge dr
-r^{-4} \left(2\Psi-r^2\right) \left(\omega^\mu{}_\nu \omega^\nu{}_\sigma dx^\sigma\right)\wedge \fu,
 \nonumber\\
 (\R{2}')^r{}_{\rho} &=&0,
 \nonumber\\
(\R{2}')^\mu{}_{\rho} &=&
r^{-2} \partial_\nu \omega^{\mu}{}_\rho dx^\nu \wedge dr \label{eq:2ndcurvaturepart}\\
& &
+r^{-2}(2\Psi-r^2)
\left[
\partial_\nu \omega^{\mu}{}_\rho dx^\nu \wedge \fu
+2\omega^{\mu}{}_\rho\fomega
-(\omega_{\rho\sigma} dx^\sigma)\wedge(\omega^\mu{}_\nu dx^\nu)
\right]\nonumber\\
 & &
-r^{-4}\left(2\Psi-r^2\right) \left(
u_\rho \omega^\mu{}_\sigma \omega^\sigma{}_\nu  dx^\nu \right)\wedge dr
-r^{-4}\left(2\Psi-r^2\right)^2 \left(
u_\rho \omega^\mu{}_\sigma \omega^\sigma{}_\nu  dx^\nu \right)\wedge \fu. \nonumber
\eea
Note that we have used $\partial_{[\sigma} \omega^\mu{}_{\nu]}=0$.
As for the rest of the contribution to $(\R{2})$ coming from the metric at the 2nd order,
we classify it in the Appendix \ref{sec:2ndOrderRiemann}.
We note that the above contribution purely coming from the 0th order metric
is special to the 2nd order curvature 2-form as can be
guessed from the fact that the curvature contains two derivatives of metric.
For the 3rd or higher order, there is no contribution coming purely from 0th order metric.

\subsection{Products of curvature 2-forms from the metric at 0th and 1st order}

This subsection focuses on the various products of the curvature 2-forms
containing only $(\R{0})$ and $(\R{1})$. These kind of products can be classified in a simple way taking into account certain observations derived from the direct computation.
For instance, on one hand, the final expression of nontrivial products of five or more curvature 2-forms
at these orders reduces to the products of four or fewer than four  curvature 2-forms
multiplied by some power of $(2\Phi_T\fomega)$. On the other hand, keeping track of $dr$ and $\fu$ in the course of
the multiplication is useful to find non-vanishing contributions; each of the $dr$ and $\fu$ is allowed to appear at most once in a given non-trivial product
of the curvature 2-forms.

\subsubsection{Products of two curvature 2-forms}
To this order, we find three kind of terms.
First, the product of two $(\R{0})$'s. One immediately finds
\be
(\R{0}\R{0})=0\,. \label{eq:R0R0}
\ee
It is worth stressing the difference between $(\R{0})$ and $0$. The former is the 0th order term of the curvature 2-form,
while the latter means zero in the usual sense. 

Secondly, the mixed products of $(\R{0})$ and $(\R{1})$ read
\bea
\left(\R{0}\R{1}\right)^r{}_r &=&
 r^{-1} \partial_r(r \Phi_T) dr\wedge \fu \wedge (2\Phi_T \fomega), \nonumber \\
\left(\R{0}\R{1}\right)^\mu{}_r &=& r^{-1}\Phi_T dx^\mu \wedge \fu \wedge (2\Phi_T \fomega),  \nonumber\\
\left(\R{0}\R{1}\right)^r{}_\rho &=&0, \\
\left(\R{0}\R{1}\right)^\mu{}_\rho &=&
- \Phi_T' u^\mu u_\rho dr\wedge \fu \wedge (2\Phi_T \fomega)
- r^{-1} \Phi_T u_\rho dx^\mu \wedge dr\wedge (2\Phi_T \fomega)
\nonumber\\
& &-2r^{-1} \Phi_T dr \wedge \fu \wedge dx^\mu \wedge (\Phi_T
\omega_{\rho\nu} dx^\nu), \nonumber
\eea
and
\bea\
\left(\R{1}\R{0}\right)^r{}_r &=& 
 r^{-1} \partial_r(r \Phi_T) dr\wedge \fu \wedge (2\Phi_T\fomega), \nonumber\\
 \left(\R{1}\R{0}\right)^\mu{}_r &=&0, \\
\left(\R{1}\R{0}\right)^r{}_\rho &=&- r\Phi_T dr\wedge (P_{\rho\nu} dx^\nu)\wedge (2\Phi_T\fomega)
+2 r \Psi \Phi_T (P_{\rho\nu }dx^\nu)\wedge \fu\wedge (2\Phi_T\fomega), \nonumber \\
\left(\R{1}\R{0}\right)^\mu{}_\rho &=&
2r^{-1}\Phi_T   (P_{\rho\nu} dx^\nu)  \wedge \fu\wedge dr \wedge (\Phi_T \omega^\mu{}_\sigma dx^\sigma)
+ r \Phi_T  u^\mu P_{\rho \nu} dx^\nu \wedge \fu \wedge (2\Phi_T \fomega)
\nonumber\\
& &
+  r^{-1}\partial_r(r \Phi_T)  u^\mu u_\rho \fu\wedge dr \wedge (2 \Phi_T\fomega). 
\eea
For future reference it is important to observe that all the terms appearing in $(\R{0}\R{1})$ and $(\R{1}\R{0})$ are proportional to either $dr$ or $\fu$ and hence vanish once multiplied with terms proportional to $dr\wedge \fu$.

A parallel construction is possible, thirdly, for the product of two $(\R{1})$'s
\bea
\left(\R{1}\R{1}\right)^r{}_r &=& (2\Phi_T \fomega)^2, \nonumber \\
\left(\R{1}\R{1}\right)^\mu{}_r &=&
-r^{-2} \fu\wedge(\Phi_T \omega^\mu{}_\nu dx^\nu )\wedge (2\Phi_T \fomega),
 \nonumber\\
\left(\R{1}\R{1}\right)^r{}_\rho &=&
  dr\wedge (\Phi_T\omega_{\rho\nu} dx^\nu)\wedge (2\Phi_T \fomega), \\
\left(\R{1}\R{1}\right)^\mu{}_\rho &=&
- u^\mu u_\rho (2 \Phi_T\fomega)^2 - 2r^{-2} \Psi  u_\rho  \fu \wedge (\Phi_T\omega^\mu{}_\nu dx^\nu)\wedge (2\Phi_T\fomega)
 \nonumber\\
 & &-  r^{-2}  u_\rho dr\wedge (\Phi_T\omega^\mu{}_\nu dx^\nu)  \wedge (2\Phi_T \fomega)
  -u^\mu \fu\wedge (\Phi_T\omega_{\rho\nu}dx^\nu) \wedge (2\Phi_T \fomega). \nonumber
\eea

A final remark concerns the wedging of any of the $(\R{1}\R{1})$ with $dr \wedge \fu$. In this cases, the only non-zero
components obey
\be
dr\wedge \fu
\wedge(\R{1}\R{1})^r{}_r=dr\wedge \fu
\wedge (2\Phi_T\fomega)^2,\quad
dr\wedge \fu
\wedge (\R{1}\R{1})^\mu{}_\rho=-u^\mu u_\rho dr\wedge \fu
\wedge (2\Phi_T\fomega)^2.
\ee

\subsubsection{Products of three curvature 2-forms}
Direct calculation shows that the products of three curvature 2-forms at the
0th or 1st order are given as follows. For the products of two $(\R{0})$'s and one $(\R{1})$,
$(\R{0}\R{0}\R{1})=0$ and $(\R{1}\R{0}\R{0})=0$ as a result of \eqref{eq:R0R0}.
Therefore the only nontrivial product is  $(\R{0}\R{1}\R{0})$ which is calculated as
\bea
\left(\R{0}\R{1}\R{0}\right)^r{}_r &=&\left(\R{0}\R{1}\R{0}\right)^r{}_\rho=\left(\R{0}\R{1}\R{0}\right)^\mu{}_r=0,
\nonumber \\
\left(\R{0}\R{1}\R{0}\right)^\mu{}_\rho&=&
2\Phi_T^2 dr\wedge \fu \wedge dx^\mu \wedge (P_{\rho\nu} dx^\nu )\wedge (2\Phi_T \fomega).
\eea
There are three kinds of products containing one $(\R{0})$ and two $(\R{1})$'s:
\bea
\left(\R{0}\R{1}\R{1}\right)^r{}_r &=&
 r^{-1}\partial_r(r \Phi_T)dr\wedge \fu \wedge (2\Phi_T\fomega)^2, \nonumber\\
\left(\R{0}\R{1}\R{1}\right)^\mu{}_r &=&
 r^{-1} \Phi_T  dx^\mu \wedge \fu \wedge(2\Phi_T \fomega)^2,
\nonumber\\
\left(\R{0}\R{1}\R{1}\right)^r{}_\rho &=&
2\Psi  r^{-1}\partial_r \left(r  \Phi_T \right) u_\rho dr\wedge \fu \wedge (2\Phi_T\fomega)^2,
\\
\left(\R{0}\R{1}\R{1}\right)^\mu{}_\rho &=&
 \Phi_T' u^\mu u_\rho dr\wedge \fu \wedge (2\Phi_T\fomega)^2
+ r^{-1} \Phi_T u_\rho dx^\mu\wedge dr\wedge (2\Phi_T\fomega)^2\nonumber\\
& &
+2 r^{-1} \Psi \Phi_T u_\rho dx^\mu \wedge \fu \wedge (2\Phi_T\fomega)^2, \nonumber
\eea
\bea
\left(\R{1}\R{0}\R{1}\right)^r{}_r &=&
 \Phi_T' dr\wedge \fu\wedge (2\Phi_T\fomega)^2,
 \nonumber\\
\left(\R{1}\R{0}\R{1}\right)^\mu{}_r& =&0, \nonumber \\
\left(\R{1}\R{0}\R{1}\right)^r{}_\rho &=&
 2\Psi \Phi_T' u_\rho dr\wedge \fu\wedge (2\Phi_T\fomega)^2,\\
\left(\R{1}\R{0}\R{1}\right)^\mu{}_\rho &=&
   \Phi_T' u^\mu u_\rho dr\wedge \fu \wedge (2\Phi_T\fomega)^2, \nonumber
\eea
\bea
\left(\R{1}\R{1}\R{0}\right)^r{}_r&=&
 r^{-1} \partial_r (r \Phi_T)dr\wedge \fu \wedge(2\Phi_T \fomega)^2, \nonumber \\
\left(\R{1}\R{1}\R{0}\right)^\mu{}_r&=&0, \\
\left(\R{1}\R{1}\R{0}\right)^r{}_\rho&=&
 2\Psi r^{-1} \partial_r (r \Phi_T ) u_\rho dr\wedge \fu \wedge(2\Phi_T \fomega)^2
+ r \Phi_T P_{\rho\nu} dx^\nu \wedge dr \wedge (2\Phi_T \fomega)^2,
\nonumber\\
\left(\R{1}\R{1}\R{0}\right)^\mu{}_\rho&=& r \partial_r(r^{-1} \Psi') u^\mu u_\rho dr\wedge \fu \wedge(2\Phi_T \fomega)^2
+r \Phi_T u^\mu \fu \wedge P_{\rho\nu} dx^\nu\wedge (2\Phi_T\fomega)^2. \nonumber
\eea
In the end, the product of the three $(\R{1})$'s is
\bea
\left(\R{1}\R{1}\R{1}\right)^r{}_r &=&(2 \Phi_T \fomega)^3,  \nonumber\\
\left(\R{1}\R{1}\R{1}\right)^\mu{}_r &=&
- r^{-2} \fu \wedge (\Phi_T  \omega^\mu{}_\nu dx^\nu )\wedge (2\Phi_T\fomega)^2,
\\
\left(\R{1}\R{1}\R{1}\right)^r{}_\rho &=&
 2\Psi  u_\rho (2\Phi_T\fomega)^3+2 \Psi  \fu \wedge(\Phi_T \omega_{\rho\nu} dx^\nu) \wedge (2\Phi_T\fomega)^2
 \nonumber\\
 & &+ dr \wedge (\Phi_T \omega_{\rho\nu} dx^\nu) \wedge
 (2\Phi_T \fomega)^2,  \nonumber\\
\left(\R{1}\R{1}\R{1}\right)^\mu{}_\rho &=&
 r^{-2}  u_\rho dr \wedge(\Phi_T\omega^\mu{}_\nu dx^\nu) \wedge  (2\Phi_T\fomega)^2\nonumber\\
 & &
-2r^{-2} dr \wedge \fu \wedge(\Phi_T \omega^\mu{}_\sigma dx^\sigma) \wedge  (\Phi_T \omega_{\rho\nu}dx^\nu)\wedge  (2\Phi_T\fomega)\nonumber\\
 & &+ u^\mu u_\rho (2\Phi_T\fomega)^3
  +u^\mu \fu \wedge( \Phi_T\omega_{\rho\nu}dx^\nu) \wedge  (2\Phi_T\fomega)^2. \nonumber
\eea
We also note that 
\be
dr\wedge \fu
\wedge(\R{1}\R{1}\R{1})^a{}_b=
dr\wedge \fu
\wedge(\R{1})^a{}_b \wedge (2\Phi_T \fomega)^2.
\ee

\subsubsection{Products of four curvature 2-forms}
\label{subsec:foruproducts}
Let us now turn to \eqref{eq:R0R0} and consider the products of four curvature 2-forms. One immediately finds that these vanish.
Hence, the product of four curvature 2-forms containing three or more $(\R{0})$'s are trivial.

For the same reason \eqref{eq:R0R0}, three possible products containing two $(\R{0})$'s and two $(\R{1})$'s,
along with $(\R{0}\R{0}\R{1}\R{1})$,  $(\R{1}\R{0}\R{0}\R{1})$ and $(\R{1}\R{1}\R{0}\R{0})$ are also trivial. 
By a direct inspection we find that the other three four curvature 2-forms containing two $(\R{0})$'s
and two $(\R{1})$'s also vanish:
\be\label{4thingszero}
(\R{0}\R{1}\R{0}\R{1})=(\R{0}\R{1}\R{1}\R{0})=(\R{1}\R{0}\R{1}\R{0})=0\,.
\ee

We consider the products containing three or four $(\R{1})$'s. The special cases of 
$(\R{0}\R{1}\R{1}\R{1})$ and $(\R{1}\R{1}\R{1}\R{0})$, can be reduced to products of two curvature 2-forms at the 0th or 1st order
wedged with $(2\Phi_T \fomega)^2$ :
\be\label{eq:0111n1110}
\left(\R{0}\R{1}\R{1}\R{1}\right)
=(2 \Phi_T \fomega)^2\wedge \left(\R{0}\R{1}\right), \qquad
\left(\R{1}\R{1}\R{1}\R{0}\right)
=(2\Phi_T\fomega)^2\wedge \left(\R{1}\R{0}\right),
\ee
where the expression such as above should be understood as $\left(\R{0}\R{1}\R{1}\R{1}\right)^a{}_b
=(2 \Phi_T \fomega)^2\wedge \left(\R{0}\R{1}\right)^a{}_b$, etc..
The remaining products containing three $(\R{1})$'s either vanish or reduce to products of three curvature 2-forms
\bea
\left(\R{1}\R{0}\R{1}\R{1}\right)^r{}_r&=& +(2\Phi_T \fomega)\wedge \left(\R{1}\R{0}\R{1}\right)^r{}_r,\qquad
\left(\R{1}\R{0}\R{1}\R{1}\right)^\mu{}_r=\left(\R{1}\R{0}\R{1}\R{1}\right)^r{}_\rho=0,\nonumber\\
\left(\R{1}\R{0}\R{1}\R{1}\right)^\mu{}_\rho&=&- (2\Phi_T \fomega)\wedge \left(\R{1}\R{0}\R{1}\right)^\mu{}_\rho,
\eea
\bea
\left(\R{1}\R{1}\R{0}\R{1}\right)^r{}_r&=& +(2\Phi_T \fomega)\wedge\left(\R{1}\R{0}\R{1}\right)^r{}_r,\qquad
\left(\R{1}\R{1}\R{0}\R{1}\right)^\mu{}_r=
\left(\R{1}\R{1}\R{0}\R{1}\right)^r{}_\rho=0,\nonumber\\
\left(\R{1}\R{1}\R{0}\R{1}\right)^\mu{}_\rho&=&- (2\Phi_T \fomega)\wedge\left(\R{1}\R{0}\R{1}\right)^\mu{}_\rho.
\eea
Among the products with four $(\R{1})$'s we find that reduce to products of $(\R{1}\R{1})$ as follows
\bea
(\R{1}\R{1}\R{1}\R{1})&=&(2\Phi_T\fomega)^2\wedge  (\R{1}\R{1}).
\eea

\subsubsection{General products of curvature 2-forms}
Having explored several situations in previous subsections, it is interesting to shift our calculations now to
the {\it general products} -- more than four products of  0th and 1st order curvature 2-forms. These can also be categorized and, denoted by $\fchi_m$ $(m\ge0)$, according to the $m$-number of $(\R{0})$'s (wedged with arbitrary number of $(\R{1})$'s) that a given product contains.
Interestingly enough, as we will show below, all general products downgrade to products of four or fewer than four curvature 2-forms
wedged with suitable powers of $(2\Phi_T \fomega)$.

\noindent
\underline{\bf Set $\fchi_0$}\\
The element of the set $\fchi_0$ is either $(\R{1}\R{1})^{k}$ or $(\R{1})\wedge (\R{1}\R{1})^{k-1}$
where $k\ge 1$.
By applying the identity
\be\label{productsof1s}
\left(\R{1}\R{1}\right)^{k+1} =\left(2 \Phi_T \fomega\right)^{2k}\wedge \left(\R{1}\R{1}\right),
\ee  we can always reduce these products to
\be
(\R{1}),\quad (\R{1}\R{1}),\quad (\R{1}\R{1}\R{1}),
\ee
wedged with an appropriate power of $(2\Phi_T \fomega)$.

\noindent
\underline{\bf Set $\fchi_1$}\\
In this case, we can have the following nontrivial possibilities:
\bea
&&\qquad\qquad\qquad\qquad\qquad
(\R{0} \R{1}^{2k}),\quad
(\R{0} \R{1}^{2k+1}),
\\
&&(\R{1}^{2k+1}\R{0}\R{1}^{2l+1}),\quad
(\R{1}^{2k+1}\R{0}\R{1}^{2l}),\quad
( \R{1}^{2k+2}\R{0}\R{1}^{2l+1}),\quad
(\R{1}^{2k+2}\R{0}\R{1}^{2l}), \nonumber
\eea
where $k,l\ge 0$.
Employing the results of the products of four curvature two-forms in subsection
\ref{subsec:foruproducts} and the subsequent relations
\bea 
\left(\R{1}\R{1}\right)^k \wedge\left(\R{1}\R{0}\R{1}\right)\wedge
\left(\R{1}\R{1} \right)^l
&=&\left(2 \Phi_T \fomega\right)^{2k+2l}\wedge
\left(\R{1}\R{0}\R{1}\right),\nonumber\\
\left(\R{1}\R{1}\R{0}\R{1}\R{1}\right)
&=&(2\Phi_T \fomega)^2\wedge \left( \R{1}\R{0}\R{1}\right) , \label{eq:101reduction}
\eea
we can reduce all the possibilities in this category into the following 8 objects
wedged with powers of $(2\Phi_T \fomega)$ to
\bea
&&\qquad (\R{0}) ,\quad
(\R{0}\R{1}),\quad
(\R{1}\R{0}),\nonumber\\
(\R{0}\R{1}\R{1}),\quad
&&(\R{1}\R{1}\R{0}),\quad
(\R{1}\R{0}\R{1}),\quad
(\R{1}\R{0}\R{1}\R{1}),\quad
(\R{1}\R{1}\R{0}\R{1}).
\eea
\noindent
\underline{\bf Set $\fchi_2$}\\
Within this subclass, there are two cases depending on whether a given product starts with $(\R{0})$ or $(\R{1})$.

First we consider the four possible products beginning with $(\R{1})$
\be
(\R{1}^{2k+2} \R{0} \R{1}^{2l+2}\R{0}),\quad
(\R{1}^{2k+2} \R{0} \R{1}^{2l+1}\R{0}),\quad
(\R{1}^{2k+1} \R{0} \R{1}^{2l+2}\R{0}),\quad
(\R{1}^{2k+1} \R{0} \R{1}^{2l+1}\R{0}),
\ee where $k,l\ge0$. Similarly to the $\fchi_1$ set, with the help of the aforementioned identities, 
these reduce to products containing either $(\R{1}\R{0}\R{1}\R{0})$ 
or $(\R{0}\R{1}\R{1}\R{0})$ which vanish.

Secondly, we consider products starting with $(\R{0})$. As in the preceding case,
there are four possibilities ($k,l\ge 0$)
\be\label{eq:fourcases}
(\R{0} \R{1}^{2k+2} \R{0} \R{1}^{2l}),\quad
(\R{0} \R{1}^{2k+2} \R{0} \R{1}^{2l+1}),\quad
(\R{0} \R{1}^{2k+1} \R{0} \R{1}^{2l}),\quad
(\R{0} \R{1}^{2k+1} \R{0} \R{1}^{2l+1}).
\ee
Via Eq.~(\ref{4thingszero}), some identities for the the $\fchi_1$ set
as well as
\be
(\R{0}\R{1}\R{1}\R{1}\R{0})=(2\Phi_T \fomega)^2\wedge (\R{0}\R{1}\R{0}),
\ee we can show that all the non-trivial elements in (\ref{eq:fourcases}) reduce to
\be
(\R{0}\R{1}\R{0}),
\ee
wedged with its corresponding power of $(2\Phi_T \fomega)$.

\noindent
\underline{\bf Set $\fchi_m$ with $m\ge 3$}\\
Recalling that $(\R{0}\R{0})=0$ for the set $\fchi_{m}$ with $m\ge 3$
we essentially have the succeeding structures (where $k,l\ge 0$)
\bea
& &(\ldots \R{0}\R{1}^{2k+1}\R{0}\R{1}^{2l+1}\R{0}\ldots),\qquad
(\ldots \R{0}\R{1}^{2k+2}\R{0}\R{1}^{2l+1}\R{0}\ldots), \nonumber\\
& &(\ldots \R{0}\R{1}^{2k+1}\R{0}\R{1}^{2l+2}\R{0}\ldots),\qquad
(\ldots \R{0}\R{1}^{2k+2}\R{0}\R{1}^{2l+2}\R{0}\ldots),
\eea where $\ldots$ denotes an arbitrary products of 0th and 1st order curvature 2-forms.
All of them reduce to strings containing either $(\R{0}\R{1}\R{0}\R{1})$ or $(\R{1}\R{0}\R{1}\R{0})$ 
as a result of \eqref{eq:101reduction} which in turn vanish. We thus conclude that $\fchi_m=\emptyset$ for $m\ge 3$ 
and, in the following arguments, we consider $\fchi_0$, $\fchi_1$ and $\fchi_2$ only.  

\subsubsection{Summary for products of curvature 2-forms at 0th and 1st order}
The building blocks for products of curvature 2-forms at 0th and 1st order
(that is, by wedging them with certain power of $(2\Phi_T \fomega)$) are
\bea
\fchi_0&:& \{(\R{1}),\quad(\R{1}\R{1}),\quad (\R{1}\R{1}\R{1})\}, \nonumber\\
\fchi_1&:& \{(\R{0}),\,(\R{0}\R{1}),\, (\R{1}\R{0}),\, (\R{0}\R{1}\R{1}),\, (\R{1}\R{1}\R{0}),\,(\R{1}\R{0}\R{1}),\,(\R{1}\R{0}\R{1}\R{1}),\,(\R{1}\R{1}\R{0}\R{1})\},\nonumber\\
\fchi_2&:& \{(\R{0}\R{1}\R{0})\},\nonumber\\
\fchi_m&:& \emptyset \quad \mbox{(for $m\ge 3$)}.
\eea
This reduction of the products allows us to evaluate various quantities in general $d+1$-dimensions,
as explained in the Appendices \ref{sec:MaxwellSources} and \ref{sec:EinsteinSources}.

\subsubsection{Traces of products of curvature 2-forms}
To evaluate the Maxwell and Einstein sources, we also encounter traces of the
products of the curvature 2-forms. Once we have the products of the curvature 2-forms,
the evaluation of the traces is straightforward. Hence, for later use in our computations we will  only consider traces of
products of even numbers of the curvature 2-forms.

With the help of \eqref{eq:R0R0} and the cyclic property of the trace,
the only non-trivial traces made of $(\R{0})$ and $(\R{1})$ are (for $k\ge 0$)
\bea\label{eq:TracesofProductsRiemanns}
  \tr\left[(\R{0}\R{1}
\left(\R{1}\R{1}\right)^{k})
\right]&=&
2   \Phi'_T dr\wedge \fu\wedge (2\Phi_T\fomega)^{2k+1}\, , 
\nonumber\\
 \tr\left[
\left(\R{1}\R{1}\right)^{k+1}
\right]&=&
2(2  \Phi_T\fomega)^{2k+2}.
\eea
Notice that Eq.~(\ref{eq:TracesofProductsRiemanns}) is (up to a factor of 2) exactly the same as products of zeroth and first order gauge field
\bea
(\F{0}\F{1} (\F{1}\F{1})^k) &=& \Phi' dr\wedge \fu \wedge(2\Phi \fomega)^{2k+1} \, , \nonumber\\
(\F{1}\F{1})^{k+1} &=& (2\Phi \fomega)^{2k+2} \, , 
\eea after we have sent $\Phi \rightarrow \Phi_T$. 

Finally, we want to remark that traces of elements of a set $\fchi_1$ are proportional to $dr\wedge\fu$
and that of the elements of $\fchi_m$ with $m\ge2$ are all zero.

\subsection{Curvature 2-forms from the metric at 2nd and 3rd order}
We consider now the curvature 2-form at the 2nd and 3rd order and
summarize their properties. For these cases, it is complicated to
calculate these higher order terms explicitly and completely from the metric.
Instead of the explicit calculation,
we focus on the contraction and form structure as well as in the symmetries of the
curvature 2-form to figure out the kind of terms appearing in the higher order curvature 2-forms.

\subsubsection{2nd order curvature 2-form}
\label{sec:2ndOrderRiemann}
To discuss $\fR$ at the 2nd order,
let us first list out the possible 2-forms at 0th and 1st orders~:
\bea
\mbox{At 0th order}&:&dr\wedge \fu,\quad  dr \wedge dx^\rho,\quad
\fu\wedge dx^\rho,\quad
dx^\rho\wedge dx^\nu, \nonumber\\
\mbox{At 1st order}&:&\fomega,\quad
\omega_{\mu\nu}\times (\mbox{zeroth order forms}).
\eea
 At 2nd order, there could be structures coming from two derivatives of the zeroth order metric. However, since we have calculated the full contribution from the 0th order metric (which is also valid to 1st order) to the curvature
 2-form in \eqref{eq:2ndcurvaturepart}, we only need to discuss the form structures constructed from those appearing in the 2nd order metric and then write the most general 2nd order curvature 2-form consistent with its symmetries. By adding to it the explicit contributions from the 0th order metric,
we can list out all the terms
that can potentially appear in the explicit calculation of the curvature 2-form at the 2nd order.

From the 2nd order metric (see Eq.~(\ref{eq:2ndOrderMetric})), the structure of the 2-forms that could appear at 2nd order are
\bea
&&\omega^{\alpha\beta}\omega_{\alpha\beta}\times (\mbox{zeroth order forms}),\\
&&\partial_\lambda \omega_{\mu}{}^\lambda\times (\mbox{zeroth order forms}),\nonumber\\
& &\omega_\mu{}^{\alpha}\omega_{\alpha\nu}\times (\mbox{zeroth order forms}).
\eea 
We would like to stress some other possibilities.
Structures such as $u^\nu \partial_{\mu}\omega^{\alpha}{}_{\nu} $ reduce to $\omega_{\mu\nu}\omega^{\alpha\nu}$ which have already been captured in the aforementioned classification.
In principle $\partial_\mu \omega_{\nu}{}^\alpha dx^\mu\wedge dx^\nu$ should also be considered, but it will not be present due to $\partial_{[\mu} \omega_{\nu]}{}^\alpha=\partial_{[\mu} \partial_{\nu]} u^\alpha=0$.
And finally, structures like $\partial^{\alpha} \omega_{\mu\nu}dx^{\nu}\wedge dx^\mu$ can be rewritten as $\partial_\mu \omega_{\nu}{}^\alpha dx^\mu\wedge dx^\nu$, which are zero.

We note that the possible zero-derivative tensor structures that one can use to construct appropriately indexed objects are
\be
u_\mu,\quad P_{\mu\nu}\, \quad \text{and, products among them.}
\ee

For later purpose, it is useful to classify the most general 0th order 2-form
consistent with the symmetry of the curvature 2-form.
The curvature 2-form at the 0th order  could have the following structures
(we ignore all non-index/non-form information)
\bea
(\oR{0})^r{}_r &=& dr\wedge \fu,\quad
(\oR{0})^\mu{}_r =
(\heartsuit) u^\mu dr\wedge \fu
+(\heartsuit) dr\wedge P^\mu{}_\sigma dx^\sigma
+\fu \wedge P^\mu{}_\sigma dx^\sigma,
\nonumber\\
(\oR{0})^r{}_\rho& =& u_\rho dr\wedge \fu
+ P_{\rho\sigma } dx^\sigma \wedge dr
+(\heartsuit) P_{\rho\sigma}  dx^\sigma\wedge \fu,
\nonumber\\
(\oR{0})^\mu{}_\rho &=&
u^\mu u_\rho dr\wedge \fu
+(\heartsuit) P^\mu{}_\rho dr\wedge \fu
-u_\rho dx^\mu \wedge dr
+(\heartsuit )u^\mu P_{\rho\sigma} dx^\sigma \wedge dr
+u_\rho dx^\mu \wedge \fu\nonumber\\
&&
+u^\mu P_{\rho\sigma} dx^\sigma \wedge \fu
 + P_{\rho\nu}dx^\nu \wedge dx^\mu.
\eea
The $(\heartsuit)$ structures do not appear in the explicit computations from the zeroth order metric.
We denote by `bar' any matrix-valued two-forms consistent with the symmetry of the curvature 2-form
including the Bianchi identity, while
the un-barred one is the explicit result computed from our metric, i.e. $(\R{n})=(\oR{n})|_{(\heartsuit)\rightarrow 0}$.
Notice that
\bea
\left(\oR{0}\oR{0}\right)^r{}_r&=&\left(\oR{0}\oR{0}\right)^r{}_\rho=\left(\oR{0}\oR{0}\right)^\mu{}_r=0,\quad
\left(\oR{0}\oR{0}\right)^\mu{}_\rho=(\heartsuit)
\left(P^\mu{}_\sigma dx^\sigma\right)\left(P_{\rho\nu} dx^\nu\right)\wedge \fu\wedge dr.\nonumber\\
\eea
In fact,  $(\R{0}\oR{0})$ and $(\oR{0}\R{0})$ have exactly the same form structures as $(\oR{0}\oR{0})$ above without any obvious simplifications. We also have
\be
(\oR{0}\oR{0}\oR{0})=0.
\ee

At second order, for example, we can have the structure of the form $\omega^{\alpha\beta}\omega_{\alpha\beta}\times (\oR{0})$.
Let $(\R{2}'')$ be a matrix two-form which include all possible structures constructed from 2nd order metric (as discussed above) {\it and} the contributions from
the 0th order metric calculated explicitly (i.e. $(\R{2}')$ in Eq.~(\ref{eq:R2prime})). In all,
\bea
(\R{2}'')^r{}_r &=&
\omega^{\alpha\beta}\omega_{\alpha\beta} (\oR{0})^r{}_r+
\partial_\lambda \omega^\lambda{}_\sigma dx^\sigma \wedge
(\fu+dr),
\nonumber\\
(\R{2}'')^\mu{}_r &=& \omega^{\alpha\beta}\omega_{\alpha\beta} (\oR{0})^\mu{}_r+\partial_\lambda \omega^{\lambda \mu} dr\wedge \fu
+ \partial_\lambda \omega^{\lambda}{}_{\sigma} dx^\sigma\wedge \left[P^\mu{}_\nu dx^\nu+u^\mu \left(\fu+ dr\right)\right]\nonumber\\
&&+\omega^{\mu\alpha}\omega_{\alpha\sigma} dx^\sigma \wedge (dr+\fu),
\nonumber\\
(\R{2}'')^r{}_\rho &=&
\omega^{\alpha\beta}\omega_{\alpha\beta} (\oR{0})^r{}_\rho
+
 \partial_\lambda \omega^{\lambda}{}_{\rho} dr\wedge \fu
 + \partial_\lambda \omega^{\lambda}{}_{\sigma} dx^\sigma\wedge (P_{\alpha \rho}dx^\alpha)
 +
 \partial_\lambda \omega^{\lambda}{}_{\sigma} dx^\sigma\wedge u_\rho
 \left[
 \fu+
dr
  \right]\nonumber\\
  & &
+\omega_{\rho}{}^{\alpha}\omega_{\alpha\sigma} dx^\sigma \wedge (dr+\fu),
\nonumber\\
(\R{2}'')^\mu{}_\rho &=&\omega^{\alpha\beta}\omega_{\alpha\beta} (\oR{0})^\mu{}_\rho
+
 \omega^{\mu\alpha}\omega_{\alpha\rho} \fu \wedge dr \nonumber\\
 & &+\omega_{\rho}{}^{\alpha}\omega_{\alpha\sigma} dx^\sigma \wedge
 \left[ P^\mu{}_\nu dx^\nu + u^\mu \left(\fu+ dr\right)\right]
+\omega^{\mu\alpha}\omega_{\alpha\sigma} dx^\sigma \wedge \left[P_{ \rho \nu}dx^\nu+
u_\rho \left(\fu +dr\right)\right]
\nonumber\\
& &
+ \partial_\lambda \omega^{\lambda\mu}\left[
u_\rho dr\wedge\fu+
(P_{\rho \nu} dx^\nu)\wedge\left(dr+\fu\right)
\right] 
+ \partial_\lambda \omega^{\lambda}{}_\rho
\left[u^\mu dr\wedge \fu
+(P^\mu{}_{\nu} dx^\nu)\wedge\left(dr+\fu\right)
\right]\nonumber\\
& &
+ \partial_\lambda \omega^\lambda{}_\sigma dx^\sigma\wedge \left[
u^\mu u_\rho(dr+\fu) +u^\mu (P_{\rho\nu} dx^\nu)
+u_\rho (P^\mu{}_{\nu} dx^\nu)
+P^\mu{}_\rho (dr+\fu)
\right]\nonumber\\
& &
-2c_{0} \omega^\mu{}_\rho \fomega
+ c_{0} \left(\omega_{\rho\sigma}dx^\sigma\right) \left(\omega^\mu{}_{\nu}dx^\nu\right)
+\partial_\nu \omega^\mu{}_\rho dx^\nu \wedge ( dr+
 \fu).
\eea
The coefficients in the terms containing $c_0$ are equal and fixed by the Bianchi identity.
We will only use this implication of the Bianchi identity in later computations.
Note that since  $(\R{2}'') \ne (\R{2})$ in general, when one computes $(\R{2})$ explicitly from the metric (up to 2nd order), some of the terms written above for $(\R{2}'')$ might be zero. Below, we will use $(\R{2}'')$ to prove statements which are also true if we replaced $(\R{2}'')$ by $(\R{2})$ since $(\R{2}'')$ contains more general terms than the actual $(\R{2})$. Hence, we will use $(\R{2})$ directly in all equations hereafter. The same comment applies to $(\R{3})$ in the subsequent subsection.

As a next step, we calculate the product of the 2nd order term in the curvature 2-form with
the lower order terms. We first calculate the product of $(\R{0})$ and $(\R{2})$ and find
\bea
\left(\R{0}\R{2}\right)^r{}_r&=&
0,
\nonumber\\
\left(\R{0}\R{2}\right)^\mu{}_r&=&
\partial_\lambda \omega^{\lambda}{}_{\sigma} dx^\sigma \wedge (P^\mu{}_\nu dx^\nu)\wedge dr \wedge \fu,
\nonumber\\
\left(\R{0}\R{2}\right)^r{}_\rho&=&
\partial_\lambda \omega^\lambda{}_\sigma dx^\sigma \wedge (P_{\rho\alpha}dx^\alpha)\wedge dr\wedge \fu,
\nonumber\\
\left(\R{0}\R{2}\right)^\mu{}_\rho
&= &
\omega^{\alpha\beta}\omega_{\alpha\beta}
\left(P^\mu{}_\sigma dx^\sigma
\right)\wedge
\left(P_{\rho\nu} dx^\nu
\right)\wedge dr\wedge \fu
+
(P^\mu{}_\sigma dx^\sigma)\wedge (\omega_\rho{}^\alpha \omega_{\alpha\sigma} dx^\sigma) \wedge dr \wedge \fu
\nonumber\\
& &+
u^\mu(\partial_\lambda \omega^\lambda{}_\sigma dx^\sigma)\wedge(P_{\rho\nu} dx^\nu) \wedge dr \wedge \fu
+
u_\rho (\partial_\lambda \omega^\lambda{}_\sigma dx^\sigma)\wedge(P^\mu{}_{\nu} dx^\nu) \wedge dr \wedge \fu
\nonumber\\
& &
+ (\partial_\lambda \omega^\lambda{}_\alpha dx^\alpha) \wedge (P^\mu{}_{\nu} dx^\nu) \wedge (P_{\rho\sigma} dx^\sigma) \wedge (dr+\fu),
\eea
and
\bea
\left(\R{2}\R{0}\right)^r{}_r&=& 0, \nonumber\\
\left(\R{2}\R{0}\right)^\mu{}_r&=&
\partial_\lambda \omega^{\lambda}{}_{\sigma} dx^\sigma \wedge (P^\mu{}_\nu dx^\nu)\wedge dr \wedge \fu,
\nonumber\\
\left(\R{2}\R{0}\right)^r{}_\rho&=&
\partial_\lambda \omega^\lambda{}_\sigma dx^\sigma \wedge (P_{\rho\alpha}dx^\alpha)\wedge dr\wedge \fu,
\nonumber\\
\left(\R{2}\R{0}\right)^\mu{}_\rho
&= &\omega^{\alpha\beta}\omega_{\alpha\beta}
\left(P^\mu{}_\sigma dx^\sigma
\right)\wedge
\left(P_{\rho\nu} dx^\nu
\right)\wedge dr\wedge \fu
+
(P_{\rho\sigma} dx^\sigma)\wedge (\omega^{\mu\alpha} \omega_{\alpha\sigma} dx^\sigma) \wedge dr \wedge \fu
\nonumber\\
& &+
u^\mu(\partial_\lambda \omega^\lambda{}_\sigma dx^\sigma)\wedge(P_{\rho\nu} dx^\nu) \wedge dr \wedge \fu
+
u_\rho (\partial_\lambda \omega^\lambda{}_\sigma dx^\sigma)\wedge(P^\mu{}_{\nu} dx^\nu) \wedge dr \wedge \fu
\nonumber\\
& &
+ (\partial_\lambda \omega^\lambda{}_\alpha dx^\alpha) \wedge (P^\mu{}_{\nu} dx^\nu) \wedge (P_{\rho\sigma} dx^\sigma) \wedge (dr+\fu).
\eea 
In the latter computation, we have explicitly used the Bianchi identity relating terms with coefficient $c_{0}$.
We note that all the terms in $(\R{0}\R{2})$ or $(\R{2}\R{0})$ are proportional to
either $dr$ or $\fu$. Thus, when multiplied with terms containing $dr\wedge \fu$, as for instance
$(\R{1}\R{0}\R{1})$ or $\tr[(\R{0}\R{1})]$, all these products vanish.
Moreover we notice that the trace of these two products is zero:
\be\label{eq:Tr02}
\tr[(\R{0}\R{2})]=0.
\ee

Let us now consider the more complicated products containing $(\R{2})$.
Using the form structures explicitly written out above, one can show that
\begin{eqnarray}
&&(\R{0}\R{2}\R{0})=(\R{0}\R{2}\R{1}\R{0})=(\R{0}\R{2}\R{1}\R{1}\R{0})=(\R{0}\R{1}\R{2}\R{0}) =
(\R{0}\R{1}\R{1}\R{2}\R{0})= 0\, ,  \nonumber  \\
&&\qquad\qquad\qquad\qquad\qquad(\R{0}\R{1}\R{0}\R{2})=(\R{2}\R{0}\R{1}\R{0})=0\, ,
\end{eqnarray}
as well as
\bea
\left(\R{0}\R{1}\R{2}\right)^r{}_r&\sim &\left(\R{0}\R{2}\right)^r{}_r \wedge \fomega
,\nonumber\\
\left(\R{0}\R{1}\R{2}\right)^r{}_\rho &\sim &\left(\R{0}\R{2}\right)^r{}_\rho \wedge \fomega
,\nonumber\\
\left(\R{0}\R{1}\R{2}\right)^\mu{}_r&\sim &\left(\R{0}\R{2}\right)^\mu{}_r \wedge \fomega
,\nonumber\\
\left(\R{0}\R{1}\R{2}\right)^\mu{}_\rho&\sim &\left(\R{0}\R{2}\right)^\mu{}_\rho \wedge \fomega
+ \left(
\omega_{\sigma}{}^{\gamma}\omega_{\gamma}{}^{\alpha} \omega_{\alpha\nu} dx^\sigma \wedge dx^\nu
\right) (P_{\rho\beta} dx^\beta)dr\wedge \fu \wedge dx^\mu,
\label{eq:012to02}
\eea where by $\sim$ we mean that tensor and form structures are the same. In the above calculation
we have used the following properties for the lower order terms:
\bea
\left(\R{0}\R{1}\oR{0}\right)^r{}_r&= &
\left(\R{0}\R{1}\oR{0}\right)^r{}_\rho=
\left(\R{0}\R{1}\oR{0}\right)^\mu{}_r=0\, ,\nonumber\\
\left(\R{0}\R{1}\oR{0}\right)^\mu{}_\rho&=&
(P_{\rho\nu} dx^\nu)\wedge dx^\mu  \wedge dr \wedge \fu\wedge \fomega.
\eea Hence, we have
\be
(\ldots \R{1}\R{0}\R{1} \ldots \R{0}\R{1}\R{2} \ldots) = \tr[(\R{0}\R{1})] \wedge\ldots \wedge (\ldots \R{0}\R{1}\R{2} \ldots)
=0.
\ee Also, since $(\R{0}\R{1})$ and $(\R{1}\R{0})$ both contain solely terms proportional to $dr$ or $\fu$,
these obey 
\bea
&&(\ldots \R{0}\R{1}\R{2} \ldots \R{1}\R{0} \ldots) \sim (\ldots \R{0}\R{2} \ldots \R{1}\R{0} \ldots), \nonumber\\
&&(\ldots \R{0}\R{1}\R{2} \ldots \R{0}\R{1} \ldots) \sim (\ldots \R{0}\R{2} \ldots \R{0}\R{1} \ldots),
\eea
by absorbing some $\fomega$ 2-forms into the `$\ldots$'). In particular, this implies
\be
(\R{0}\R{1}\R{2}\R{1}\R{0})\sim(\R{0}\R{2}\R{1}\R{0})\wedge \fomega=0,\quad
(\R{0}\R{1}\R{2}\R{1}\R{1}\R{0})\sim(\R{0}\R{2}\R{1}\R{1}\R{0})\wedge \fomega=0.
\ee

Additionally, employing the relations
\bea
\left(\R{0}\R{1}\R{1}\oR{0}\right)^r{}_r&= &
\left(\R{0}\R{1}\R{1}\oR{0}\right)^\mu{}_r=
\left(\R{0}\R{1}\R{1}\oR{0}\right)^r{}_\rho=
0, \nonumber\\
\left(\R{0}\R{1}\R{1}\oR{0}\right)^\mu{}_\rho&=&
 (P_{\rho\nu} dx^\nu ) \wedge dx^\mu\wedge  dr\wedge \fu  \wedge \fomega^2,
\eea one can confirm that
\be
(\R{0}\R{1}\R{1}\R{2})\sim (\R{0}\R{2})\wedge \fomega^2,
\ee which immediately leads to
\bea
(\R{0}\R{1}\R{1}\R{2}\R{0})\sim (\R{0}\R{2}\R{0}) &=&0,\nonumber\\
(\R{0}\R{1}\R{1}\R{2}\R{1}\R{0})\sim (\R{0}\R{2}\R{1}\R{0}) &=&0,\nonumber\\
(\R{0}\R{1}\R{1}\R{2}\R{1}\R{1}\R{0})\sim (\R{0}\R{2}\R{1}\R{1}\R{0})&=&0.
\eea
Similarly, one can show that
\be
\tr[\R{1}\R{0}\R{2}]
=\tr[\R{1}\R{1}\R{0}\R{2}]=0,
\ee which entails, through Eq.~(\ref{eq:012to02}),
\be
\tr[\R{1}\R{0}\R{1}\R{2}]
=\tr[\R{1}\R{1}\R{0}\R{1}\R{2}]=0\,.
\ee

Collecting all the above results, the products containing $(\R{2})$ condense to
\be\label{traceRiemann2}
(\fchi_1 \R{2} \fchi_1)=0, \qquad \tr[\fchi_1 \R{2}] =0, \qquad (\fchi_2 \R{2}) = (\R{2} \fchi_2)=0.
\ee

\subsubsection{3rd order curvature 2-form}
\label{sec:ThirdOrderRiemann}

For the 3rd order terms in the curvature 2-form merely a few identities, which we are going to elaborate along this subsection, will be relevant for the arguments in this paper. 

The first equality to show is
\be \label{eq01030or03010}
(\R{0}\R{1}\R{0}\R{3}\R{0})=(\R{0}\R{3}\R{0}\R{1}\R{0})=0.
\ee
Before jumping into the classification of the most general form structures of ($\R{3}$), let us discuss the structure of $(\R{0}\R{1}\R{0})$ and $(\R{0})$ in more detail.
Given a matrix-valued 2-form $(\form{M})$, we have the following relations:
\bea
\left(\R{0}\R{1}\R{0}\form{M}\R{0}\right)^r{}_r &=&\left(\R{0}\R{1}\R{0}\form{M}\R{0}\right)^r{}_\rho =\left(\R{0}\R{1}\R{0}\form{M}\R{0}\right)^\mu{}_r=0\, ,
\nonumber\\
\left(\R{0}\R{1}\R{0}\form{M}\R{0}\right)^\mu{}_\rho&=&
\left(\R{0}\R{1}\R{0}\right)^\mu{}_\sigma \wedge (\form{M})^\sigma{}_\nu \wedge (\R{0})^\nu{}_\rho.
\eea  and
\bea
\left(\R{0}\form{M}\R{0}\R{1}\R{0}\right)^r{}_r &=&\left(\R{0}\form{M}\R{0}\R{1}\R{0}\right)^r{}_\rho  =
\left(\R{0}\form{M}\R{0}\R{1}\R{0}\right)^\mu{}_r=0\, ,
\nonumber\\
\left(\R{0}\form{M}\R{0}\R{1}\R{0}\right)^\mu{}_\rho&=&
(\R{0})^\mu{}_\nu \wedge (\form{M})^\nu{}_\sigma \wedge \left(\R{0}\R{1}\R{0}\right)^\sigma{}_\rho,
\eea where we have used the fact that, except of one term in $(\R{0})^\rho{}_\mu$,
all the terms in $(\R{0})$  are proportional to either $dr$ or $\fu$. Then the explicit calculation
shows that
\bea\label{eq:0103}
(\R{0}\R{1}\R{0})^\mu{}_\rho (\form{M})^\rho{}_\nu (\R{0})^\nu_{\ \alpha}&\sim &\fu\wedge dr\wedge dx^\mu \wedge
( P_{\rho \beta} (\form{M})^\beta{}_\nu
dx^\rho\wedge   dx^\nu)\wedge (P_{\alpha\sigma} dx^\sigma ), \nonumber\\
(\R{0})^\mu{}_\nu (\form{M})^\nu{}_\mu\left(\R{0}\R{1}\R{0}\right)^\rho{}_\alpha&\sim &
\fu\wedge dr \wedge  dx^\mu\wedge  (
P_{\rho \beta}(\form{M})^{\beta}{}_{\nu }
dx^\rho \wedge dx^\nu)\wedge( P_{\alpha \sigma} dx^\sigma)
.
\eea

Now, let us construct the relevant case where $(\form{M})=(\R{3})$. Since we are wedging $(\R{3})$ with
$dr\wedge \fu$,
the only possible form structure for $(\R{3})^\alpha{}_\beta$ that survives is $dx^\alpha\wedge (P_{\beta\sigma} dx^\sigma)$.
From above, we see that if the indices of $(\form{M})^\rho{}_\nu$ are carried by
$dx^\rho,~P_{\nu\sigma} dx^\sigma,~u^\rho$ or $u_\nu$,
they will be zero after contraction with the projection matrix $P$ or after wedging with $\fu$.
Therefore, the free indices on $(\form{M})$ must come from the different powers (and appropriate contractions) of $\omega_{\alpha\beta}$ or derivatives of $\omega_{\alpha\beta}$. At 3rd order, the possible two-forms can be split in
\be
\mbox{Case I}: \omega_{\alpha_1 \beta_1} \omega_{\alpha_2 \beta_2} \omega_{\alpha_3 \beta_3}dx^\mu\wedge dx^\nu, \quad 
\mbox{Case II}: \omega_{\alpha_1 \beta_1} \partial_\gamma \omega_{\alpha_2 \beta_2}dx^\mu\wedge dx^\nu.
\ee

In Case I, the two form indices can contract in two ways with $ \omega_{\alpha_1 \beta_1} \omega_{\alpha_2 \beta_2} \omega_{\alpha_3 \beta_3}$ and then we need to further contract two more free indices to have only two free tensor indices. After the contraction, we are left with
\be
\omega^{\beta}{}_{\lambda} \omega^\lambda{}_{\nu}\fomega ,\quad \text{or}\quad 
\omega^{\beta\lambda} \omega_{\lambda\sigma} \omega_{\nu\alpha}dx^\alpha \wedge dx^\sigma.
\ee where we have used $\omega_{\sigma\alpha} \omega^\alpha{}_{\beta}dx^\beta \wedge dx^\sigma=0$.
However, we  see that these terms vanish by inserting them into Eq.~(\ref{eq:0103}).

On the other hand, in Case II there is an odd number of free indices (hence cannot reduce to an 2-index object by contracting the indices within that term). Thus
at least a contraction with $u^\mu$ is necessary leaving two possibilities, either
\be
\omega^{\alpha_1}{}_{\alpha_2}u^{\lambda} \partial_\gamma \omega_{\lambda \beta_2}dx^\mu\wedge dx^\nu, \quad\text{or}\quad
\omega^{\alpha_1}{}_{\alpha_2} u^\gamma (\partial_\gamma \omega_{\alpha_2 \beta_2})dx^\mu\wedge dx^\nu.
\ee
However, since $u^{\lambda} \partial_\gamma \omega_{\lambda \beta_2}=-\omega_\gamma{}^{\lambda}\omega_{\lambda \beta_2}$ and $u^\gamma \partial_\gamma \omega_{\alpha_2 \beta_2}
=u^\gamma \partial_{\alpha_2} \omega_{\gamma \beta_2}=-\omega_{\alpha_2}{}^\gamma \omega_{\gamma\beta_2}$ , Case II reduces to Case I.
We thus confirmed \eqref{eq01030or03010}
 which leads to 
\begin{eqnarray}
(\fchi_2 \R{3} \fchi_1) = (\fchi_1 \R{3} \fchi_2) =0\, . 
\end{eqnarray}

Actually, the above argument is pertinent also when we evaluate 
$\tr[(\R{0}\R{1}\R{0}\R{3})]$ and $\tr[(\R{0}\R{3})]\wedge dr \wedge \fu$. Due to
\bea
\tr[(\R{0}\R{1}\R{0}\R{3})]&=&(\R{0}\R{1}\R{0})^\mu{}_\rho (\R{3})^\rho{}_\mu
\sim dr\wedge \fu \wedge \fomega  \left(P_{\beta\mu}(\R{3})^\beta{}_{\nu}dx^\mu \wedge dx^\nu\right), \nonumber\\
\tr[(\R{0}\R{3})]\wedge dr \wedge \fu&=&
(\R{0})^\mu{}_\nu (\R{3})^\nu{}_\mu  \wedge dr \wedge \fu
\sim dr \wedge \fu \wedge \left(P_{\beta\mu}(\R{3})^\beta{}_{\nu}dx^\mu \wedge dx^\nu\right),\nonumber
\eea
and using analogous arguments as before, we conclude
\be\label{trace0103or03}
\tr[(\R{0}\R{1}\R{0}\R{3})]=\tr[(\R{0}\R{3})]\wedge dr \wedge \fu=0.
\ee
Similarly, we can also confirm that $(\R{0}\R{3}\R{0})\wedge dr\wedge \fu=0$, 
which leads to
\begin{eqnarray}
(\fchi_1 \R{3} \fchi_1) \wedge dr\wedge \fu =0\, . 
\end{eqnarray}

In summary, we have shown that
\be\label{traceRiemann3}
(\fchi_2 \R{3} \fchi_1) = (\fchi_1 \R{3} \fchi_2) =(\fchi_1 \R{3} \fchi_1) \wedge dr\wedge \fu =\tr[\fchi_2 \R{3}]=0.
\ee

\subsection{General structures of trace  with higher-order curvature 2-form}
\label{sec:generalstructureTrace}

In the computations of Maxwell and Einstein sources, we often encounter 
\be
{\cal T}\equiv  \tr\left[\form{R}^{2k}\right]\, . 
\ee  
In this subsection, we will prove that when 2nd or higher order terms of $\fR$ is contained, 
$\mathcal{T}$ vanishes at sufficiently low orders. In other words, to have a nonzero result, 
there is a constraint
on the total number of the derivatives in $\mathcal{T}$ when it contains 
2nd or higher order terms of $\fR$.  As we will see in the computation of the sources, this constraint makes 
the classification of the sources with 2nd and/or higher order terms of $\fR$ (as well as $\fF$) easy.

Let us notice that the trace ${\cal T}$ can take one of the following three forms
(by using the cyclic property of the trace):
\bea
&&
TR_{(1)} \equiv  \tr[\fchi],  \nonumber \\
&&TR_{(2)} \equiv
\tr[\form{\upsilon}^{(1)}\fchi\form{\upsilon}^{(2)}\fchi\cdots
\form{\upsilon}^{(j)}\fchi\cdots
\form{\upsilon}^{(I)}\fchi], \quad (I\ge1),  \nonumber \\
&&TR_{(3)} \equiv  \tr[\form{\upsilon}^{(1)}],
\eea
where the symbol $\fchi$ represents one of the elements in
 $\fchi_0\cup \fchi_1\cup\fchi_2$,
that is, it is a string made of $(\R{0})$'s and $(\R{1})$'s only. 
Another symbol $\form{\upsilon}^{(j)}$ is defined to represent a string made of 2nd or higher order terms,
$(\R{m})$ with $m\ge2$ (for example, $(\R{2}^2\R{5}\R{3})$).
In the following argument, we sometimes use $\fchi_m$ to denote
an element of $\fchi_m$ for simplicity.
Let us define $N_{Ti}$ as the number of derivatives for the case of ${\cal T}=TR_{(i)}$ (for $i=1,2,3$). 
Then we have the following constraints on $N_{Ti}$, depending on which form $\mathcal{T}$ takes: 

\begin{itemize}
\item 
${\cal T}=TR_{(1)}:$ \\
In this case, since $\tr[\fchi_2]=0$, ${\cal T} = \tr[\fchi_0]$ or $\tr[\fchi_1]$, which means 
that $N_{T1}$ for nontrivial ${\cal T}$ can take two values:
\begin{itemize}
\item $N_{T1}=2k:~TR_{(1)}=\tr[\fchi_0]\,$. 
\item  $N_{T1}=2k-1:~TR_{(1)}=\tr[\fchi_1]\sim dr\wedge \fu$\,.
\end{itemize}
\item 
${\cal T}=TR_{(2)}:$ \\ 
For ${\cal T}=TR_{(2)}$, we will show that ${\cal T}$ vanishes for $N_{T2} \le 2k$ and thus 
$N_{T2}> 2k$ is required for non-trivial ${\cal T}$. 
To see this, we consider the following three cases separately: 
\begin{itemize}
\item $N_{T2}\le 2k-2:$ \\
We always encounter two or more $\fchi_2$'s and thus $\cal{T}$ vanishes. 
\item $N_{T2}=2k-1:$ \\
To avoid having more than one $\fchi_2$, we must set all $\form{\upsilon}^{(j)}$'s to be $(\R{2})$, 
one $\fchi$ to be $\fchi_2$ and the rest of $\fchi$'s to be $\fchi_1$. However, ${\cal T}$ in this case 
also vanishes as a result of $(\fchi_2\R{2})=(\R{2}\fchi_2)=0$ (see Eq.~(\ref{traceRiemann2})).
\item $N_{T2}=2k:$\\  
In order not to have more than one $\fchi_2$,  there are two sub-cases:
\begin{itemize}
\item One of the $\form{\upsilon}^{(j)}$'s is $(\R{3})$, the rest of the $\form{\upsilon}^{(j)}$'s are $(\R{2})$'s, 
one of the $\fchi$'s is $\fchi_2$ and the rest of the $\fchi$'s are $\fchi_1$'s. In this case, we always encounter 
$(\R{2}\fchi_2)=(\fchi_2 \R{2})=0$ (see Eq.~(\ref{traceRiemann2})) for $I\ge2$. For $I=1$ (that is, 
when there is one $(\R{3})$ and no $(\R{2})$'s), $TR_{(2)}=\tr[\R{3}\fchi_2]$ which also 
vanishes thanks to Eq.~(\ref{traceRiemann3}). 

\item All $\form{\upsilon}^{(j)}$'s are $(\R{2})$'s.  When one of $\fchi$'s is $\fchi_2$ and the rest are  $\fchi_1$'s, 
we always encounter $(\R{2}\fchi_2)=(\fchi_2 \R{2})=0$.  On the other hand, 
when all $\fchi$'s are  $\fchi_1$'s, ${\cal T}$ contains $(\fchi_1\R{2} \fchi_1)$ or $\tr[\fchi_1 \R{2}]$ and both of them 
are zero due to Eq.~(\ref{traceRiemann3}). 
\end{itemize}
\end{itemize}
\item ${\cal T}=TR_{(3)}:$ \\
It is easy to see that $N_{T3}$ satisfies $N_{T3}\ge 4 k$. 
The equality holds only when $\form{\upsilon}^{(1)}$ contains $(\R{2})$ only. 
\end{itemize}

We will use these bounds on $N_{Ti}$'s repeatedly in our arguments in the subsequent sections. 


\section{Maxwell sources}
\label{sec:MaxwellSources}
In this Appendix, we present the computations of the Maxwell sources
at the leading order in the fluid/gravity derivative expansion using the relations summarized in the Appendix \ref{sec:productRiemanns}.
The simplest example are the anomaly polynomials containing purely the $U(1)$ field strength, $\fP_{CFT} = c_{_A} \fF^{n+1}$,
on AdS$_{2n+1}$, which we have already explained in Sec.~\ref{sec:abelianCS}. We will consider the mixed Chern-Simons terms here. We will do so by
starting with the computations of the Maxwell sources for specific examples on
AdS$_{5}$ and AdS$_{9}$ to complement the case of AdS$_{7}$ in Sec.~\ref{sec:ExplicitAdS7MaxwellText}.
We then provide a general argument for any dimensions
and any mixed Chern-Simons terms to show the complete agreement with the replacement rule. Interestingly enough, as we will show, the 2nd and higher order terms in the
$U(1)$ field strength $\fF$ and curvature 2-form $\fR$ do not contribute to the
leading order computations of the Maxwell sources. Remarkably, in AdS$_{2n+1}$ the leading
nontrivial result for the Maxwell sources is proportional to $\fomega^{n-1}$.

\subsection{AdS$_5$}
Let us start with AdS$_5$ case and consider the mixed term of the form
\be
\fP_{CFT_4}=  c_{_M} \form{F} \wedge \tr[\form{R}^2]\, ,
\ee
where $c_{_M}$ is a constant (in the following examples, we also introduce
constant $c_{...}$ or simply $c$ to denote coefficients
in the expression of the anomaly polynomials).
From this, we get the following expression for the Maxwell source:
\be
{}^\star \form{J}_H=- c_{_M}  \tr[\form{R}^2]\, .
\ee
We first restrict ourselves to the 0th and 1st order terms of the curvature 2-forms.
As shown in Appendix \ref{sec:productRiemanns} the product $(\R{0}\R{0}) =0$, therefore the leading order contribution is
\be\label{eq:5DcMSourece}
{}^\star\form{J}_H=-  2c_{_M} \tr[(\R{0}\R{1})]
=   \frac{d}{dr}(2c_{_M}\Phi^2_T) (\fu\wedge(2\fomega) \wedge dr),\quad
J_H^{(2\omega)}=  \frac{1}{r} \frac{d}{dr}\frac{\partial (2c_{_M}\Phi \Phi^2_T)}{\partial \Phi}
,
\ee
where ${}^\star \form{J}_H$ is indeed of order $\fomega^1$.

Let us then take into account the higher order terms in the curvature 2-form.
In this case, it is obvious that the higher order terms
in the curvature 2-form  do not contribute at order $\fomega^1$ or lower.

For the Abelian Chern-Simons term discussed in Sec.~\ref{sec:abelianCS} and the mixed 
Chern-Simons term above, 
a straightforward comparison with the conventions of  \cite{Megias:2013joa} gives the following dictionary:
\begin{equation}\label{eq:ads5compareMegias}
-16\pi G_{_N} c_{_A}   =  \frac{4\kappa}{3}  \ ,\qquad
-16\pi G_{_N} c_{_M}  =   4\lambda \, ,  \nonumber
\end{equation}
\begin{equation}
\mathbb{J} = \frac{1}{\rH^2} 16 \pi G_{_N} r^3 \THall^{(2\omega)}\,,\qquad \mathbb{A} =-\frac{1}{\rH} 16 \pi G_{_N} r \JH^{(2\omega)} \,.
\end{equation}
Using these relations, one can show that our sources  in Eq.~(\ref{eq:AbelianSources}) and Eq.~(\ref{eq:5DcMSourece}) are the same as those found by authors of \cite{Megias:2013joa}
(assuming sign typos in $W_{4i}$ term in \cite{Megias:2013joa} ).

\subsection{AdS$_9$}
As a final explicit example for the Maxwell source, we consider AdS$_9$ case.
There are three kinds of
the mixed terms in the anomaly polynomial
\be
\fP_{CFT_8}= c_{_M} \form{F}^3 \wedge \tr[\form{R}^2]
+ \tilde{c}_{_M} \form{F} \wedge \tr[\form{R}^2] \wedge \tr[\form{R}^2]
+ \tilde{\tilde{c}}_{_M} \form{F} \wedge \tr[\form{R}^4]\,.
\ee
The corresponding Maxwell source is given by
\be
{}^{\star}\form{J}_H=
- 3c_{_M} \form{F}^2 \wedge \tr[\form{R}^2]
- \tilde{c}_{_M}  \tr[\form{R}^2] \wedge \tr[\form{R}^2]
- \tilde{\tilde{c}}_{_M}  \tr[\form{R}^4]\,.
\ee

For simplicity, we will evaluate the three terms separately.
\begin{itemize}
\item \underline{\bf First term (proportional to $c_{_M}$)}  \\
Let us first ignore the 2nd and higher order
terms in $\fF$ and $\fR$. We then find that the leading contribution to this term is
\bea\label{eq:AdS9Maxwell}
{}^\star\form{J}_H&=&- 6 c_{_M} (\F{0}\F{1})\wedge \tr[(\R{1}\R{1})]
- 6 c_{_M} (\F{1}^2)\wedge \tr[(\R{0}\R{1})]\nonumber\\
&&=
\frac{d}{dr}\left(
6 c_{_M}\Phi^2\Phi^2_T\right)(\fu \wedge (2\fomega)^3\wedge dr)\, , 
\eea
which is of order $\fomega^3$. Lower orders terms such as
\begin{eqnarray}
&&(\F{0}^2)\wedge \tr[(\R{1}\R{1})]\, , \quad (\F{0}\F{1})\wedge \tr[(\R{0}\R{1})]\, , \quad (\F{1}^2)\wedge \tr[(\R{0}\R{0})]\, , 
\nonumber \\
&&(\F{0}^2)\wedge \tr[(\R{0}\R{1})]\, , 
\quad (\F{0}\F{1})\wedge \tr[(\R{0}\R{0})]\, ,
\quad (\F{0}\F{0})\wedge \tr[(\R{0}\R{0})]\, ,
\end{eqnarray}
will all vanish as a result of 
$(\R{0}\R{0})=0$, $(\F{0})\sim dr\wedge \fu$ and $\tr[(\R{0}\R{1})]\sim dr\wedge \fu$. 
Therefore (\ref{eq:AdS9Maxwell}) is indeed the leading order contribution to the Maxwell source proportional to $c_M$.

Having identified already the leading contributions we can consider 2nd and higher order terms of $\fF$ and $\fR$. 
At order $\fomega^3$ or lower, we encounter the following configurations 
containing 2nd or higher order terms: 
\begin{eqnarray}
&&(\F{0}^2)\wedge \tr[(\R{1}\R{2})]\, , \quad (\F{0}\F{1})\wedge \tr[(\R{0}\R{2})]\, , \quad 
(\F{0}\F{2})\wedge \tr[(\R{0}\R{1})]\, , \nonumber \\
&&(\F{1}\F{2})\wedge \tr[(\R{0}\R{0})]\, , \quad (\F{0}^2)\wedge \tr[(\R{0}\R{3})]\, , \quad (\F{0}\F{3})\wedge \tr[(\R{0}\R{0})]\, , \nonumber \\
&&\qquad\qquad (\F{0}^2)\wedge \tr[(\R{0}\R{2})]\, , \quad (\F{0}\F{2})\wedge \tr[(\R{0}\R{0})]\,.
\end{eqnarray}
One can direct identify that all these vanish since 
$(\R{0}\R{0})=0$, $\tr[(\R{0}\R{2})]=0$, $(\F{0})\sim dr\wedge \fu$ and $\tr[(\R{0}\R{1})]\sim dr\wedge \fu$. 
We therefore determine that there is no $\fomega^3$ or lower order contribution containing 
2nd or higher order terms of $\fF$ or $\fR$. 

\item \underline{\bf Second term (proportional to $\tilde{c}_{_M}$)} \\
By considering the 0th and 1st order terms
in $\fR$ only, we uncover that the leading order term of order $\fomega^3$ turns out to be
 \be
{}^\star\form{J}_H= -4\, \tilde{c}_{_M} \tr[(\R{0}\R{1})]\wedge \tr[(\R{1}\R{1})]
= \frac{d}{dr}(2^2  \tilde{c}_{_M}\Phi_T^4) (\fu\wedge(2\fomega)^3\wedge dr)\,.
\ee  
Meanwhile, $(\R{0}\R{0})=0$ and $\tr[(\R{0}\R{1})]\sim dr\wedge \fu$, renders trivially all lower orders terms 
\begin{eqnarray}
&&\tr[(\R{0}\R{0})]\wedge \tr[(\R{1}\R{1})]\, , \qquad \tr[(\R{0}\R{1})]\wedge \tr[(\R{0}\R{1})]\, , 
\nonumber \\
&& \tr[(\R{0}\R{0})]\wedge \tr[(\R{0}\R{1})]\, , 
\qquad \tr[(\R{0}\R{0})]\wedge \tr[(\R{0}\R{0})]\, .
\end{eqnarray} 
Once we take into account the 2nd and higher order contributions to $\fR$,  at $\fomega^3$ or lower order, we are faced with 
several possibilities
\begin{eqnarray}
&&\tr[(\R{0}\R{0})]\wedge \tr[(\R{1}\R{2})]\, , \qquad \tr[(\R{0}\R{1})]\wedge \tr[(\R{0}\R{2})]\, ,  
\quad \tr[(\R{0}\R{0})]\wedge \tr[(\R{0}\R{3})]\, , 
\nonumber \\
&&\qquad\qquad\qquad\,\,\,\,\qquad\qquad\qquad \tr[(\R{0}\R{0})]\wedge \tr[(\R{0}\R{2})]\, . 
\end{eqnarray}
However, all of them are zero as a result of  $(\R{0}\R{0})=0$ and $\tr[(\R{0}\R{2})]=0$.  
Consequently there is no $\fomega^3$ or lower order contributions containing 
2nd or higher order terms of $\fR$. 

\item \underline{\bf Third term (proportional to $ \tilde{\tilde{c}}_{_M} $)}\\
Disregarding the 2nd and higher order terms in $\fF$ and $\fR$,
we find the leading $\fomega^3$ order contribution to the third term 
\be
{}^\star\form{J}_H
= -4 \tilde{\tilde{c}}_{_M}  \tr[(\R{0}\R{1}\R{1}\R{1})]
=  \frac{d}{dr}(2 \tilde{\tilde{c}}_{_M} \Phi_T^4)
(\fu\wedge (2\fomega)^3\wedge dr)\,.
\ee
As for the lower order terms, which from $(\R{0}\R{0})=0$ and $(\R{0}\R{1}\R{0}\R{1})=0$ we will conclude that these vanish, we can have terms involving
\begin{eqnarray}
\tr[(\R{0}\R{0}\R{1}\R{1})]\,, \quad \tr[(\R{0}\R{1}\R{0}\R{1})]\, , 
\quad \tr[(\R{0}\R{0}\R{0}\R{1})]\, , 
\quad \tr[(\R{0}\R{0}\R{0}\R{0})]\,.
\end{eqnarray}
From the 2nd and higher order terms in $\fR$, many configurations at $\fomega^3$ or lower order are possible 
\begin{eqnarray}
&&\tr[(\R{0}\R{0}\R{1}\R{2})]\, , \quad \tr[(\R{0}\R{0}\R{2}\R{1})]\, ,\quad \tr[(\R{0}\R{1}\R{0}\R{2})]\, , \quad \tr[(\R{0}\R{0}\R{0}\R{3})]\, , \nonumber \\
&&\qquad\qquad\qquad\qquad\qquad\qquad\tr[(\R{0}\R{0}\R{0}\R{2})]\,.
\end{eqnarray}
The entire set will be irrelevant when we considering $(\R{0}\R{2}\R{0})=$$(\R{0}\R{1}\R{0}\R{2})$$=0$ and $(\R{0}^2)=0$.
In this way, we can say that there is no $\fomega^3$ or lower order contributions containing 
2nd or higher order terms of $\fR$. 

\end{itemize}

Adding the contributions computed for each of the three terms of the AdS$_9$ Maxwell source, yields
\bea
{}^\star\form{J}_H &=&
\frac{d}{dr}\frac{\partial}{\partial\Phi}\left(
 2c_{_M}\Phi^3\Phi^2_T + 2^2  \tilde{c}_{_M} \Phi \Phi_T^4+2 \tilde{\tilde{c}}_{_M}  \Phi \Phi_T^4\right)(\fu \wedge (2\fomega)^3\wedge dr), \label{eq:ads9maxwellsource}\\
J_H^{(2\omega)^3}&=&
=  \frac{1}{r^5} \frac{d}{dr}\frac{\partial}{\partial \Phi}\left(
2c_{_M} \Phi^3\Phi_T^2
+(2^2\tilde{c}_{_M} +2\tilde{\tilde{c}}_{_M} ) \Phi \Phi_T^4
\right),
\eea
where ${}^\star\form{J}_H$ is of order $\fomega^3$.


\subsection{General argument for Maxwell sources}
\label{appendix:generalMaxwell}
Previously in Appendix \ref{sec:MaxwellSources}
and Sec.~\ref{sec:abelianCS},
we recorded explicit examples and calculated the Maxwell sources
for the mixed terms in the anomaly polynomials for dimension $n\le4$, corresponding to solutions up to AdS$_9$.
Now, we consider more general cases for AdS$_{2n+1}$ in any dimensions.

Let us then study the mixed term in the bulk anomaly polynomial
\be
\fP_{CFT_{2n+1}}= c_{_M} \fF^{l+1}\wedge \tr\left[\form{R}^{2k_1}\right]\wedge  \tr\left[\form{R}^{2k_{2}}\right]
\wedge \ldots \wedge \tr\left[\form{R}^{2k_{p}}\right],
\ee 
where $l\ge0$, $p\ge1$ and $k_i>0$ ($i=1, 2,\cdots,  p$) 
which might be considered in AdS$_{2n+1}$ for $n=2\, k_{tot}+l$ (here $k_{tot}=\sum_{i=1}^p k_i$).
We will argue that the general Maxwell source is
\bea
{}^\star\form{J}_H&=& - c_{_M} (l+1) \fF^{l} \wedge \tr\left[\fR^{2k_1}\right]\wedge  \tr\left[\fR^{2k_{2}}\right]
\wedge \ldots \wedge \tr\left[\fR^{2k_{p}}\right]
\equiv- c_{_M} (l+1) \fF^{l} \wedge {\cal T},   \nonumber \\
\qquad && {\rm where\qquad }  {\cal T} = \tr\left[\fR^{2k_1}\right]\wedge  \tr\left[\fR^{2k_{2}}\right]
\wedge \ldots \wedge \tr\left[\fR^{2k_{p}}\right].  \label{eq:maxwell_general}
\eea

\noindent
\underline{\bf (1) 0th and 1st order terms only} \\
To start we will overlook the 2nd and higher order contributions to
$\fF$ and $\fR$. For the leading order contribution to the Maxwell source \eqref{eq:maxwell_general}, since
both $(\F{0})$ and $\tr[\fchi_1]$ are proportional to $dr\wedge\fu$ and $\tr[\fchi_2]=0$, we obtain
{\small{
\bea
{}^\star\form{J}_H&=&-c_{_M}(l+1)(\F{1}^{l}) \wedge\sum_{i=1}^{p} (2k_i) \tr\left[(\R{1}^{2k_1})\right]\wedge\ldots \wedge\tr\left[(\R{0}\R{1})\wedge(\R{1}^{2k_{i}-2})\right]\wedge \ldots \wedge
\tr\left[(\R{1}^{2k_p})\right]\nonumber\\	
&& -c_{_M} l(l+1)(\F{0})\wedge (\F{1}^{l-1}) \wedge \tr\left[(\R{1}^{2k_1})\right]\wedge\ldots  \wedge
\tr\left[(\R{1}^{2k_p})\right]   \nonumber \\
&=&\frac{d}{dr}\frac{\partial}{\partial \Phi}\left[c_{_M} \Phi^{l+1}(2^p\Phi_T^{2k _{tot}})\right](\fu\wedge(2\fomega)^{n-1}\wedge dr),
\label{eq:mawell_general_result}
\eea
}}
which is of order $\fomega^{n-1}$.
Therefore $J^{(2\omega)^{n-1}}_H $ is given by
\be
J_H^{(2\omega)^{n-1}}
 =\frac{1}{r^{d-3}} \frac{d}{dr}\frac{\partial
\mathbb{G}^{(2\omega)^{n-1}}}{\partial \Phi}
=\frac{1}{r^{d-3}} \frac{d}{dr}
\frac{\partial}{\partial \Phi}\left(c_{_M}\Phi^{l+1}  (2^p\Phi_T^{2k_{tot}})\right),
\ee
where we have used $d=2n$ and verified the replacement rule for $\mathbb{G}^{(2\omega)^{n-1}}$.

\noindent
\underline{\bf (2) 2nd and higher order terms} \\
The next step is to show that there is no contribution
containing $(\R{2}),  (\F{2})$ or higher order terms at order $\fomega^{n-1}$ or lower in the computation of
the Maxwell source. The underlying idea, as the number of the 2nd and higher order terms
is increased, is to add more 0th order terms to compensate which will entail the classification of the contribution with 2nd and/or higher order terms tractable.

Recall from Appendix.~\ref{sec:generalstructureTrace} that the trace $\tr[\fR^{2k_i}]$ can take one of the following three forms
(by using the cyclic property of the trace):
\bea
&&TR_{(1)} \equiv  \tr[\fchi]\,,  \nonumber \\
&&TR_{(2)} \equiv
\tr[\form{\upsilon}^{(1)}\fchi\form{\upsilon}^{(2)}\fchi\cdots
\form{\upsilon}^{(j)}\fchi\cdots
\form{\upsilon}^{(I)}\fchi]\,, \quad (I\ge1)\,,  \nonumber \\
&&TR_{(3)} \equiv  \tr[\form{\upsilon}^{(1)}]\,,
\eea
where the symbol $\fchi$ represents one of the elements in
 $\fchi_0\cup \fchi_1\cup\fchi_2$, while $\form{\upsilon}^{(j)}$ is used to represent a string made of 2nd or higher order terms,
 e.g. $(\R{2}),(\R{3})$ or higher order terms.

Without loss of generality, let us suppose that contribution from first $p_1$ traces is of the form $TR_{(1)}$
while for $p_1 +1 \le i \le p_2$ it is of the form $TR_{(2)}$. Then the remaining $\tr[\fR^{2k_i}]$ contribution for $p_2+1 \le i \le p$ is of the form $TR_{(3)}$. 
With this ordering, after dividing ${\cal T}$ into three parts, ${\cal T} ={\cal T}_1\wedge{\cal T}_2\wedge{\cal T}_3$, where
${\cal T}_1$ (${\cal T}_2$, ${\cal T}_3$, respectively)
is made of the traces of the form $TR_{(1)}$ ($TR_{(2)}$, $TR_{(3)}$, respectively) only, we have
\bea
&&{\cal T}_{1} \equiv\  \tr[\fR^{2k_1}] \wedge\ldots \wedge \tr[\fR^{2k_{p_1}}] \,  ,  \nonumber \\
&&{\cal T}_{2} \equiv\ \tr[\fR^{2k_{p_1+1}}] \wedge\ldots \wedge \tr[\fR^{2k_{p_2}}]\, ,  \nonumber \\
&&{\cal T}_{3} \equiv\ \tr[\fR^{2k_{p_2+1}}] \wedge\ldots \wedge \tr[\fR^{2k_{p}}]\, .
\eea We denote the number of derivatives contained in ${\cal T}_i$  $(i=1,2,3)$ by $N_{T_i}$.

Let us first consider the case where either $p_1<p$ or $p_2<p$, i.e. there exists at least one $TR_{(2)}$ or $TR_{(3)}$. The non-trivial contributions for each ${\cal T}_i$ will satisfy a lower bound on 
the number of derivatives $N_{T_i}$. Results from Appendix~\ref{sec:generalstructureTrace} imply that ${\cal T}_i$ is zero unless
\begin{itemize}
\item ${\cal T}_1:~ N_{T_1} = \sum_{i=1}^{p_1} (2k_i) $ or 
$\sum_{i=1}^{p_1} (2k_i)-1$\,. \\
In the first case, all the traces in ${\cal T}_1$ are $\tr[\fchi_0]$'s.
In the second case, ${\cal T}_1$ contains exactly one $\tr[\fchi_1]\sim dr\wedge \fu$ 
while the rest of the traces are equal to $\tr[\fchi_0]$'s.
Since we cannot have more than one $\tr[\fchi_1]$, these are the only two non-trivial cases.
\item ${\cal T}_2:~N_{T_2} > \sum_{i=p_1+1}^{p_2} (2k_i)$.
\item ${\cal T}_3:~ N_{T_3} \ge \sum_{i=p_2+1}^{p} (4k_i)>\sum_{i=p_2+1}^{p} (2k_i)$\,.
\end{itemize}

Taking the number of the derivatives in $\fF^l$ to be $N_F$, the total number of derivatives in the Maxwell source is given by $
N_{tot} = N_F + N_{T_1} + N_{T_2} + N_{T_3}$. In turn, two separate cases have to be analyzed. Namely, $N_F \ge l$ or $N_F <l$.
On one hand, for $N_F \ge l$,
\be\label{eq:MWGenSourceInEq}
N_{tot} > \sum_{i=1}^{p} (2 k_i) +l-1 = 2 k_{tot}+l -1= n-1\, .
\ee 
Therefore this case does not add to the Maxwell source at $\fomega^{n-1}$ or lower order.
On the other, for $N_F <l$, since $(\F{0}^2)=0$, the only potentially non-trivial configurations are the ones with 
$(\fF^l)=(\F{0}\F{1}^{l-1})$ and thus $N_F=l-1$. 
All satisfy the inequality in Eq.~(\ref{eq:MWGenSourceInEq}), except 
for the case where $N_{T_1}=\sum_{i=1}^{p_1} (2k_i)-1$. 
In this exceptional case, ${\cal T}_1$ and $\fF^l$ contain  $\tr[\fchi_1]\sim dr\wedge \fu$ and $\F{0}\sim dr\wedge \fu$ respectively. 
Thus, this possibility is trivial.

The previous argument assumes that $p_1$ or $p_2$ are simultaneously equal to $p$, i.e. there exists at least one $TR_{(2)}$ or $TR_{(3)}$. When $p_1=p_2=p$, 
the trace becomes ${\cal T}={\cal T}_1$, thus $\fF^l$ needs to contain 2nd and/or higher order terms of $\fF$. 
However, since $(\F{0}^2)$=0, nontrivial contributions to $\fF^l$ containing 2nd and/or higher order terms 
must satisfy $N_F\ge l$. This implies $N_{tot}\ge 2k_{tot}+l-1$.   
For the $N_{tot}=2k_{tot}+l-1$ case,
we necessarily have one $\tr[\fchi_1] \sim dr\wedge \fu$ in ${\cal T}_1$ and also one $(\F{0})$ in $\fF^l$, 
and thus this product is zero.

To summarize, we have confirmed that 2nd and higher order terms of
$\fF$ and $\fR$ do not add to the Maxwell sources at
$\fomega^{n-1}$ or lower order.

\section{Einstein sources}\label{sec:EinsteinSources}

The computation of the Einstein sources is the main subject of this section.
For this purpose, we have to consider solely the $(r, \mu)$-compontent of  $(\THall)_{ab}$. 
As we will explain, only the $(\SpH)_{(r\mu)}{}^r$ components of $(\SpH)_{(ab)}{}^c$ will contribute to $(\THall)_{r\mu}$.
Surprisingly, $(\SpH)_{(r\mu)}{}^r$ satisfies a replacement rule which subsequently implies a replacement rule for 
$(\THall)_{r\mu}$ in Eq.~(\ref{eq:SourceGen}). We will discuss how this works in Appendix~\ref{sec:RepRuleSigma}.  
We then scan specific examples in AdS$_{2n+1}$ ($n=1,2,3,4$) in Appendix~\ref{sec:Ads3} to
Appendix~\ref{sec:Ads9} where such replacement 
rules are verified and consider a more general replacement rule for  $(\SpH)_{(r\mu)}{}^r$ in AdS$_{2n+1}$ in 
Appendix~\ref{section:GeneralEinsteinsource} .
At the end of this section, we provide a proof of some 
symmetry properties of the spin Hall current
used in Appendix~\ref{sec:RepRuleSigma} to demonstrate that only
$(\SpH)_{(r\mu)}{}^r$ contributes to $(\THall)_{r\mu}$.

\label{section:EinsteinSources}

\subsection{Replacement rule for $(\SpH)_{(r\mu)}{}^r$}
\label{sec:RepRuleSigma}
To evaluate the Einstein source, we consider the $(r, \mu)$-component of $(\THall)_{ab}$ defined by
\be\label{eq:Trmu}
(\THall)_{r\mu}\equiv\nabla_c (\SpH)_{(r\mu)}{}^c
=\partial_c (\SpH)_{(r\mu)}{}^c
-\Gamma^d{}_{rc} (\SpH)_{(d\mu)}{}^c
-\Gamma^d{}_{\mu c} (\SpH)_{(rd)}{}^c
+\Gamma^c{}_{cd} (\SpH)_{(r\mu)}{}^d\,\,.
\ee \
For a moment, let us assume that the only  nontrivial contribution to $(\SpH)_{(ab)}{}^c$ are those of the  0th and 1st order terms
of  $\fF$ and $\fR$. 
Moreover, we also assume that the leading order contribution to $(\THall)_{r\mu}$ is of order $\fomega^{n-1}$
and comes only from  $(\SpH)_{(ab)}{}^c$ at $\fomega^{n-1}$ order. 
These two at first ad hoc assumptions will be justified in the upcoming sections by  proving two statements: 
(1) when only 0th and 1st order terms of $\fF$, $\fR$ and the Christoffel symbol are considered, the leading order contribution 
to $(\THall)_{r\mu}$ is of order $\fomega^{n-1}$ and comes from 
$(\SpH)_{(ab)}{}^c$ at order $\fomega^{n-1}$;
(2) 2nd or higher order terms of $\fF$, $\fR$ 
and the Christoffel connection can generate nontrivial contribution to $(\THall)_{r\mu}$
only at $\fomega^{n}$ or higher order.    
Then
\bea
(\THall)_{r\mu}&=&\partial_r (\SpH)_{(r\mu)}{}^r
+r^{-1}(d-1) (\SpH)_{(r\mu)}{}^r
-r^{-1} P^\nu{}_\mu (\SpH)_{(r\nu)}{}^r \nonumber\\
&&
-\Gamma^r{}_{\mu c} (\SpH)_{(rr)}{}^c
-r^{-1} P^\nu{}_\rho (\SpH)_{(\nu\mu)}{}^\rho
-\Gamma^\nu{}_{\mu \rho} (\SpH)_{(r\nu)}{}^\rho
-\Gamma^r{}_{r\rho} (\SpH)_{(r\mu)}{}^\rho,\quad\quad
\eea where we have used the fact  that $\Gamma^a{}_{a r}=r^{-1} (d-1)$ and $\Gamma^a{}_{a\mu}=0$ 
at the 0th order.

From Appendix~\ref{sec:proveAssumptions}, we know that the leading order contributions  to $(\SpH)_{ab}{}^c$ from zeroth and first order objects satisfy several symmetry properties
\bea\label{eq:Assumptions}
(\SpH)_{(rr)}{}^c= (\SpH)_{(r\nu)}{}^\rho=P^\nu{}_\rho (\SpH)_{(\nu\mu)}{}^\rho=0, \qquad 
P^\nu{}_\mu (\SpH)_{(r\nu)}{}^r=(\SpH)_{(r\mu)}{}^r\,,
\eea 
yielding an expression for $(\THall)_{r\mu}$ which depends only on $(\SpH)_{(r\mu)}{}^r$, namely,
\be
(\THall)_{r\mu}=\frac{1}{r^{d-2}}\frac{d}{dr}\left[
r^{d-2}(\SpH)_{(r\mu)}{}^r
\right]\, .
\ee

Based on the prior observations, we claim that 
 \be\label{eq:repSigma}
(\SpH)_{(r\mu)}{}^r=-\frac{1}{2 r^{d-2}} \frac{d}{dr}\left[ r\frac{\partial  \mathbb{G}^{\tV}
}{\partial \Phi_T}
\right]V_\mu \equiv\SpH^\tV V_\mu\, .
\ee  This leads to
\be
\THall^\tV
= 
-\frac{1}{2r^{d-1}} r\frac{d}{dr}\left[
 \frac{d}{dr}\left( r\frac{\partial  \mathbb{G}^{\tV}
}{\partial \Phi_T}
\right)\right]
=-\frac{1}{2r^{d-1}}
\frac{d}{dr}\left[r^2
 \frac{d}{dr}\left(\frac{\partial  \mathbb{G}^{\tV}}{\partial \Phi_T}
\right) \right], 
\ee which is the replacement rule for the Einstein source in Eq.~(\ref{eq:SourceGen}).

In the next sections, we will first work out some examples to verify Eq.~(\ref{eq:repSigma}) 
and justify the assumptions we have made above. 
We then prove this new replacement rule for the most general AdS$_{2n+1}$ case.
%

\subsection{AdS$_3$}
\label{sec:Ads3}

The simplest situation is AdS$_3$ with the anomaly polynomial given by
\be
\fP_{CFT_2}= c_{_g} \tr [\fR^2]\, .
\ee
The leading order contribution to $(\SpH)^{ab}{}_c$ can be easily found
\be
(\SpH)^{ab}{}_c
=
-2c_{_g} \varepsilon^{a\,p_1p_2} (\R{0})^b{}_{c\,p_1 p_2}\, ,
\label{eq:ads3_sigma_leading}
\ee which represents a $\fomega^0$ order.
Evaluating
\be
\label{eq:SigmarmrAdS3}
(\SpH)_{r\mu}{}^r
= 
+4c_{_g}  \Psi'  V_\mu
\, ,\quad
(\SpH)_{\mu r}{}^r
=-4 c_{_g} ( r \Psi'')  V_\mu \, . 
\ee  we find
\be
\SpH^\tV=-\frac{1}{2}
\frac{d}{dr}\left[ r \frac{\partial}{\partial \Phi_T} \left(2 c_{_g}\Phi_T^2\right) \right]\, , 
\ee which is consistent with $ \mathbb{G}^{\tV}=2 c_{_g} \Phi_T^2$ for the gravitational Chern-Simons term in AdS$_3$ and confirms
 the replacement rule in Eq.~(\ref{eq:repSigma}).

We note that even if we take into account 2nd and higher order terms in the Christoffel connection
or the curvature 2-form, it is obvious that there is no extra contribution at the leading order
to the Einstein source.

\subsection{AdS$_5$}
The mixed term in the anomaly polynomial in AdS$_5$ spacetimes is
\be
\fP_{CFT_4}= c_{_M} \form{F} \wedge \tr [\fR^2]\,,
\ee
which in turn gives
\be
(\SpH)^{ab}{}_c
=
-c_{_M} \varepsilon^{a\,p_1p_2 p_3 p_4}F_{p_1 p_2} R^b{}_{c\,p_3 p_4}\, .
\ee

The leading order term of $(\SpH)^{ab}{}_c$ is of order $\fomega^0$ such that
\be\label{eq:0F0Rzero}
(\SpH)'{}^{ab}{}_c\equiv
 -c_{_M}\varepsilon^{a\,p_1p_2 p_3 p_4} \wedge (\F{0})_{p_1 p_2} (\R{0})^b{}_{c\,p_3 p_4}
\sim \varepsilon^{a\,p_3p_4 \nu r}u_\nu (\R{0})^b{}_{c\,p_3 p_4}\,.
\ee
The only non-zero component of $dr\wedge\fu\wedge (\R{0})^b{}_{c}$ is
$dr\wedge\fu\wedge (\R{0})^\beta{}_{\gamma} \sim dr\wedge\fu\wedge dx^\beta \wedge (P_{\gamma \delta}dx^\delta)$ thus the only nontrivial components of $(\SpH)'{}^{ab}{}_c$ are
\be
(\SpH)'{}^{\alpha \beta}{}_\gamma \sim
\varepsilon^{\alpha \beta \nu \delta r}u_\nu P_{\gamma\delta}\, .
\ee 
This implies that
$(\SpH)'_{rb}{}^c=(\SpH)'_{ar}{}^c=(\SpH)'_{ab}{}^r=0,$ and hence the only potentially non-trivial contribution 
to $(\THall)_{r\mu}$ from $(\SpH)'{}^{ab}{}_c$ is the second term in Eq.~(\ref{eq:Trmu}), which involves
$(\SpH)'_{(\delta \mu)}{}^\gamma$. Conversely,
\be
(\SpH)'_{\delta \mu}{}^\gamma
\sim
\varepsilon^{\alpha \beta \gamma  \nu r} P_{\alpha \delta} P_{\beta \mu}u_\nu\quad \Rightarrow\quad
(\SpH)'_{(\delta \mu)}{}^\gamma=0\, ,
\ee 
and, we conclude that $(\SpH)'{}^{ab}{}_c$ does not contribute to 
$(\THall)_{r\mu}$ at any order. 
 
It is worth stressing that the aforementioned arguments are actually more general, in the sense that the arguments follow (and play no role) whenever the spin Hall current contains a structure of the form 
$(\SpH)^{ab}{}_{c} \sim  \varepsilon^{a p_1 p_2\cdots } \left(
 dr\wedge \fu \wedge(\R{0})^b{}_c  \right)_{p_1 p_2\cdots}$\,.

To find the first non-trivial contribution to $(\THall)_{r\mu}$ we will have
to calculate the subleading term of the spin Hall current given by
 \be
(\SpH)^{ab}{}_c=
-c_{_M} \varepsilon^{a\,p_1p_2 p_3 p_4}
\left[
(\F{0})_{p_1 p_2} (\R{1})^b{}_{c\,p_3 p_4}
+(\F{1})_{p_1 p_2} (\R{0})^b{}_{c\,p_3 p_4}
\right]
\equiv
(\SpH^{(1)})^{ab}{}_c+ (\SpH^{(2)})^{ab}{}_c\, ,
\ee
where we have divided this contribution into two pieces for later use.
The first term $(\SpH^{(1)})^{ab}{}_c$, considering $(\F{0})= \Phi' dr\wedge \fu$, becomes
\be
(\SpH^{(1)})_{\mu r}{}^r
=-4 c_{_M} r^{-2} \Phi' (r\Phi_T) V_\mu
 \, ,\quad
  \qquad\qquad (\SpH^{(1)})_{r\mu}{}^r=0\,,
\ee leading to
\be
(\SpH^{(1)}){}^\tV=-2 c_{_M} r^{-2} \Phi'(r \Phi_T)\,.
\ee
The second term  $(\SpH^{(2)})_{(\mu r)}{}^r$ can be evaluated easily 
by {\it attaching} $(\F{1})=2\Phi \fomega$ to the AdS$_3$ computation in Appendix~\ref{sec:Ads3} and hence
\be
(\SpH^{(2)}){}^\tV=-2 c_{_M} r^{-2}\Phi \partial_r(r  \Phi_T)\,.
\ee
Combining the two terms yields\footnote{
To compare with \cite{Megias:2013joa}, we refer the readers back to
the discussion around
Eq.~(\ref{eq:ads5compareMegias}).
}
\be
\SpH^\tV=- \frac{1}{2r^2}\frac{d}{dr} \left[
r \frac{\partial}{\partial \Phi_T} \left(2 c_{_M} \Phi   \Phi_T^2\right)
\right]\, ,
\ee showing an agreement with Eq.~(\ref{eq:repSigma}).

As in the case of AdS$_3$, it is obvious that 2nd and higher order terms in
$\fF$, $\fR$ and the Christoffel connection add nontrivially
to the Einstein source at the $\fomega^1$ or lower orders.

\subsection{AdS$_7$}
\label{sec:appendixHallAdS7Einstein}
Now we consider AdS$_7$ case with three different types of terms in the anomaly polynomial
with potential to  contribute nontrivially to the Einstein source.
We deal with each term separately.\\

\noindent
\underline{\bf (1) $\fP_{CFT_6}= c_{_M} \form{F}^2 \wedge \tr [\fR^2]$} \\
The spin Hall current is denoted by 
\be
(\SpH)^{ab}{}_c
=
-\frac{1}{2}c_{_M} \varepsilon^{ap_1p_2 p_3 p_4 p_5 p_6 }F_{p_1 p_2} F_{p_3 p_4} R^b{}_{cp_5 p_6}\, .
\ee

We concentrate on the 0th and 1st order terms in $\fF$, $\fR$ and the Christoffel connection and ignore the higher orders to start.
By noticing that  $(\F{0}^2)=0$, the leading order contribution to the spin Hall current is
of order $\fomega^1$ reduces to
\begin{eqnarray} \label{eq:spp7d1st}
(\SpH)^{'ab}{}_c \equiv -c_{_M}\varepsilon^{ap_1p_2 p_3 p_4 p_5 p_6}
(\F{0})_{p_1 p_2} (\F{1})_{p_3 p_4} (\R{0})^b{}_{cp_5 p_6}\, . 
\end{eqnarray}
However, as in the case of $(\F{0})\wedge(\R{0})$ term in AdS$_5$ (see Eq.~(\ref{eq:0F0Rzero}) and below)
these do not contribute to the Einstein source at any order of the derivative expansion. 
Therefore, we will only need to consider terms of order $\fomega^2$ in $(\SpH)^{ab}{}_c$ of the form
\be
(\SpH)^{ab}{}_c=
-\half c_{_M} \varepsilon^{ap_1p_2 p_3 p_4 p_5 p_6}
\left[
2\times (\F{0})_{p_1 p_2} (\F{1})_{p_3 p_4} (\R{1})^b{}_{cp_5 p_6}
+(\F{1})_{p_1 p_2} (\F{1})_{p_3 p_4} (\R{0})^b{}_{cp_5 p_6}
\right].
\ee
This structure is essentially the same as the AdS$_5$ case with extra $(\F 1)=(2\Phi\fomega)$ {\it attached}. 
It follows that
\be
(\SpH^{(1)}){}^\tV=-2 c_{_M} r^{-4} (2\Phi \Phi' )(r\Phi_T)\,,\quad
(\SpH^{(2)}){}^\tV=-2 c_{_M} r^{-4}\Phi^2 \partial_r(r  \Phi_T)\,,
\ee which add up to
\be
\SpH^\tV= -\frac{1}{2 r^{4}} \frac{d}{dr}
\left[
r \frac{\partial}{\partial \Phi_T}
\left( 2 c_{_M} \Phi^2 \Phi_T^2\right)
\right]\, ,
\ee and also agrees with Eq.~(\ref{eq:repSigma}).

It is interesting to see that when 2nd and higher order terms of $\fF$, $\fR$ and the Christoffel connection 
are taken into account, no extra contributions to the Einstein source exist at 
$\fomega^2$ or lower orders. Mainly, due to $(\F{0}^2)=(\F{0}\F{2})=0$ as well as the fact that 
the leading order contribution to the spin Hall current is of order $\fomega^1$. 

Observe that we can generalize this line of thoughts to the case of 
$\fP_{CFT_{d+1}}= c_{_M} \form{F}^l \wedge \tr[\fR^2]$ in AdS$_{d+1}$ (with $d+1=2l+3$): (1)
the leading order term of the spin Hall current does not contribute at any order
to the Einstein source, (2) the two subleading order contributions
(which are of order $\fomega^l$) to $(\SpH)^{ab}{}_c$ are
\be
(\SpH^{(1)}){}^\tV=-   r^{2-d} (l\Phi^{l-1} \Phi') (2c_{_M} r\Phi_T)\, ,\quad
(\SpH^{(2)}){}^\tV=-  r^{2-d}\Phi^l \partial_r(2c_{_M} r  \Phi_T)\, ,
\ee which add up to
\be
\SpH^\tV= -\frac{1}{2 r^{d-2}} \frac{d}{dr}
\left[
r \frac{\partial}{\partial \Phi_T}
\left( 2 c_{_M} \Phi^l \Phi_T^2\right)
\right]\, ,  
\ee and represent to final result which is as well in agreement with Eq.~(\ref{eq:repSigma}), and finally (3) the 2nd and higher order terms in
$\fF$, $\fR$ and the Christoffel connection do not generate extra contribution 
to the Einstein source at $\fomega^{l}$ or lower orders, 
since $(\F{0}^2)=(\F{0}\F{2})=0$ and the fact that 
the leading order contribution to the spin Hall current is of order $\fomega^{l-1}$. \\

\noindent
\underline{\bf (2) $\fP_{CFT_6}= c_{_g} \tr [\fR^2] \wedge \tr [\fR^2]$} \\
The spin Hall current is 
\be
(\SpH)^{ab}{}_c
=
-c_{_g} \varepsilon^{a\,p_1p_2 p_3 p_4 p_5 p_6 }
R^f{}_{g p_1 p_2} R^g{}_{f p_3 p_4}
R^b{}_{c\,p_5 p_6}\,.
\ee

As in the previous case,
let us for a while consider 0th and 1st order terms in $\fR$ and the Christoffel connection only.
The leading order contribution to the spin Hall current is of order $\fomega^1$ and is given by
\be\label{eq:lowerorder}
(\SpH)^{'ab}{}_c\equiv -2c_{_g} \varepsilon^{ap_1p_2 p_3 p_4 p_5 p_6}
\left[
 \tr[(\R{0}\R{1})]_{p_1 p_2 p_3 p_4} (\R{0})^b{}_{cp_5 p_6}
\right]\, . 
\ee 
Notice that
 $\tr[(\R{0}\R{1})]\wedge (\R{0})\sim dr\wedge \fu \wedge(\R{0})$
and thus we can apply analogous arguments as in the AdS$_5$ case 
(see Eq.~\eqref{eq:0F0Rzero} and below). Therefore, such a term does not contribute 
to the Einstein source at any order of the derivative expansion.
The next  to leading order term, comprises the $\fomega^2$ terms of the spin Hall current, reduces to
\bea
(\SpH)^{ab}{}_c&=&
- c_{_g} \varepsilon^{ap_1p_2 p_3 p_4 p_5 p_6}
\left[
2\times \tr[(\R{0}\R{1})]_{p_1 p_2 p_3 p_4} (\R{1})^b{}_{cp_5 p_6}
+\tr[(\R{1}\R{1})]_{p_1 p_2 p_3 p_4}(\R{0})^b{}_{cp_5 p_6}
\right] \nonumber \\
&\equiv&
(\SpH^{(1)})^{ab}{}_c+(\SpH^{(2)})^{ab}{}_c\, ,
\eea
where $(\SpH^{(1)})^{ab}{}_c$ and $(\SpH^{(2)})^{ab}{}_c$ respectively denote the
first and second terms of the right hand side of the first line.

We first evaluate $(\SpH^{(1)})^\tV$ and find
\be
(\SpH^{(1)})^\tV
=-8  r^{-4} (2 \Phi_T\Phi_T' ) (r c_{_g} \Phi_T)\, . 
\ee
The second term $(\SpH^{(2)})^\tV$ is essentially the same as  the computation in the AdS$_3$ case, with extra $\tr[(\R{1}\R{1})]=2(2\fomega \Phi_T)^2$
{\it attached} so that
\be
(\SpH^{(2)})^\tV 
=-8 r^{-4} \Phi_T^2  \partial_r (r c_{_g} \Phi_T)\, . 
\ee
Combining the above two contributions, we finally obtain
\be
\SpH^\tV =-\frac{1}{2r^{4}} \frac{d}{dr}\left[
r (c_{_g} 16 \Phi_T^3 )
\right]\, ,
\ee which again agrees with Eq.~(\ref{eq:repSigma}).

Alike other cases in AdS$_7$ the 2nd and higher order terms in $\fR$ and the Christoffel connection, do not contribute to $(\THall)_{r\mu}$ at $\fomega^2$ or lower orders,
as of $(\R{0}^2)=0$ and $\tr[(\R{0}\R{2})]=0$ and the fact that the leading order contribution to the spin Hall current is of order $\fomega^1$.\\

\noindent
\underline{\bf (3)  $\fP_{CFT_6}= \tilde{c}_{_g} \tr [\fR^4]$}\\
The spin Hall current in this case is given by 
\be
(\SpH)^{ab}{}_c
=
-\tilde{c}_{_g} \varepsilon^{ap_1 p_2 \ldots p_6 }
R^b{}_{e p_1 p_2} R^e{}_{f p_3 p_4}
R^f{}_{cp_5 p_6}
\,.
\ee

The leading contribution to the spin Hall current is of order $\fomega^{1}$ while considering only the 0th and 1st order terms in $\fR$ and Christoffel connection
\be
(\SpH)'{}^{ab}{}_c\equiv
-\tilde{c}_{_g} \varepsilon^{ap_1p_2\ldots p_6}
 (\R{0}\R{1}\R{0})^b{}_{cp_1 p_2\ldots p_6}.
\ee
Identical arguments as in the previous case,  since
$(\R{0}\R{1}\R{0})\sim \tr[(\R{0}\R{1})]\wedge(\R{0})$. 
indicate that $(\SpH)'{}^{ab}{}_c$ generates no contribution to $(\THall)_{r\mu}$ independently of its order.   
The next order terms in $(\SpH)^{ab}{}_c$ are given by
\be
(\SpH)^{ab}{}_c=
-\tilde{c}_{_g} \varepsilon^{ap_1p_2\ldots p_6}
\left[
 (\R{0}\R{1}\R{1})^b{}_{cp_1p_2\ldots p_6}
+(\R{1}\R{0}\R{1})^b{}_{cp_1p_2\ldots p_6}
+(\R{1}\R{1}\R{0})^b{}_{cp_1p_2\ldots p_6}
\right],
\ee from which a few components are relevant to the Einstein source computation, specifically
\bea
(\SpH)_{r\mu}{}^r
=8 \tilde{c}_{_g} (r^{-4}\Phi_T ^3) V_\mu\,,\quad
(\SpH)_{\mu r}{}^r
=-8\tilde{c}_{_g} r^{-4} \left[
\partial_r\left( r\Phi_T^3\right)
+\Phi_T^3
\right]V_\mu\, . 
\eea In the end, we find
\be
\SpH^\tV=-\frac{1}{2r^4}  \frac{d}{dr}\left[r
\frac{\partial}{\partial \Phi_T} 
\left(
2 \tilde{c}_{_g}\Phi_T^ 4
\right)\right]\, , 
\ee which coincides with Eq.~(\ref{eq:repSigma}).

Two results, $(\R{0}^2)=0$ and $(\R{0}\R{2}\R{0})=0$ and the fact that 
the spin Hall current starts at $\fomega^1$ order, imply that the 2nd and higher order terms in $\fR$ and Christoffel connection,
support no further contributions to the Einstein source at $\fomega^2$ or lower orders.

\subsection{AdS$_9$}
\label{sec:Ads9}
For AdS$_9$, we consider the following two cases that have not been discussed yet. \\

\noindent
\underline{\bf (1) $\fP_{CFT_8}= \tilde{c}_{_M} \form{F}\wedge\tr [\fR^2] \wedge \tr [\fR^2]$}\\
The corresponding spin Hall current is given by 
\be
(\SpH)^{ab}{}_c
=
-\half  \tilde{c}_{_M} \varepsilon^{a\,p_1p_2 p_3 p_4 p_5 p_6 p_7 p_8}
F_{p_1 p_2}
R^f{}_{gp_3 p_4} R^g{}_{f p_5 p_6}
R^b{}_{c\,p_7 p_8}\, .
\ee

Taking the 0th and 1st order terms in the derivative expansion of
$\fF$, $\fR$ and Christoffel connection we find the leading order contribution to $(\SpH)^{ab}{}_c$
\bea
(\SpH)'{}^{ab}{}_c&\equiv&
- \half \tilde{c}_{_M}  \varepsilon^{a\,p_1p_2 p_3 p_4 p_5 p_6 p_7 p_8}  \\
&&\times
\left[
2\times (\F{1})_{p_1p_2}\tr[(\R{0}\R{1})]_{p_3 p_4 p_5 p_6} (\R{0})^b{}_{c\,p_7 p_8}
+(\F{0})_{p_1p_2}\tr[(\R{1}\R{1})]_{p_3 p_4 p_5 p_6} (\R{0})^b{}_{c\,p_7 p_8}
\right],\nonumber
\eea which is of order $\fomega^{2}$.
Since $(\F{0})\wedge(\R{0})\sim \tr[(\R{0}\R{1})]\wedge(\R{0})$, the statements  employed for AdS$_5$ and AdS$_7$ are applicable here as well implying that the $(\SpH)'{}^{ab}{}_c$ does contribute trivially to the Einstein source at all orders. 
The next leading order (i.e. $\fomega^3$ order) of the spin Hall current becomes
\bea\label{eq:SigmaAdS9}
(\SpH)^{ab}{}_c&=&
- \half \tilde{c}_{_M}  \varepsilon^{a\,p_1p_2 p_3 p_4 p_5 p_6 p_7 p_8}
\left[
2\times (\F{1})_{p_1p_2}\tr[(\R{0}\R{1})]_{p_3 p_4 p_5 p_6} (\R{1})^b{}_{c\,p_7 p_8}\right.\\
&&\qquad \left.
+ (\F{1})_{p_1p_2} \tr[(\R{1}\R{1})]_{p_3 p_4 p_5 p_6}(\R{0})^b{}_{c\,p_7 p_8}
 +
(\F{0})_{p_1p_2} \tr[(\R{1}\R{1})]_{p_3 p_4 p_5 p_6}(\R{1})^b{}_{c\,p_7 p_8}
\right]. \nonumber
\eea
The contribution from the first two terms can be easily evaluated by attaching $(\F{1})=(2\Phi \fomega)$ 
to the $c_{_M}$ term calculation of the AdS$_7$. The resultant contribution to $\SpH^\tV $ is
\be
-8 \tilde{c}_{_M} r^{-6} \Phi \partial_r\left(
r \Phi_T^3 \right) \, . 
\ee The third term in Eq.~(\ref{eq:SigmaAdS9}) could be simply calculated by
attaching $\tr[(\R{1}\R{1})]=2(2\Phi_T\fomega)^2$ to the AdS$_5$ result ($(\SpH^{(1)}){}^\tV$ for the $c_{_M}$ term). 
Then the contribution to $\SpH^\tV$ from this term is 
\be
-8 \tilde{c}_{_M} r^{-6} \Phi' (r\Phi_T^3)\, . 
\ee By combining them, the overall contribution takes the form
\be
\SpH^\tV= - \frac{1}{2 r^{6}} \frac{d}{dr}\left[r  
\frac{\partial}{\partial \Phi_T} 
(4\, \tilde{c}_{_M} \Phi \Phi_T^4 )
\right]\, , 
\ee
in accord with Eq.~(\ref{eq:repSigma}). 

Let us take into account 2nd and higher order terms in $\fF$, $\fR$ and Christoffel connection. 
We consider $\fF \wedge \tr[\fR^2]\wedge \fR$ from which the spin Hall current is constructed. 
Then we classify the $\fomega^3$ and lower order contributions to $\fF \wedge \tr[\fR^2]\wedge \fR$ 
containing 2nd or higher order terms as follows:
\begin{itemize}
\item 
$(\F{0})\wedge \tr[(\R{0}\R{0})]\wedge (\R{2})$,\, $(\F{2})\wedge \tr[(\R{0}\R{0})]\wedge (\R{0})$,  \,
$(\F{1})\wedge \tr[(\R{0}\R{0})]\wedge (\R{2})$, \\$(\F{2})\wedge \tr[(\R{0}\R{0})]\wedge (\R{1})$, \,
$(\F{0})\wedge \tr[(\R{0}\R{0})]\wedge (\R{3})$,\, $(\F{3})\wedge \tr[(\R{0}\R{0})]\wedge (\R{0})$ \\
$\to$ vanish because $\tr[(\R{0}\R{0})]=0$\,. 
\item 
$(\F{0})\wedge \tr[(\R{0}\R{2})]\wedge (\R{0})$,\, $(\F{0})\wedge \tr[(\R{0}\R{2})]\wedge (\R{1})$,  \,
$(\F{1})\wedge \tr[(\R{0}\R{2})]\wedge (\R{0})$ \\
$\to$ vanish because $\tr[(\R{0}\R{2})]=0$\,. 
\item 
$(\F{0})\wedge \tr[(\R{0}\R{1})]\wedge (\R{2})$\\ $\to$ vanish because 
both $(\R{0})$ and $\tr[(\R{0}\R{1})]$ are proportional to $dr\wedge \fu$. 
\item 
$(\F{0})\wedge \tr[(\R{0}\R{3})]\wedge (\R{0})$\\ $\to$ vanish because 
$(\F{0})\sim dr\wedge \fu$ and then $\tr[(\R{0}\R{3})]\wedge dr\wedge \fu =0$. 
\item 
$(\F{2})\wedge \tr[(\R{0}\R{1})]\wedge (\R{0})$,\, $(\F{0})\wedge \tr[(\R{1}\R{2})]\wedge (\R{0})$  \\
$\to$ do not contribute to the Einstein source since both of them are proportional to \\
\quad \,\,$dr\wedge \fu \wedge (\R{0})^b{}_c$\,. 
\end{itemize}
In addition to this, since the derivative expansion of the spin Hall current starts at $\fomega^2$, 
the 2nd and higher order terms of the Christoffel connection do not  generate new contribution to the Einstein 
source at $\fomega^3$ or lower order.  We thus conclude that 2nd and higher order terms in $\fF$, $\fR$ and Christoffel connection do not contribute to the Einstein source at $\fomega^3$ or lower order. \\

\noindent
\underline{(2) $\fP_{CFT_8}= \tilde{\tilde{c}}_{_M}\, \form{F}\wedge\tr [\fR^4]$}\\
The spin Hall current in this case is 
\be
(\SpH)^{ab}{}_c
=
-\half \tilde{\tilde{c}}_{_M} \varepsilon^{a\,p_1p_2 \ldots p_8 } F_{p_1 p_2}
R^b{}_{f p_3 p_4} R^f{}_{g p_5 p_6} R^g{}_{ c p_7 p_8}
\, .
\ee

 We notice that both $(\F{0})$ and $(\R{0}\R{1}\R{0})$
are proportional to $dr\wedge \fu$ and all the components of $(\R{0}\R{1}\R{1})$ ,
$(\R{1}\R{0}\R{1})$ and $(\R{1}\R{1}\R{0})$ are linear combinations of terms proportional to
$dr$ or $\fu$.
Then the leading order term of $(\SpH)^{ab}{}_c$ -- restricting ourselves to the 0th and 1st order terms
in $\fF$, $\fR$ and the Christoffel connection -- is given by
\bea
(\SpH)'{}^{ab}{}_c&\equiv&
-\half \tilde{\tilde{c}}_{_M} \varepsilon^{ap_1\ldots p_8}(\F{1})_{p_1 p_2}
\left[
 (\R{0}\R{1}\R{0})^b{}_{cp_3\ldots p_8}
\right],
\eea
which is of order $\fomega^2$. 
As in the argument for the $\tilde{c}_g$ term in the AdS$_7$ case, $(\SpH)'{}^{ab}{}_c$ does not contribute
to the Einstein source at any order. We thus consider the next order terms, of order $\fomega^3$, in $(\SpH){}^{ab}{}_c$:
\bea \label{eq:subshcads92}
(\SpH)^{ab}{}_c&=&
-\half \tilde{\tilde{c}}_{_M} \varepsilon^{ap_1\ldots p_8}(\F{1})_{p_1 p_2}
\left[
 (\R{0}\R{1}\R{1})^b{}_{cp_3\ldots p_8}
+(\R{1}\R{0}\R{1})^b{}_{cp_3\ldots p_8}
+(\R{1}\R{1}\R{0})^b{}_{cp_3\ldots p_8}
\right]\nonumber\\
& &-\half \tilde{\tilde{c}}_{_M} \varepsilon^{ap_1\ldots p_8}(\F{0})_{p_1 p_2}
\left[
 (\R{1}\R{1}\R{1})^b{}_{cp_3\ldots p_8}
\right].
\eea
The contribution to the Einstein source from the first line (labeled by $ (\THall^{(1)})_{r\mu}$) is evaluated essentially
in the same way as the $\tilde{c}_g$ term in the AdS$_7$ case with extra $(\F{1})=(2\Phi\omega)$ attached
\be \label{eq:sigma1V}
(\SpH^{(1)} ){}^\tV=-\frac{1}{2 r^6} (8 \tilde{\tilde{c}}_{_M})\left[ \Phi \partial_r \left(
 r \Phi_T^3
\right)\right]\, . 
\ee
To evaluate the second line of  $\eqref{eq:subshcads92}$
(labeled by $(\SpH^{(2)} )^{ab}{}_c$), we first note that $(\SpH^{(2)} )^{rb}{}_c=u_\alpha (\SpH^{(2)})^{\alpha b}{}_c =0$, 
since $(\F{0})\sim dr \wedge \fu$ which produces
\be
(\SpH^{(2)} )_{\mu r}{}^r= -\frac{1}{2r^6} 16 \tilde{\tilde{c}}_{_M} [\Phi' (r\Phi_T^3)]
V_\mu
\quad
(\SpH^{(2)} )_{r\mu}{}^r=0\, , 
\ee and henceforth
\be \label{eq:sigma2V}
(\SpH^{(2)} ){}^\tV=- \frac{1}{2r^6}(8 \tilde{\tilde{c}}_{_M}) \left[ \Phi' \left( r\Phi_T^3 \right)\right]\, .
\ee Combining  (\ref{eq:sigma1V}) and  (\ref{eq:sigma2V}) ultimately add to
\be
(\SpH ){}^\tV=
-\frac{1}{2r^6}\frac{d}{dr} \left[r  \frac{\partial}{\partial \Phi_T}
\left(2 \tilde{\tilde{c}}_{_M} \Phi\Phi_T^4\right) \right]\, , 
\ee
which agrees with Eq.~(\ref{eq:repSigma}).

Once we take into account 2nd and higher order terms in $\fF$, $\fR$ and Christoffel symbol,
then we encounter many potential nontrivial terms. Let us consider $\fF\wedge\fR^3$ from
which we construct the spin Hall current. Then we can classify the contribution of order $\fomega^2$ or
$\fomega^3$
with 2nd or higher order terms of $\fF$ and $\fR$ contained as follows :
\begin{itemize}
\item ${(\F{0})\wedge(\R{0}\R{2}\R{0})}$, ${(\F{1})\wedge(\R{0}\R{2}\R{0})}$
$\to$ vanish because of $(\R{0}\R{2}\R{0}) = 0$.
\item ${(\F{2})\wedge(\R{0}\R{1}\R{0})}$
$\to$ vanish because $(\F{2})\sim dr\wedge \fu$.
\item ${(\F{0})\wedge(\R{0}\R{3}\R{0})}$
$\to$ vanish because of $(\R{0}\R{3}\R{0})\wedge dr\wedge \fu =0$.
\item ${(\F{0})\wedge(\R{2}\R{1}\R{0})}$,\, ${(\F{0})\wedge(\R{2}\R{0}\R{1})}$,\, 
${(\F{0})\wedge(\R{0}\R{2}\R{1})}$,\, ${(\F{0})\wedge(\R{1}\R{2}\R{0})}$, \\
${(\F{0})\wedge(\R{1}\R{0}\R{2})}$,\, ${(\F{0})\wedge(\R{0}\R{1}\R{2})}$ \\
$\to$ vanish because  all the terms appearing in $(\R{0}\R{1})$, $(\R{1}\R{0})$, $(\R{0}\R{2})$ and $(\R{2}\R{0})$
are \\ \quad\, proportional to $dr$ or $\fu$. 
\end{itemize}
We also notice that 2nd and higher order terms in the Christoffel connection do not give nontrivial contribution
to the Einstein source at $\fomega^3$ or lower order, since
the spin Hall current does not contain nontrivial $\fomega^1$ or lower order terms.
We thus conclude that 2nd and higher order terms do not contribute to
the Einstein source at the leading order or lower.

\subsection{General argument for Einstein sources}
\label{section:GeneralEinsteinsource}

In this subsection, we provide a general argument to calculate the leading order contribution to
the Einstein source. In particular, we will show that 2nd and higher order terms in $\fR$, $\fF$ as
well as the Christoffel connection do not contribute at this or lower orders.  The proof of this
straightforward argument involves many steps. We will initially resolve the most simple cases and construct the more elaborated and most general arguments towards the end.

\vspace{0.5cm}
\noindent
\underline{\bf  Case I: $\fP_{CFT}= {c_{_g}}\, \tr[\fR^{2k+4}] \,\,\,\, {\rm with}\,\,\,\ k\ge 0$}\\
We first consider the case with the single trace of the curvature 2-forms.
This anomaly polynomial is allowed on AdS$_{d+1}$ with $d=4k+6$.
Since the case of $\tr[\fR^{2}]$ has been covered in details in the AdS$_3$ example, we shall consider $k\ge0$ here.

\noindent
\underline{\bf (1) 0th and 1st order terms only}\\
For a while, we assume that there is no contribution containing 2nd or higher order
terms of $\fR$ and Christoffel connection.  By noticing that more than two $(\R{0})$'s
are not allowed in a given wedge product of $\fR$,
the derivative of the anomaly polynomial $\partial \fP_{CFT}/\partial \fR^{a}{}_b$ at the leading order is proportional to $(\R{0}\R{1}\R{0})\wedge(2\Phi_T\fomega)^{2k}$.
As in the previous arguments for AdS$_9$,  it does not generate
any contribution to the Einstein source at any order.
We thus consider the next leading contribution to the derivative:
\bea\label{eq:orderEinstein1}
\frac{\partial \fP_{CFT}}{\partial \fR^a{}_b}&=&
{c_{_g}}(2k+4)\sum_{i=0}^{2k+2} (\R{1}^{i}\R{0}\R{1}^{2k+2-i})^b{}_a \\
&=&
{c_{_g}}(2k+4)\left[
 (\R{0}\R{1}\R{1})^b{}_a
+(\R{1}\R{1}\R{0})^b{}_a
+(2k+1)(\R{1}\R{0}\R{1})^b{}_a
\right]\wedge (2\Phi_T \fomega)^{2k}, \nonumber
\eea
which is of order $\fomega^{2k+2}$.
In the second line, we have used the identities in the Appendix \ref{sec:productRiemanns}.
Repeating the same calculation as in the $\tilde{c}_g$ term in AdS$_{7}$ case, we obtain
the corresponding contribution to the spin Hall current as follows:
\bea
(\SpH)_{r\mu}{}^r
&=&
+2 (2k+4){c_{_g}} r^{2-d}(\Phi_T ^{2k+3}) V_\mu \, , 
\nonumber\\
(\SpH)_{\mu r}{}^r
&=&-2(2k+4){c_{_g}} r^{3-d}(\Phi_T^{2k+2}  )\left[
(2k+2)\Phi_T'+r^{-1}\Phi_T
+r^{-1} \partial_r(r \Phi_T)
\right]V_\mu \, . 
\eea This leads to
\bea
\SpH^\tV
&=&-
\frac{1}{2 r^{d-2}}
(2{c_{_g}})(2k+4) \left[
(2k+2)(\Phi_T^{2k+1} \Phi_T') (r\Phi_T)
+ \Phi_T^{2k+2} \partial_r(r \Phi_T)\right] \nonumber\\
&=&-
\frac{1}{2 r^{d-2}}
 \frac{d}{dr} \left[r 
\frac{\partial}{\partial \Phi_T}(2{c_{_g}} \Phi_T^{2k+4})
\right],
\eea which agrees with Eq.~(\ref{eq:repSigma}).

\noindent
\underline{\bf (2) 2nd and higher order terms}\\
Now let us take into account the 2nd and higher order terms in $\fR$ and Christoffel connection
to show that these do not contribute at $\fomega^{2k+2}$ or lower order
of the Einstein source.
By using the same notation as in the general argument for Maxwell source,
we classify the wedge products of $(2k+3)$ curvature 2-forms into the following five cases
($I\ge1$):
 \bea\label{eq:casestructures}
\mbox{Case A }&:& \left(\form{\upsilon}^{(1)}\fchi\form{\upsilon}^{(2)}\fchi
\ldots \form{\upsilon}^{(I)}\fchi \form{\upsilon}^{(I+1)}\right), \nonumber \\
\mbox{Case B }&:& \left(\form{\upsilon}^{(1)}\fchi\form{\upsilon}^{(2)}\fchi
\ldots \form{\upsilon}^{(I)}\fchi\right),\quad
 \left(\fchi\form{\upsilon}^{(1)}\fchi\form{\upsilon}^{(2)}
\ldots \fchi\form{\upsilon}^{(I)}\right)\, , 
\nonumber\\
\mbox{Case C }&:&  \left(\fchi\form{\upsilon}^{(1)}\fchi\form{\upsilon}^{(2)}
\ldots \fchi\form{\upsilon}^{(I)}\fchi^{(I+1)}\right),   \nonumber \\
\mbox{Case D }&:&  \left(\form{\upsilon}^{(1)}\right), \nonumber \\
\mbox{Case E }&:&  \left(\fchi\right) .
\eea
Since Case E does not contain 2nd or higher order terms, 
we consider the rest of the cases, Case A, B, C and D, one by one to show that
2nd and higher order terms do not contribute to the Einstein source
at the leading ($\fomega^{2k+2}$) or lower order.
Let us denote the number of derivatives in the $ (\fR^{2k+3})^b{}_a$ by $N_R$.
To consider contributions up to the order that we are interested in, we have $N_R \le 2k+2$.

\begin{itemize}
\item \underline{\bf Case  A} \\
In this case, there are one more $\form{\upsilon}^{(j)}$'s than $\fchi$'s.
Let us suppose that all the  $\form{\upsilon}^{(j)}$'s are $(\R{2})$.
Then, for the product to be $\fomega^{2k+2}$ or lower order, at least two of $\fchi$'s
are $\fchi_2$ and thus there is no contribution from this case.
When some or all of $\form{\upsilon}^{(j)}$'s contain 3rd or higher order terms of $\fR$
or are products of the 2nd or higher order terms, we need to introduce more  $\fchi_2$
and thus this case does not provide any $\fomega^{2k+2}$ or lower order contribution to
the derivative of the anomaly polynomial $\partial \fP_{CFT}/\partial \fR^a{}_b$\, .

\item \underline{\bf Case  B} \\
For Case B, the product contains the same number of $\form{\upsilon}^{(j)}$'s and
 $\fchi$'s. Let us consider order $\fomega^{2k+2}$ contribution first.
 Except for the case in which all the  $\form{\upsilon}^{(j)}$'s are $(\R{2})$,
 we need to have more than one $\fchi_2$ which vanishes as in the previous case.
Then, the only nontrivial possibility is when all the $\form{\upsilon}^{(j)}$'s are $(\R{2})$, while
one of $\fchi$'s is  $\fchi_2$ and the rest are all $\fchi_1$.
This contribution vanishes nevertheless as a result of $(\fchi_2 \R{2})=(\fchi_2 \R{2})=0$.

 For lower order than $\fomega^{2k+2}$, we need to
 add more than one $\fchi_2$'s and we also have vanishing contribution.

\item \underline{\bf Case  C} \\
This third case requires more careful treatment since the number of $\fchi$'s
is one more than that of $\form{\upsilon}^{(j)}$'s. We call the total number of
the derivative in $\form{\upsilon}^{(j)}$'s as $N_{\upsilon}$ and carry out the classification
depending on the value of $N_{\upsilon}$:
\begin{itemize}
\item \underline{\bf (1) $N_\upsilon = 2I $} \\
The only allowed configuration is that all the $\form{\upsilon}^{(j)}$'s are $(\R{2})$.
To have order $\fomega^{2k+2}$ contribution to $\partial \fP_{CFT}/\partial \fR^a{}_b$, all the $\fchi$'s need to be $\fchi_1$.
Since $(\fchi_1\R{2}\fchi_1)=0$, the product vanishes in this case. For lower-order contributions, there always exists at least one $\fchi_2$. This means that the product contains 
$(\R{2}\fchi_2)$ or $(\fchi_2 \R{2})$ which are zero as in Case B.

\item \underline{\bf (2) $N_\upsilon = 2I+1 $} \\
The only choice of $\form{\upsilon}^{(j)}$'s is that
there is one $(\R{3})$ and the rest are $(\R{2})$. In this case,
to have the contribution of order $\fomega^{2k+2}$ and
to avoid the appearance of more than one $\fchi_2$'s,
we need to have one $\fchi_2$ and the rest of the $\fchi$'s need to be $\fchi_1$.
We first note that $(\R{2}\fchi_2)=  (\fchi_2\R{2})=0$ and thus we cannot put 
$\fchi_2$ next to $(\R{2})$ to have a nontrivial result. Then, 
we instead encounter the structure of the form $(\fchi_1\R{3}\fchi_2)$ or  $(\fchi_2\R{3}\fchi_1)$, 
which is also zero because of Eq.~(\ref{traceRiemann3}).

For the product to be lower order than $\fomega^{2k+2}$, we need to introduce more than one $\fchi_2$'s
and thus these contributions vanish.

\item \underline{\bf (3) $N_\upsilon \ge 2I+2 $} \\
In any of this case, we need to have more than one $\fchi_2$'s, which leads to vanishing
contribution.

\end {itemize}

\item \underline{\bf Case  D} \\
Obviously, this case does not give any contribution of order $\fomega^{2k+2}$ or lower.

\end{itemize}

To summarize, we confirmed that 2nd and higher order terms of $\fR$ do not give any
contribution to the derivative of the anomaly polynomial $\partial \fP_{CFT}/\partial \fR^a{}_b$ at  $\fomega^{2k+2}$ or lower order. We also notice that the derivative expansion of the 
spin Hall current (containing $\fR$ at any order) starts at $\fomega^{2k+1}$ and thus 
2nd and higher order terms of the Christoffel connection do not  generate new contribution 
to the Einstein source at  $\fomega^{2k+2}$ or lower order. 
Therefore, the Einstein source at  $\fomega^{2k+2}$ or lower order 
does not contain the 2nd or higher order terms of $\fR$ as well as Christoffel connection.

Together with the result from the AdS$_3$  case (with $\tr[\fR^2]$), we have  also shown that for a product of $(2k+1)$ curvature 2-forms $(\fR^{2k+1})$, the contribution which contains 2nd and/or higher order terms of $\fR$
can become nontrivial only at order $\fomega^{2k+1}$ or higher, i.e. $N_R >  2k$. 

\vspace{0.5cm}
\noindent
\underline{\bf  Case II: $\fP_{CFT}= {c_{_M}}\, \form{F}^l\wedge\tr[\fR^{2k+4}] \,\,\,\,{\rm with} \,\,\,\,k,l\ge 0$}\\
Let us now take into account $U(1)$ gauge field and consider the mixed term
with single trace of the product of curvature 2-forms.
This term in the anomaly polynomial is admitted in AdS$_{d+1}$ with $d=4k+2l+6$.
Since the case of $\fF^l\wedge \tr[\fR^{2}]$ has been covered in details in Appendix~\ref{sec:appendixHallAdS7Einstein}, we shall consider $k\ge0$ here.

\noindent
\underline{\bf (1) 0th and 1st order terms only}\\
As in the previous case, we just consider the 0th and 1st order terms in $\fR$, $\fF$ and the
Christoffel connection only. Then there are two potential
 leading contribution to $\partial \fP_{CFT}/\partial \fR^a{}_b$ at the order $\fomega^{2k+l+1}$.
The first one is a linear combination of  the terms of the form
$(\R{0}\R{1}\R{0})\wedge\fomega^{2k+l}$. As in the argument
for AdS$_9$, this kind of terms does not contribute to the Einstein source at any order. The second one is proportional to
$(\F{0})\wedge(\R{0}\R{1})\wedge\fomega^{2k+l}$ or $(\F{0})\wedge(\R{1}\R{0})\wedge\fomega^{2k+l}$.
These however vanish
since $(\F{0})\sim dr\wedge \fu$ and all the terms in  $(\R{0}\R{1})$ and $(\R{1}\R{0})$
are proportional to $dr$ or $\fu$.
Therefore we focus on the subleading contribution (order $\fomega^{2k+l+2}$)
to the derivative $\partial \fP_{CFT}/\partial \fR^a{}_b$:
\bea\label{eq:orderEinstein2}
\frac{\partial \fP_{CFT}}{\partial \fR^a{}_b}&=&
 {c_{_M}}(2k+4)\left[
(\F{1}^l)\wedge \sum_{i=0}^{2k+2} (\R{1}^{i}\R{0}\R{1}^{2k+2-i})^b{}_a
+l\, (\F{0})\wedge (\F{1}^{l-1})\wedge (\R{1}\R{1}^{2k+2})^b{}_a
\right]\nonumber\\
&=& {c_{_M}}(2k+4)\left\{
\left[
(\R{0}\R{1}\R{1})^b{}_a
+(\R{1}\R{1}\R{0})^b{}_a
+(2k+1)(\R{1}\R{0}\R{1})^b{}_a
\right]\wedge (2\Phi \fomega)^l \wedge  (2\Phi_T \fomega)^{2k}
\right.
\nonumber\\
&& \left.
+ l (\F{0})\wedge (\R{1})^b{}_a\wedge(2\Phi \fomega)^{l-1}\wedge(2\Phi_T \fomega)^{2k+2}
\right\}.  
\eea
In the second line, we have used the identities in the Appendix \ref{sec:productRiemanns}.

The evaluation of the contribution to the Einstein source from the terms in the first line is the same as in Case I, 
but with an extra $(2\Phi \fomega)^l$ attached.
Therefore, the contribution from the first line to $\SpH^\tV$ is evaluated as
\be
-
\frac{1}{2 r^{d-2}}(2k+4)(2{c_{_M}}) \Phi^l
 \partial_r\left(r 
 \Phi_T^{2k+3}
\right).\ee 
Similarly, the second line is essentially the same as the contribution encountered 
 in the AdS$_7$ case (see $(\SpH^{(1)} ){}^\tV$ term in the $c_M$ case) but with extra $(2\Phi_T\fomega)$'s and $(2\Phi\fomega)$'s. Then we have
\be
-\frac{1}{2r^{d-2}}(2k+4) (2{c_{_M}})\left[ ( r\Phi_T^{2k+3} ) \partial_r (\Phi^l)\right].
\ee Combining both terms gives
\be
\SpH^\tV=-\frac{1}{2r^{d-2}}\frac{d}{dr}\left[  r \frac{\partial}{\partial \Phi_T} (({c_{_M}} \Phi^l (2\Phi_T^{2k+4} ) ) \right],
\ee which agrees with Eq.~(\ref{eq:repSigma}).

\noindent
\underline{\bf (2) 2nd and higher order terms}\\
We call the number of the derivative contained in $\fF^l$ as $N_F$.
Similarly as before, the number of derivatives in the $ (\fR^{2k+3})^b{}_a$ is defined as $N_R$.
Since we are considering up to order $\fomega^{2k+l+2}$ contribution, we have that $N_F + N_R \le  2k+l+2$.
and then carry out the classification depending of its value. For the curvature 2-form part,
we use the classification, Case A, B, C, D and E, that we introduced in Case I.
Then the classification goes as follows:
\begin{itemize}
\item
\underline{\bf $N_F < l$} \\
Since $(\F{0}^2)=0$,  the only case which can give nontrivial contribution is when $N_F=l-1$ (and hence $N_R\le 2k+3$) where $\fF^l =(\F{0}\F{1}^{l-1})\sim dr\wedge \fu$. Then, to have a nontrivial result, 
no $\fchi_2\sim dr\wedge \fu$ is allowed in the product of the curvature 2-forms.
For the 2nd and/or higher order terms to be contained
we have only to consider Case A,B,C and D for the curvature 2-form part.
Then the classification goes as follows:
\begin{itemize}
\item \underline{(1) Case A} \\
To have the contribution of order $\fomega^{2k+l+2}$ or lower,
we need to have at least one $\fchi_2$. We thus conclude that
this type of contribution vanishes.
\item \underline{(2) Case B} \\
In order not to have any $\fchi_2$, we need to set
all the $\form{\upsilon}^{(j)}$'s to be $(\R{2})$ and all the $\fchi$'s to
be $\fchi_1$. Then 
we always encounter  $(\R{2}\fchi_1)$ or $(\fchi_1\R{2})$ which contains 
terms proportional to $dr$ or $\fu$ only. 
Therefore this contribution vanishes when it is wedged with $(\F{0})$. 
\item \underline{(3) Case C} \\
Let us first consider the contribution with $N_R=2k+3$.  
There are two possible nontrivial contributions:

The first one is when all the $\form{\upsilon}^{(j)}$'s are $(\R{2})$,
while there is only one $\fchi_0$ and all the rest of $\fchi$'s are $\fchi_1$. 
We then always encounter $(\R{2}\fchi_1)$ or $(\fchi_1\R{2})$, which means
that the product vanishes when it is wedged with $(\F{0})$ as in Case B.

Another case is when one of $\form{\upsilon}^{(j)}$'s is $(\R{3})$, the rest are
$(\R{2})$ and all the $\fchi$'s are $\fchi_1$. In this case, if there exist
at least one $(\R{2})$, then this product vanishes since $(\fchi_1\R{2}\fchi_1)=0 $. 
On the other hand, when there is no $(\R{2})$,
we have the structure of the form $(\fchi_1\R{3}\fchi_1)$, 
which also vanishes when it is wedged with $(\F{0})\sim dr\wedge \fu$  as a result of Eq.~(\ref{traceRiemann3}).

When lower order contribution is  considered, there is only one case in which no $\fchi_2$ is contained:
all the $\form{\upsilon}^{(j)}$'s are $(\R{2})$ and all the $\fchi$'s are $\fchi_1$. This case
turns out to be zero since the product contains $(\R{2}\fchi_1)$ or $(\fchi_1\R{2})$, 
which vanishes when it is wedged with $(\F{0})$ as in Case B.

\item \underline{(4) Case D} \\
Obviously this case does not give any nontrivial contribution to the Einstein source
at $\fomega^{2k+l+2}$ or lower order.
\end{itemize}

\item
\underline{\bf $N_F = l$} \\
Due to $(\F{0}^2)=(\F{0}\F{2})=0$, 
it follows that $\fF^l=(\F{1}^l)$. The classification of the product of the curvature 2-form part is exactly the same as in Case I. 

\item
\underline{\bf $N_F\ge l+1$} \\
Previously in Case I, we showed that second and higher order objects
do not contribute to $(\fR^{2k+3})$ for $N_R\le 2k+2$. Since in this case $N_R \le 2k+1$ these will similarly be trivial.

\end{itemize}

To summarize we have shown that 2nd and higher order terms in $\fF$ and $\fR$ do not
generate extra contribution to $\partial \fP_{CFT}/\partial \fR^a{}_b$ at $\fomega^{2k+l+2}$ or lower order.
In addition to this, we also notice that the derivative expansion of the 
spin Hall current (containing $\fR$ and $\fF$ at any order) starts at $\fomega^{2k+l+1}$ and thus 
2nd and higher order terms of the Christoffel connection do not  generate new contribution 
to the Einstein source at  $\fomega^{2k+l+2}$ or lower order. 
Therefore, 2nd and higher order terms in $\fF$ and $\fR$ as well as the Christoffel connection
do not contribute to the $\fomega^{2k+l+2}$ or lower order of the Einstein source.

In particular, together with the analysis
from $\fF^l\wedge \tr[\fR^{2}]$, we have also
 shown that in $dr\wedge \fu \wedge (\fR^{2k+1})$, the contribution which contains at least one
2nd or higher order term of $\fR$ can become nontrivial only at order $\fomega^{2k+2}$ or higher, i.e. $N_R > 2k+1$.

\vspace{0.5cm}
\noindent
\underline{\bf  Case III: $\fP_{CFT}= c_{_g}\, \tr[\fR^{2k_1}]\wedge\ldots \wedge\tr[\fR^{2k_p}]$
\,\,\,\, {\rm with}\,\,\,\, $k_i\ge1, p\ge2$.}  \\
We next turn to the multiple traces case. We first start with the purely gravitational term
admitted in AdS$_{d+1}$ with $d= 4k_{tot}-2$, where $k_{tot}\equiv \sum_{i=1}^p k_i$.

\noindent
\underline{\bf (1) 0th and 1st order terms only} \\
To start with this computation, we ignore all terms aside from the 0th and 1st order terms in $\fR$ and the Christoffel connection.
And, first consider the case with more than two $(\R{0})$'s in the derivative 
$\partial \fP_{CFT}/\partial \fR^a{}_b$.
We always encounter $\tr[\fchi_1]\wedge  \tr[\fchi_1]$ or $(\fchi_2)\wedge \tr[\fchi_1]$, 
both of which are zero since  $\fchi_2\sim dr\wedge \fu$ and  $\tr[\fchi_1]\sim dr\wedge \fu$. 
Secondly, we consider the case with two $(\R{0})$'s and find that
the only nontrivial contribution to $\partial \fP_{CFT}/\partial \fR^a{}_b$ is of the form 
$(\fchi_1)\wedge \tr[\fchi_1]$ or $\fchi_2$ multiplied by an appropriate
power of $\fomega$. 
We notice that all the terms appearing in $(\R{0}\R{1})$ and  $(\R{1}\R{0})$
are proportional to $dr$ or $\fu$ and thus the only nontrivial contribution in $(\fchi_1)\wedge \tr[\fchi_1]$
is proportional to $(\R{0})\wedge \tr[(\R{0}\R{1})] $.  Henceforth, the arguments used in the AdS$_7$
and AdS$_9$ cases pertain here too, and we can conclude that these only contribute trivially to the Einstein source at any order.  

Therefore, to calculate the Einstein source at the leading order, we need only to consider the case involving just one $(\R{0})$.
The derivative $\partial \fP_{CFT}/\partial \fR^a{}_b$ in this case is given by 
\bea
\frac{\partial \fP_{CFT}}{\partial \fR^a{}_b}
&= & {c_{_g}}\, \sum_{i=1}^{p}(2 k_i)2^{p-1}
\form{S}_{(i)} ^b{}_a
\wedge (2\Phi_T\fomega)^{2k_{tot}-4}
\nonumber\\
& & +\,
{c_{_g}}\, \sum_{i,j; i\neq j}^{p}(2 k_i)(2k_j)2^{p-1}(\R{1})^b{}_a\wedge\tr[(\R{0}\R{1})]
\wedge (2\Phi_T\fomega)^{2k_{tot}-4},  \label{eq:twoterms_case3} 
\eea
where
\bea
\form{S}_{(i)}^b{}_a&=& [(\R{0}\R{1}\R{1})+(\R{1}\R{1}\R{0})+(2k_i-3)(\R{1}\R{0}\R{1})]^b{}_a,
\,\,\,\, {\rm for}\,\,\,\, k_i\ge2\,,   \nonumber \\
\form{S}_{(i)}^b{}_a&=& (\R{0})^b{}_a \wedge (2\Phi_T \fomega)^2,
\,\,\,\, {\rm for}\,\,\,\, k_i=1\,.
\eea
The first line of the expression for $\partial \fP_{CFT}/\partial \fR^a{}_b$ is from the cases where $(\R{0})$ is located at the product of
$\fR$'s (that is, not in the traces), while the second one corresponds to the cases
where $(\R{0})$ is inside of one of the traces. 

Without loss of generality, let $\form{S}_{(i)}^b{}_a=(\R{0})^b{}_a\wedge (2\Phi_T \fomega)^2$ for $1 \le i\le p_0$ 
(i.e. these are the terms generated when $\partial /\partial \fR^a{}_b$ acts on a $\tr [\fR^2]$ in $\fP_{CFT}$) and $\form{S}_{(i)}^b{}_a= [(\R{0}\R{1}\R{1})+(\R{1}\R{1}\R{0})+(2k_i-3)(\R{1}\R{0}\R{1})]^b{}_a,
$ for $p_0 +1\le i\le p$ 
(i.e. these are the terms generated 
when $\partial /\partial \fR^a{}_b$ acts on a $\tr [\fR^{2k_i}]$ with $k_i\ge2$ in $\fP_{CFT}$). Then, we can further simplify the derivative $\partial \fP_{CFT}/\partial \fR^a{}_b$ as 
\bea \label{eq:derivativePcase3simplified}
\frac{\partial \fP_{CFT}}{\partial \fR^a{}_b}
&= &
{c_{_g}}\, \left(\sum_{i=1}^{p_0} k_i\right)2^{p}
(\R{0})^b{}_a
\wedge (2\Phi_T\fomega)^{2k_{tot}-2}\nonumber\\
&&
+{c_{_g}}\, \left( \sum_{i=p_0+1}^{p}k_i\right)2^{p}
[(\R{0}\R{1}\R{1})+(\R{1}\R{1}\R{0})+(2k_i-3)(\R{1}\R{0}\R{1})]^b{}_a
\wedge (2\Phi_T\fomega)^{2k_{tot}-4}
\nonumber\\
& &
+4{c_{_g}}\, \left(k_{tot}^2- \sum_{i=1}^pk^2_i \right)2^{p-1}(\R{1})^b{}_a\wedge\tr[(\R{0}\R{1})]
\wedge (2\Phi_T\fomega)^{2k_{tot}-4}\, .   \label{eq:twoterms_case3v2} 
\eea
The contributions to $\SpH^\tV$ from the first two lines  
can be obtained from the results of  the AdS$_3$ case as well as Case I 
by attaching appropriate powers of $\fomega$. The results combine to give
\bea
&& -\frac{1}{2 r^{d-2}}({c_{_g}}\, 2^p) \sum_{i=1}^{p}(2k_i)
\left[
 \partial_r\left(r 
 \Phi_T^{2k_i-1}
\right)\right] (\Phi_T)^{2k_{tot}-2k_i } \nonumber\\
&=& -\frac{1}{2 r^{d-2}}({c_{_g}}\, 2^p) \sum_{i=1}^{p}(2k_i)
\left[
\left(
 \partial_r(r \Phi_T) 
 +r(2k_i-2)  \Phi_T'
\right)\right] 
(\Phi_T)^{2k_{tot}-2} \nonumber\\
&=& -\frac{1}{2 r^{d-2}}({c_{_g}}\, 2^p)
\left\{
 2k_{tot}\partial_r(r \Phi_T) 
 +4\left[
\left( \sum_{i=1}^{p}k_i^2 \right) - k_{tot}\right] r\Phi_T'
\right\}
\Phi_T^{2k_{tot}-2}.
\eea 
For the third line of \eqref{eq:derivativePcase3simplified} the contributions to $\SpH^\tV$ 
can be calculated by using the result of $(\SpH^{(1)})^\tV$ in the $c_g$  term of AdS$_7$ 
by attaching appropriate power of $\fomega$ and one immediately finds 
\be
-\frac{1}{2r^{d-2}} ({c_{_g}}\, 2^{p})\left[(2k_{tot})^2- 4\sum_{i=1}^pk^2_i \right]   (r\Phi_T' )\Phi^{2k_{tot}-2}_T.
\ee By adding them up, we finally have
\bea
\SpH^\tV&=&
-\frac{1}{2 r^{d-2}}({c_{_g}}\, 2^p)(2k_{tot})
\left[
 \partial_r(r \Phi_T) 
 +\left[ 2k_{tot}-2\right] r\Phi_T'
\right]
\Phi_T^{2k_{tot}-2} \nonumber\\
&=&
-\frac{1}{2 r^{d-2}}
\frac{d}{dr}
\left[
 r  \frac{\partial }{\partial \Phi_T} ({c_{_g}}\, 2^p \Phi_T^{2k_{tot}})
\right],
\eea which agrees with Eq.~(\ref{eq:repSigma}).

\noindent
\underline{\bf (2) 2nd and higher order terms} \\
Now we consider 2nd and higher order terms in $\fR$ and Christoffel connection to
show that these do not generate contributions to the Einstein source at
order $\fomega^{n-1}=\fomega^{2k_{tot}-2}$ or lower orders. 
We note that terms in $\partial \fP_{CFT}/\partial \fR^a{}_b$ have the following form
\be
{\cal T}\wedge \left( \fR^{2k_1-1}\right)\equiv
{\cal T}\wedge {\cal R}.
\ee
As in the general argument for the Maxwell sources,
we divide the trace part ${\cal T}$ as ${\cal T}={\cal T}_1\wedge {\cal T}_2\wedge {\cal T}_3$,
where ${\cal T}_1$ (${\cal T}_2$, ${\cal T}_3$, respectively) is the wedge product of the trace of the form
$TR_{(1)}$ ($TR_{(2)}$, $TR_{(3)}$, respectively) only.
Without loss of generality, we have considered the term generated by acting
 the derivative $\partial /\partial\fR^a{}_{b}$ on the first trace $\tr[\fR^{2k_1}]$.
We also let each trace $\tr[\fR^{2k_i}]$ 
to be of the form $TR_{(1)}$ for $2\le i \le p_1$,  $TR_{(2)}$ for $p_1 +1 \le i \le p_2$ 
and $TR_{(3)}$ for $p_2+1 \le i \le p$:
\bea
&&{\cal T}_1 \equiv \ \tr[\fR^{2k_2}] \wedge\ldots \wedge \tr[\fR^{2k_{p_1}}] \,  ,  \nonumber \\
&&{\cal T}_{2} \equiv\ \tr[\fR^{2k_{p_1+1}}] \wedge\ldots \wedge \tr[\fR^{2k_{p_2}}]\, ,  \nonumber \\
&&{\cal T}_{3} \equiv\ \tr[\fR^{2k_{p_2+1}}] \wedge\ldots \wedge \tr[\fR^{2k_{p}}] \,. 
\eea
We note that in contrast with the Maxwell case, the $k_i$  in ${\cal T}_1$ starts from $i=2$.
We then define $N_{T_i}$ to be the number of derivatives contained in ${\cal T}_i$ and $N_R$ the number of derivatives 
in ${\cal R}$.

We first study the case where either $p_1 < p$ or $p_2<p$ is satisfied 
(i.e. ${\cal T}_2\wedge{\cal T}_3$ contains at least one nontrivial trace and thus 
${\cal T}$ contains 2nd and/or higher order terms of $\fR$). 
 Similarly to the general argument for the Maxwell source, we know that
${\cal T}_i$ is zero unless :
\begin{itemize}
\item ${\cal T}_1:~ N_{T_1} = \sum_{i=2}^{p_1} (2k_i) $ or $\sum_{i=2}^{p_1} (2k_i)-1$ \\
\qquad(we notice that, in the latter case, ${\cal T}_1$  contains one $\tr[\chi_1]\sim dr\wedge \fu$ ).
\item ${\cal T}_2:~N_{T_2} > \sum_{i=p_1+1}^{p_2} (2k_i)$.
\item ${\cal T}_3:~ N_{T_3} \ge \sum_{i=p_2+1}^{p} (4k_i)$.
\end{itemize} 
From the results in Case I, it is also known that 
 the contributions to ${\cal R}$ containing 2nd and/or higher order terms of $\fR$
  become nontrivial only when $N_R \ge 2k_1-1$. 
When only the 0th and 1st order terms of $\fR$ are considered, the 
possible nontrivial configurations are ${\cal R}=\fchi_0,\fchi_1$ or $\fchi_2$,
 which correspond to $N_R=2k_1,2k_1-1$ or $2k_1-2$, respectively.

Combining these results on ${\cal T}$ and ${\cal R}$, for $N_R \ge 2k_1-1$, we deduce that 
\be \label{eq:inequ_case3}
N_{tot} > 2k_{tot}-2=n-1\, . 
\ee  Therefore the contributions to $\partial \fP_{CFT}/\partial \fR^a{}_b$ are of order higher than $\fomega^{n-1}$.
Let us consider another case, that of $N_R=2k_1-2$. 
For $N_{T_1} = \sum_{i=2}^{p_1} (2k_i) $,  the above inequality \eqref{eq:inequ_case3} still holds while for $N_{T_1}=\sum_{i=2}^{p_1} (2k_i)-1$, 
the contribution vanishes since ${\cal R}=\fchi_2\sim dr\wedge \fu$ and 
${\cal T}_1$ contains one  $\tr[\fchi_1]\sim dr\wedge \fu$.

Comments on the special case with $p_1=p_2=p$ are in order (i.e. 
${\cal T}$ does not contain any 2nd or higher order term of $\fR$). 
In this case, ${\cal R}$ needs to have 2nd and/or higher terms of $\fR$ but
from the result of Case I,  $N_R \ge 2k_1-1$ is required to have a nontrivial result. 
For $N_{T_1} = \sum_{i=2}^p (2k_{i})$, this leads to $N_{tot} \ge 2k_{tot}-1>n-1$. 
For $N_{T_1} = \sum_{i=2}^p (2k_{i})-1$, we first notice that ${\cal T}_1$ contains a $\tr[\fchi_1]\sim dr\wedge\fu$. 
In addition to this, from Case II, we know that $dr\wedge \fu \wedge (\fR^{2k_1-1})$ with 2nd and/or higher order term of $\fR$ 
contained in $(\fR^{2k_1-1})$ can become nontrivial only when $N_R > 2k_1-1$. Then we finally have 
$N_{tot} > 2k_{tot}-2=n-1$. 

From the above argument we conclude that there is no nontrivial contribution
to the derivative $\partial \fP_{CFT}/\partial \fR^a{}_b$ at order $\fomega^{2k_{tot}-2}$ (or lower) coming from 2nd or higher order terms of $\fR$. 
We also notice that the Christoffel connection at the 2nd or higher order do not 
generate new contribution to $\partial \fP_{CFT}/\partial \fR^a{}_b$ at order $\fomega^{2k_{tot}-2}$ (or lower), 
since the derivative expansion of the spin Hall current starts at the order $\fomega^{2k_{tot}-3}$.
 
In all, we have shown that 2nd and higher order terms in $\fR$ as well as the Christoffel connection
do not give nonzero contribution to the Einstein source at the $\fomega^{2k_{tot}-2}$ or
lower order .

\vspace{0.5cm} 
\noindent
\underline{\bf  Case IV: $\fP_{CFT}= c_{_M}\, \fF^l\wedge \tr[\fR^{2k_1}]\wedge\ldots \wedge\tr[\fR^{2k_p}]$
\,\,\,\,{\rm with}\,\,\,\, $l\ge1$, $k_i\ge1$ {\rm and} $p\ge2$} \\
Lastly we consider the most general form of the terms in the anomaly polynomial.
This anomaly polynomial is admitted in AdS$_{d+1}$ with $d= 4k_{tot}+2l-2$

\noindent
\underline{\bf (1) 0th and 1st order terms only} \\
Ignoring the 2nd or higher terms in $\fR$, $\fF$ and the Christoffel
symbol we will now consider the derivative $\partial \fP_{CFT}/\partial \fR^a{}_b$. 
We first show that there is no contribution to the Einstein source at order lower than $\fomega^{2k_{tot}+l-2}$.
Since $(\F{0}^2)=0$, $\fF^l$ can contain one or no $(\F{0})$. When $\fF^l$  contains no $(\F{0})$ (i.e. $\fF^l=(\F{1}^l)$), 
the argument is the same
as Case III. We next consider the case when $\fF^l$ contains only one $(\F{0})$.
In this case, for the curvature 2-form part, 
since $\tr[\fchi_1]\sim dr\wedge \fu$ and $\tr[\fchi_2]=0$,
no $(\R{0})$ can be put in any of the traces. Hence, all the $(\R{0})$'s need to be located
in the non-trace part. Moreover, since $(\R{0}\R{1}\R{0})\sim dr\wedge \fu$,
only one or no $(\R{0})$ can be put in that part. In the former case 
(in which $\partial \fP_{CFT}/\partial \fR^a{}_b$ is of order $\fomega^{2k_{tot}+l-3}$), 
the only one nontrivial possibility is when $\partial \fP_{CFT}/\partial \fR^a{}_b$ is proportional to
$(\F{0})\wedge (\R{0})$ wedged by an appropriate power of $\fomega$.
As we have seen in the AdS$_5$ case,  this does not give any contribution to the Einstein source
at any order. Therefore, we conclude that there is no $(\R{0})$ in the non-trace part, i.e. all the $\fR$'s are $(\R{1})$.
Hence, the first non-trivial contribution to the Einstein source is at order $\fomega^{2k_{tot}+l-2}$ and is given by
$\fomega^{2k_{tot}+l-2}$ contribution
\bea \label{eq:derivative01only_case4}
& &- {c_{_M}}\, \sum_{i=1}^{p}(2 k_i)2^{p-1}
\form{S}_{(i)} ^b{}_a
\wedge (2\Phi_T\fomega)^{2k_{tot}-4}\wedge (2\Phi\fomega)^{l}
\nonumber\\
& &
- {c_{_M}}\sum_{i,j; i\neq j}^{p}(2 k_i)(2k_j)2^{p-1}(\R{1})^b{}_a\wedge\tr[(\R{0}\R{1})]
\wedge (2\Phi_T\fomega)^{2k_{tot}-4}\wedge (2\Phi\fomega)^{l} \nonumber \\
&& - {c_{_M}} l(\F{0})\wedge\sum_{i=1}^{p} (2k_i)2^{p-1}(\R{1})^b{}_a\wedge (2\Phi_T\fomega)^{2k_{tot}-2}
\wedge(2\Phi\fomega)^{l-1},
  \label{eq:threeterms}
\eea
where
\bea
\form{S}_{(i)}^b{}_a&=& [(\R{0}\R{1}\R{1})+(\R{1}\R{1}\R{0})+(2k_i-3)(\R{1}\R{0}\R{1})]^b{}_a,
\,\,\,\, {\rm for}\,\,\,\, k_i\ge2,   \nonumber \\
\form{S}_{(i)}^b{}_a&=& (\R{0})^b{}_a,
\,\,\,\, {\rm for}\,\,\,\, k_i=1.
\eea
In \eqref{eq:derivative01only_case4}, the first two lines are contributions coming from the cases 
with $\fF^l = (\F{1}^l)$, while the third one is from the cases with $\fF^l = (\F{0}\F{1}^{l-1})$.

On the one hand, the contributions  to  $\SpH^\tV$ from the first two lines
can be obtained by attaching $(2\fomega \Phi)^l$
to the result of Case III. We find
\be
-\frac{1}{2 r^{d-2}} (2k_{tot})( 2^p {c_{_M}})\Phi^l
\partial_r
\left[
 r ( \Phi_T^{2k_{tot}-1})
\right].
\ee On the other hand, we can evaluated the contribution from third line to $\SpH^\tV$  by attaching appropriate powers of $(2\Phi\fomega)$ and $(2 \Phi_T \fomega) $ 
to the result for the term proportional to $(\F{0})\wedge(\R{0})^b{}_a$ 
of $\partial \fP_{CFT}/\partial \fR^a{}_b$ in Case II (see the final line of Eq. \eqref{eq:orderEinstein2})
\be
-\frac{1}{2r^{d-2}}(2k_{tot})(2^p {c_{_M}}  )(r \Phi_T^{2k_{tot}-1}) \partial_r (\Phi^l)\,.
\ee Combining both contributions, that also yields an agreement with Eq.~(\ref{eq:repSigma}), we have
\be
\SpH^\tV=-\frac{1}{2 r^{d-2}} \frac{d}{dr}\left[
r \frac{\partial}{\partial \Phi_T}
({c_{_M}}
 \Phi^l   2^p \Phi^{2k_{tot}})
\right] \, .
\ee

\noindent
\underline{\bf (2) 2nd and higher order terms} \\
Our aim is to show that  2nd and higher order terms of $\fR$, $\fF$ as well as 
the Christoffel connection do not contribute to the Einstein source at order $\fomega^{2k_{tot}+l-2}$ or lower.   
For this purpose, we carry out the classification based on the number of the derivatives contained in $\fF^l$ 
(we denote it by $N_F$).
For the classification of the curvature 2-from part, we use the same notations and strategy
as in Case III. Similarly to that in Case III, let us start with the case 
where either $p_1<p$ or $p_2<p$ is satisfied.
For $N_F \ge l$, the arguments go through exactly as in Case III.
For $N_F < l$, since $(\F{0}^2)=0$, the only one potential nontrivial case is 
when $\fF^l = (\F{0}^l\F{1})\sim dr\wedge \fu$ ($N_F=l-1$).  
To have a nonzero result, ${\cal T}_1$ must not contain any $\tr[\fchi_1]\sim dr\wedge \fu$ and hence the
$N_{T_1}=\sum_{i=2}^{p_1}(2k_i)-1$ case gives no contribution to $\partial \fP_{CFT}/\partial \fR^a{}_b$. For $N_{T_1}=\sum_{i=2}^{p_1}(2k_i)$, since $\partial \fP_{CFT}/\partial \fR^a{}_b$ contains 
$dr\wedge\fu\wedge {\cal R}$ with ${\cal R}=\fR^{2k_1-1}$, it is known from the result of Case II 
that this contribute can become nontrivial only when $N_R > 2k_1-1$. 
Then we have
\be
N_{tot} >2k_{tot}+l-2=n-1.
\ee 

For $p_1=p_2=p$, the same argument above goes through exactly.

Therefore, we conclude that there is no contribution to the derivative $\partial \fP_{CFT}/\partial \fR^a{}_b$ at the $\fomega^{2k_{tot}+l-2}$ or lower order with 2nd or higher order terms of $\fR$ or $\fF$. 
In addition to this, since the derivative expansion of the spin Hall current start with $\fomega^{2k_{tot}+l-3}$, 
the 2nd and higher order terms in the Christoffel connection do not generate any contribution to 
the Einstein source at $\fomega^{2k_{tot}+l-2}$ or lower order. 

To recapitulate, we confirmed that the leading order contribution to the Einstein source is of order $\fomega^{2k_{tot}+l-2}$
and does not contain 2nd or higher order terms of $\fR$ and $\fF$ or any Christoffel connection.

 \subsection{Symmetry properties of the spin Hall current}
\label{sec:proveAssumptions}
This subsection is devoted to the prove of the symmetry properties of the Hall current summarized in  Eqs.~(\ref{eq:Assumptions}). 
For the benefit of the reader, we quote Eqs ~(\ref{eq:Assumptions}) below:
\bea\label{eq:Assumptions2}
(\SpH)_{(rr)}{}^c= (\SpH)_{(r\nu)}{}^\rho=P^\nu{}_\rho (\SpH)_{(\nu\mu)}{}^\rho=0, \qquad 
P^\nu{}_\mu (\SpH)_{(r\nu)}{}^r=(\SpH)_{(r\mu)}{}^r\, . 
\eea 
As confirmed in the previous sections, we know that  the leading order contribution to $(\SpH)_{(ab)}{}^c$ 
is at order $\fomega^{n-1}$ and is composed of 0th and 1st order objects only. 
Therefore, in this part,  we will prove Eqs.~(\ref{eq:Assumptions}) for such contributions.
For concreteness, we will explicitly mention the AdS$_3$ case to illustrate how the proofs go. 
The generalization to higher dimensions is straightforward and we highlight  
the analogous computations/arguments.
 
For AdS$_3$ with a gravitational Chern-Simons term, the anomaly polynomial reads:
\be
\fP_{CFT_2}= c_{_g} \tr [\fR^2]\, .
\ee
To calculate the Einstein source, what we need to evaluate is
\be
(\THall)_{ab} = \nabla_c (\SpH)_{(ab)}{}^c\, , \qquad {\rm where\quad}
(\SpH)^{ab}{}_c =
-2\,c_{_g} \varepsilon^{a\,p_1p_2} R^b{}_{c\,p_1 p_2}\,.
\ee
The leading order contribution to $(\SpH)^{ab}{}_c$ is given by
\be
(\SpH)^{ab}{}_c
=
-2c_{_g} \varepsilon^{a\,p_1p_2} (\R{0})^b{}_{c\,p_1 p_2}\, ,
\label{eq:ads3_sigma_leadingv2}
\ee which is of order $\fomega^0$. 
For computational purpose, it is useful to notice that
\be\label{eq:switchorderSigma}
(\SpH)^{a}{}_b{}^c
=+2c_{_g} \varepsilon^{a\,p_1p_2} (\R{0})^c{}_{b\,p_1 p_2}\, , 
\ee 
which results from the anti-symmetric property of $(\SpH)^{abc}$ : $(\SpH)^{abc} = -(\SpH)^{acb} $. 

For higher-dimensional AdS$_{2n+1}$, as we have seen in the previous sections,  
in the computations of the $\fomega^{n-1}$ order contribution to $(\SpH)^{ab}{}_c$, 
we encounter similar expressions for $(\SpH)^{ab}{}_c$ as above but with $(\R{0})$ replaced by either of 
$(\R{0}\R{1}\R{1}),(\R{1}\R{0}\R{1})$ or $(\R{1}\R{1}\R{0})$ or  $dr\wedge \fu \wedge (\R{1})$
or by  a linear combination of them (wedged by an appropriate power of $\fomega$).
 In all of these cases, we will show  Eqs.~(\ref{eq:Assumptions2}) hold.

Then the proof of Eqs.~(\ref{eq:Assumptions2}) is as follows:
\begin{enumerate}
\item \underline{ $ (\SpH)_{(r r)}{}^c=0$} \\
Since $(\R{0})^c{}_r$ is proportional to $\fu$, we have 
\be
 (\SpH)_{r r}{}^c=
 -u_\alpha  (\SpH)^\alpha{}_{r}{}^c
 \sim u_\alpha \varepsilon^{\alpha p_1 p_2} 
 (\R{0})^c{}_{rp_1 p_2}=0\, , 
\ee 
and thus $ (\SpH)_{(r r)}{}^c=0$ for the AdS$_3$ case. 

Let us next explain generalization to the higher dimensional cases. For the case with
$(\R{0})$ replaced by $(\R{0}\R{1}\R{1}),(\R{1}\R{0}\R{1})$, $(\R{1}\R{1}\R{0})$, or $dr\wedge \fu \wedge (\R{1}\R{1}\R{1})$, 
this relation holds,  since $(\R{0}\R{1}\R{1})^c{}_r$, $(\R{1}\R{0}\R{1})^c{}_r$, $(\R{1}\R{1}\R{0})^c{}_r$ 
and $dr\wedge \fu \wedge (\R{1}\R{1}\R{1})^c{}_r$ are all proportional to $\fu$, leading to 
 $ (\SpH)_{r r}{}^c\sim u_\alpha u_{\beta}\epsilon^{\alpha\beta\cdots}=0$. 

\item \underline{$(\SpH)_{(r\mu)}{}^\nu=0$} \\
To show the relation, we first evaluate the terms before symmetrization. The direct calculation shows
\bea
(\SpH)_{r\mu}{}^\nu&=&
 - u_\alpha (\SpH)^\alpha{}_{\mu}{}^\nu
 =-2 c_{_g} u_\alpha \varepsilon^{\alpha p_1 p_2} (\R{0})^\nu{}_{\mu p_1 p_2} 
 =-4 c_{_g} (r^{-1} \Psi' u_\mu)( \varepsilon^{\alpha \nu r}  u_\alpha )\, , \nonumber\\
 (\SpH)_{\mu r}{}^\nu&=&
 -u_\mu  (\SpH)^r{}_{r}{}^\nu
 =-2 c_{_g} u_\mu \varepsilon^{r p_1 p_2} (\R{0})^\nu{}_{r p_1 p_2}
 =4 c_{_g} (r^{-1} \Psi' u_\mu) (\varepsilon^{\alpha \nu r}u_\alpha )\, , 
\nonumber\\
\eea 
from which we have  $(\SpH)_{(r\mu)}{}^\nu=0$. 

For higher-dimensional computation, when $(\R{0})$ is replaced by  $(\R{1}\R{0}\R{1})$ or $(\R{1}\R{1}\R{0})$, 
the components before the symmetrization are already zero. When the replacement is done by $(\R{0}\R{1}\R{1})$, 
$(\SpH)_{r\mu}{}^\nu$ and $(\SpH)_{\mu r}{}^\nu$ are nonzero but $(\SpH)_{(r\mu)}{}^\nu=0$.  
In the cases of $dr\wedge \fu \wedge (\R{1}) $, it is straightforward to show 
$(\SpH)_{r\mu}{}^\nu\sim u_\alpha u_\beta \varepsilon^{\alpha \beta \cdots} =0$, 
while $(\SpH)_{\mu r}{}^\nu =0$ holds since $ (\R{1})^\nu{}_r$ is proportional to $\fu$. 

\item \underline{$P^\nu{}_\rho (\SpH)_{(\nu\mu)}{}^\rho=0$}\\
Before the symmetrization, we first have
\be
P^\nu{}_\rho (\SpH)_{\nu\mu}{}^\rho
\sim
 P_{\rho\alpha}(\SpH)^\alpha{}_{\mu}{}^\rho
 \sim
  P_{\rho\alpha}\varepsilon^{\alpha p_1 p_2} (\R{0})^\rho{}_{\mu p_1 p_2}=0\, , 
\ee since the upper index $\rho$ of $ (\R{0})^\rho{}_{\mu}$ is carried either by $u^\rho$ or $dx^\rho$. 
We also have
\be
P^\nu{}_\rho (\SpH)_{\mu\nu}{}^\rho
\sim P^\nu{}_\rho (\R{0})^\rho{}_\nu=0\, ,
\ee since either $u^\rho, u_\nu$ or $dx^\rho\wedge (P_{\nu \sigma} dx^\sigma)$ appears in 
all the terms in $(\R{0})^\rho{}_\nu$. Thus we conclude $P^\nu{}_\rho (\SpH)_{(\nu\mu)}{}^\rho=0$ holds. 

For higher dimensions, when $(\R{0})$ is replaced by either of $(\R{0}\R{1}\R{1}),(\R{1}\R{0}\R{1})$ or $(\R{1}\R{1}\R{0})$, 
the same arguments go through and we have 
$P^\nu{}_\rho (\SpH)_{\nu\mu}{}^\rho=P^\nu{}_\rho (\SpH)_{\mu\nu}{}^\rho=0$. 
In the case of the replacement by $dr\wedge \fu \wedge (\R{1}) $,
$P^\nu{}_\rho (\SpH)_{\nu\mu}{}^\rho=P^\nu{}_\rho (\SpH)_{\mu\nu}{}^\rho=0$ hold as a result of 
$dr\wedge \fu \wedge (\R{1})^\rho{}_\mu
\sim u^\rho u_\mu$. 
Therefore, the relation
$P^\nu{}_\rho (\SpH)_{(\nu\mu)}{}^\rho=0$ is true for higher dimensions as well. 

\item \underline{$P^\nu{}_\mu (\SpH)_{(r\nu)}{}^r=(\SpH)_{(r\mu)}{}^r$} \\ 
Lastly, from Eq.~(\ref{eq:SigmarmrAdS3}) in the computation of the Einstein source in AdS$_3$, 
we see that both the lower index $\nu$ 
of $(\SpH)_{r\nu}{}^r$ and $(\SpH)_{\nu r}{}^r$
are carried by $P_{\nu \gamma}$. Therefore the equality $P^\nu{}_\mu (\SpH)_{(r\nu)}{}^r=(\SpH)_{(r\mu)}{}^r$ holds.

In higher dimensions, when  $(\R{0})$ is replaced by $(\R{0}\R{1}\R{1}),(\R{1}\R{0}\R{1})$ or $(\R{1}\R{1}\R{0})$, 
the lower index $\nu$ of $(\SpH)_{r\nu}{}^r$ and $(\SpH)_{\nu r}{}^r$ is still carried by $P_{\nu\gamma}$ and thus the same computations go through. 
In the case of the replacement by $dr\wedge \fu \wedge (\R{1}) $,
we have
$ (\SpH)_{r \nu}{}^r \sim u_\alpha u_\beta \varepsilon^{\alpha\beta\ldots}=0$ as well as 
 $(\SpH)_{\nu r}{}^r =r^2 P_{\nu \alpha} (\SpH)^\alpha{}_{ r}{}^r$.  
Consequently $P^\nu{}_\mu (\SpH)_{(\nu r)}{}^r=(\SpH)_{(\mu r)}{}^r$ holds even in higher dimensions. 
  
\end{enumerate}

 \bibliographystyle{utphys}
\bibliography{fluids-bib}

\providecommand{\href}[2]{#2}\begingroup\raggedright\begin{thebibliography}{10}

\bibitem{Loganayagam:2012zg}
R.~Loganayagam, ``{Anomalies and the Helicity of the Thermal State},''
\href{http://arxiv.org/abs/1211.3850}{{\ttfamily arXiv:1211.3850 [hep-th]}}.

\bibitem{Bhattacharyya:2007vs}
S.~Bhattacharyya, S.~Lahiri, R.~Loganayagam, and S.~Minwalla, ``{Large rotating
  AdS black holes from fluid mechanics},''
  \href{http://dx.doi.org/10.1088/1126-6708/2008/09/054}{{\em JHEP} {\bfseries
  0809} (2008) 054},
\href{http://arxiv.org/abs/0708.1770}{{\ttfamily arXiv:0708.1770 [hep-th]}}.

\bibitem{Erdmenger:2008rm}
J.~Erdmenger, M.~Haack, M.~Kaminski, and A.~Yarom, ``{Fluid dynamics of
  R-charged black holes},''
  \href{http://dx.doi.org/10.1088/1126-6708/2009/01/055}{{\em JHEP} {\bfseries
  0901} (2009) 055},
\href{http://arxiv.org/abs/0809.2488}{{\ttfamily arXiv:0809.2488 [hep-th]}}.

\bibitem{Banerjee:2008th}
N.~Banerjee, J.~Bhattacharya, S.~Bhattacharyya, S.~Dutta, R.~Loganayagam, {\em
  et al.}, ``{Hydrodynamics from charged black branes},''
  \href{http://dx.doi.org/10.1007/JHEP01(2011)094}{{\em JHEP} {\bfseries 1101}
  (2011) 094},
\href{http://arxiv.org/abs/0809.2596}{{\ttfamily arXiv:0809.2596 [hep-th]}}.

\bibitem{Torabian:2009qk}
M.~Torabian and H.-U. Yee, ``{Holographic nonlinear hydrodynamics from AdS/CFT
  with multiple/non-Abelian symmetries},''
  \href{http://dx.doi.org/10.1088/1126-6708/2009/08/020}{{\em JHEP} {\bfseries
  08} (2009) 020},
\href{http://arxiv.org/abs/0903.4894}{{\ttfamily arXiv:0903.4894 [hep-th]}}.

\bibitem{Son:2009tf}
D.~T. Son and P.~Surowka, ``{Hydrodynamics with Triangle Anomalies},''
  \href{http://dx.doi.org/10.1103/PhysRevLett.103.191601}{{\em Phys.Rev.Lett.}
  {\bfseries 103} (2009) 191601},
\href{http://arxiv.org/abs/0906.5044}{{\ttfamily arXiv:0906.5044 [hep-th]}}.

\bibitem{Kharzeev:2009p}
D.~E. Kharzeev and H.~J. Warringa, ``{Chiral Magnetic conductivity},''
  \href{http://dx.doi.org/10.1103/PhysRevD.80.034028}{{\em Phys.Rev.}
  {\bfseries D80} (2009) 034028},
  \href{http://arxiv.org/abs/0907.5007}{{\ttfamily arXiv:0907.5007 [hep-ph]}}.

\bibitem{Lublinsky:2009wr}
M.~Lublinsky and I.~Zahed, ``{Anomalous Chiral Superfluidity},''
  \href{http://dx.doi.org/10.1016/j.physletb.2010.01.015}{{\em Phys.Lett.}
  {\bfseries B684} (2010) 119--122},
\href{http://arxiv.org/abs/0910.1373}{{\ttfamily arXiv:0910.1373 [hep-th]}}.

\bibitem{Neiman:2010zi}
Y.~Neiman and Y.~Oz, ``{Relativistic Hydrodynamics with General Anomalous
  Charges},'' \href{http://dx.doi.org/10.1007/JHEP03(2011)023}{{\em JHEP}
  {\bfseries 1103} (2011) 023},
\href{http://arxiv.org/abs/1011.5107}{{\ttfamily arXiv:1011.5107 [hep-th]}}.

\bibitem{Bhattacharya:2011tra}
J.~Bhattacharya, S.~Bhattacharyya, S.~Minwalla, and A.~Yarom, ``{A Theory of
  first order dissipative superfluid dynamics},''
\href{http://arxiv.org/abs/1105.3733}{{\ttfamily arXiv:1105.3733 [hep-th]}}.

\bibitem{Kharzeev:2011ds}
D.~E. Kharzeev and H.-U. Yee, ``{Anomalies and time reversal invariance in
  relativistic hydrodynamics: the second order and higher dimensional
  formulations},'' \href{http://dx.doi.org/10.1103/PhysRevD.84.045025}{{\em
  Phys.Rev.} {\bfseries D84} (2011) 045025},
\href{http://arxiv.org/abs/1105.6360}{{\ttfamily arXiv:1105.6360 [hep-th]}}.

\bibitem{Loganayagam:2011mu}
R.~Loganayagam, ``{Anomaly Induced Transport in Arbitrary Dimensions},''
\href{http://arxiv.org/abs/1106.0277}{{\ttfamily arXiv:1106.0277 [hep-th]}}.

\bibitem{Neiman:2011mj}
Y.~Neiman and Y.~Oz, ``{Anomalies in Superfluids and a Chiral Electric
  Effect},'' \href{http://dx.doi.org/10.1007/JHEP09(2011)011}{{\em JHEP}
  {\bfseries 1109} (2011) 011},
\href{http://arxiv.org/abs/1106.3576}{{\ttfamily arXiv:1106.3576 [hep-th]}}.

\bibitem{Dubovsky:2011sk}
S.~Dubovsky, L.~Hui, and A.~Nicolis, ``{Effective field theory for
  hydrodynamics: Wess-Zumino term and anomalies in two spacetime dimensions},''
\href{http://arxiv.org/abs/1107.0732}{{\ttfamily arXiv:1107.0732 [hep-th]}}.

\bibitem{Kimura:2011ef}
T.~Kimura and T.~Nishioka, ``{The Chiral Heat Effect},''
\href{http://arxiv.org/abs/1109.6331}{{\ttfamily arXiv:1109.6331 [hep-th]}}.

\bibitem{Lin:2011aa}
S.~Lin, ``{An anomalous hydrodynamics for chiral superfluid},''
  \href{http://dx.doi.org/10.1103/PhysRevD.85.045015}{{\em Phys.Rev.}
  {\bfseries D85} (2012) 045015},
\href{http://arxiv.org/abs/1112.3215}{{\ttfamily arXiv:1112.3215 [hep-ph]}}.

\bibitem{Manes:2012hf}
J.~L. Manes and M.~Valle, ``{Parity violating gravitational response and
  anomalous constitutive relations},''
  \href{http://dx.doi.org/10.1007/JHEP01(2013)008}{{\em JHEP} {\bfseries 1301}
  (2013) 008},
\href{http://arxiv.org/abs/1211.0876}{{\ttfamily arXiv:1211.0876 [hep-th]}}.

\bibitem{Banerjee:2012iz}
N.~Banerjee, J.~Bhattacharya, S.~Bhattacharyya, S.~Jain, S.~Minwalla, {\em et
  al.}, ``{Constraints on Fluid Dynamics from Equilibrium Partition
  Functions},'' \href{http://dx.doi.org/10.1007/JHEP09(2012)046}{{\em JHEP}
  {\bfseries 1209} (2012) 046},
\href{http://arxiv.org/abs/1203.3544}{{\ttfamily arXiv:1203.3544 [hep-th]}}.

\bibitem{Jensen:2012jy}
K.~Jensen, ``{Triangle Anomalies, Thermodynamics, and Hydrodynamics},''
  \href{http://dx.doi.org/10.1103/PhysRevD.85.125017}{{\em Phys.Rev.}
  {\bfseries D85} (2012) 125017},
\href{http://arxiv.org/abs/1203.3599}{{\ttfamily arXiv:1203.3599 [hep-th]}}.

\bibitem{Jain:2012rh}
S.~Jain and T.~Sharma, ``{Anomalous charged fluids in 1+1d from equilibrium
  partition function},'' \href{http://dx.doi.org/10.1007/JHEP01(2013)039}{{\em
  JHEP} {\bfseries 1301} (2013) 039},
\href{http://arxiv.org/abs/1203.5308}{{\ttfamily arXiv:1203.5308 [hep-th]}}.

\bibitem{Valle:2012em}
M.~Valle, ``{Hydrodynamics in 1+1 dimensions with gravitational anomalies},''
  \href{http://dx.doi.org/10.1007/JHEP08(2012)113}{{\em JHEP} {\bfseries 1208}
  (2012) 113},
\href{http://arxiv.org/abs/1206.1538}{{\ttfamily arXiv:1206.1538 [hep-th]}}.

\bibitem{Jensen:2012kj}
K.~Jensen, R.~Loganayagam, and A.~Yarom, ``{Thermodynamics, gravitational
  anomalies and cones},''
\href{http://arxiv.org/abs/1207.5824}{{\ttfamily arXiv:1207.5824 [hep-th]}}.

\bibitem{Golkar:2012kb}
S.~Golkar and D.~T. Son, ``{Non-Renormalization of the Chiral Vortical Effect
  Coefficient},''
\href{http://arxiv.org/abs/1207.5806}{{\ttfamily arXiv:1207.5806 [hep-th]}}.

\bibitem{Bhattacharyya:2012xi}
S.~Bhattacharyya, S.~Jain, S.~Minwalla, and T.~Sharma, ``{Constraints on
  Superfluid Hydrodynamics from Equilibrium Partition Functions},''
  \href{http://dx.doi.org/10.1007/JHEP01(2013)040}{{\em JHEP} {\bfseries 1301}
  (2013) 040},
\href{http://arxiv.org/abs/1206.6106}{{\ttfamily arXiv:1206.6106 [hep-th]}}.

\bibitem{Valle:2013aia}
M.~Valle, ``{Kinetic theory and evolution of cosmological fluctuations with
  neutrino number asymmetry},''
\href{http://arxiv.org/abs/1307.0392}{{\ttfamily arXiv:1307.0392 [hep-th]}}.

\bibitem{Bhattacharyya:2013ida}
S.~Bhattacharyya, J.~R. David, and S.~Thakur, ``{Second order transport from
  anomalies},''
\href{http://arxiv.org/abs/1305.0340}{{\ttfamily arXiv:1305.0340 [hep-th]}}.

\bibitem{Megias:2013uua}
E.~Megias and F.~Pena-Benitez, ``{Fluid/Gravity Correspondence, Second Order
  Transport and Gravitational Anomaly},''
\href{http://arxiv.org/abs/1307.7592}{{\ttfamily arXiv:1307.7592 [hep-th]}}.

\bibitem{Jensen:2013kka}
K.~Jensen, R.~Loganayagam, and A.~Yarom, ``{Anomaly inflow and thermal
  equilibrium},''
\href{http://arxiv.org/abs/1310.7024}{{\ttfamily arXiv:1310.7024 [hep-th]}}.

\bibitem{Jensen:Nov2013}
K.~Jensen, R.~Loganayagam, and A.~Yarom, ``{To appear},''
\href{http://arxiv.org/abs/13xx.xxxx}{{\ttfamily arXiv:13xx.xxxx [hep-th]}}.

\bibitem{Hou:2012xg}
D.-F. Hou, H.~Liu, and H.-c. Ren, ``{A Possible Higher Order Correction to the
  Vortical Conductivity in a Gauge Field Plasma},''
  \href{http://dx.doi.org/10.1103/PhysRevD.86.121703}{{\em Phys.Rev.}
  {\bfseries D86} (2012) 121703},
\href{http://arxiv.org/abs/1210.0969}{{\ttfamily arXiv:1210.0969 [hep-th]}}.

\bibitem{Landsteiner:2011cp}
K.~Landsteiner, E.~Megias, and F.~Pena-Benitez, ``{Gravitational Anomaly and
  Transport},'' \href{http://dx.doi.org/10.1103/PhysRevLett.107.021601}{{\em
  Phys.Rev.Lett.} {\bfseries 107} (2011) 021601},
\href{http://arxiv.org/abs/1103.5006}{{\ttfamily arXiv:1103.5006 [hep-ph]}}.

\bibitem{Loganayagam:2012pz}
R.~Loganayagam and P.~Surowka, ``{Anomaly/Transport in an Ideal Weyl gas},''
  \href{http://dx.doi.org/10.1007/JHEP04(2012)097}{{\em JHEP} {\bfseries 1204}
  (2012) 097},
\href{http://arxiv.org/abs/1201.2812}{{\ttfamily arXiv:1201.2812 [hep-th]}}.

\bibitem{Landsteiner:2011iq}
K.~Landsteiner, E.~Megias, L.~Melgar, and F.~Pena-Benitez, ``{Holographic
  Gravitational Anomaly and Chiral Vortical Effect},''
  \href{http://dx.doi.org/10.1007/JHEP09(2011)121}{{\em JHEP} {\bfseries 1109}
  (2011) 121},
\href{http://arxiv.org/abs/1107.0368}{{\ttfamily arXiv:1107.0368 [hep-th]}}.

\bibitem{Chapman:2012my}
S.~Chapman, Y.~Neiman, and Y.~Oz, ``{Fluid/Gravity Correspondence, Local Wald
  Entropy Current and Gravitational Anomaly},''
  \href{http://dx.doi.org/10.1007/JHEP07(2012)128}{{\em JHEP} {\bfseries 1207}
  (2012) 128},
\href{http://arxiv.org/abs/1202.2469}{{\ttfamily arXiv:1202.2469 [hep-th]}}.

\bibitem{Azeyanagi:2013}
T.~Azeyanagi, R.~Loganayagam, G.~S. Ng, and M.~J. Rodriguez.
In preparation.

\bibitem{Kraus:2006wn}
P.~Kraus, ``{Lectures on black holes and the AdS(3) / CFT(2) correspondence},''
  {\em Lect.Notes Phys.} {\bfseries 755} (2008) 193--247,
\href{http://arxiv.org/abs/hep-th/0609074}{{\ttfamily arXiv:hep-th/0609074
  [hep-th]}}.

\bibitem{Alday:2009qq}
L.~F. Alday, F.~Benini, and Y.~Tachikawa, ``{Liouville/Toda central charges
  from M5-branes},''
  \href{http://dx.doi.org/10.1103/PhysRevLett.105.141601}{{\em Phys.Rev.Lett.}
  {\bfseries 105} (2010) 141601},
\href{http://arxiv.org/abs/0909.4776}{{\ttfamily arXiv:0909.4776 [hep-th]}}.

\bibitem{Bah:2012dg}
I.~Bah, C.~Beem, N.~Bobev, and B.~Wecht, ``{Four-Dimensional SCFTs from
  M5-Branes},'' \href{http://dx.doi.org/10.1007/JHEP06(2012)005}{{\em JHEP}
  {\bfseries 1206} (2012) 005},
\href{http://arxiv.org/abs/1203.0303}{{\ttfamily arXiv:1203.0303 [hep-th]}}.

\bibitem{Banerjee:2012cr}
N.~Banerjee, S.~Dutta, S.~Jain, R.~Loganayagam, and T.~Sharma, ``{Constraints
  on Anomalous Fluid in Arbitrary Dimensions},''
  \href{http://dx.doi.org/10.1007/JHEP03(2013)048}{{\em JHEP} {\bfseries 1303}
  (2013) 048},
\href{http://arxiv.org/abs/1206.6499}{{\ttfamily arXiv:1206.6499 [hep-th]}}.

\bibitem{Stone:2012ud}
M.~Stone, ``{Gravitational Anomalies and Thermal Hall effect in Topological
  Insulators},'' \href{http://dx.doi.org/10.1103/PhysRevB.85.184503}{{\em
  Phys.Rev.} {\bfseries B85} (2012) 184503},
\href{http://arxiv.org/abs/1201.4095}{{\ttfamily arXiv:1201.4095
  [cond-mat.mes-hall]}}.

\bibitem{Deser:2003vh}
S.~Deser and B.~Tekin, ``{Energy in topologically massive gravity},'' {\em
  Class.Quant.Grav.} {\bfseries 20} (2003) L259,
\href{http://arxiv.org/abs/gr-qc/0307073}{{\ttfamily arXiv:gr-qc/0307073
  [gr-qc]}}.

\bibitem{Solodukhin:2005ah}
S.~N. Solodukhin, ``{Holography with gravitational Chern-Simons},''
  \href{http://dx.doi.org/10.1103/PhysRevD.74.024015}{{\em Phys.Rev.}
  {\bfseries D74} (2006) 024015},
\href{http://arxiv.org/abs/hep-th/0509148}{{\ttfamily arXiv:hep-th/0509148
  [hep-th]}}.

\bibitem{Bouchareb:2007yx}
A.~Bouchareb and G.~Clement, ``{Black hole mass and angular momentum in
  topologically massive gravity},''
  \href{http://dx.doi.org/10.1088/0264-9381/24/22/018}{{\em Class.Quant.Grav.}
  {\bfseries 24} (2007) 5581--5594},
\href{http://arxiv.org/abs/0706.0263}{{\ttfamily arXiv:0706.0263 [gr-qc]}}.

\bibitem{Compere:2008cv}
G.~Compere and S.~Detournay, ``{Semi-classical central charge in topologically
  massive gravity},'' \href{http://dx.doi.org/10.1088/0264-9381/26/1/012001,
  10.1088/0264-9381/26/13/139801}{{\em Class.Quant.Grav.} {\bfseries 26} (2009)
  012001},
\href{http://arxiv.org/abs/0808.1911}{{\ttfamily arXiv:0808.1911 [hep-th]}}.

\bibitem{Skenderis:2009nt}
K.~Skenderis, M.~Taylor, and B.~C. van Rees, ``{Topologically Massive Gravity
  and the AdS/CFT Correspondence},''
  \href{http://dx.doi.org/10.1088/1126-6708/2009/09/045}{{\em JHEP} {\bfseries
  0909} (2009) 045},
\href{http://arxiv.org/abs/0906.4926}{{\ttfamily arXiv:0906.4926 [hep-th]}}.

\bibitem{Tachikawa:2006sz}
Y.~Tachikawa, ``{Black hole entropy in the presence of Chern-Simons terms},''
  \href{http://dx.doi.org/10.1088/0264-9381/24/3/014}{{\em Class.Quant.Grav.}
  {\bfseries 24} (2007) 737--744},
\href{http://arxiv.org/abs/hep-th/0611141}{{\ttfamily arXiv:hep-th/0611141
  [hep-th]}}.

\bibitem{Bonora:2011gz}
L.~Bonora, M.~Cvitan, P.~Dominis~Prester, S.~Pallua, and I.~Smolic,
  ``{Gravitational Chern-Simons Lagrangians and black hole entropy},''
  \href{http://dx.doi.org/10.1007/JHEP07(2011)085}{{\em JHEP} {\bfseries 1107}
  (2011) 085},
\href{http://arxiv.org/abs/1104.2523}{{\ttfamily arXiv:1104.2523 [hep-th]}}.

\bibitem{Loganayagam:2008is}
R.~Loganayagam, ``{Entropy Current in Conformal Hydrodynamics},''
  \href{http://dx.doi.org/10.1088/1126-6708/2008/05/087}{{\em JHEP} {\bfseries
  0805} (2008) 087},
\href{http://arxiv.org/abs/0801.3701}{{\ttfamily arXiv:0801.3701 [hep-th]}}.

\bibitem{Megias:2013joa}
E.~Megias and F.~Pena-Benitez, ``{Holographic Gravitational Anomaly in First
  and Second Order Hydrodynamics},''
\href{http://arxiv.org/abs/1304.5529}{{\ttfamily arXiv:1304.5529 [hep-th]}}.

\end{thebibliography}\endgroup

\end{document}